\documentclass{aa} 
\usepackage{graphicx}
\usepackage{url}
\usepackage[breaklinks=true]{hyperref}
\hypersetup{colorlinks=true,linkcolor=blue,citecolor=blue,filecolor=blue,urlcolor=blue}
\usepackage{twoopt}
\usepackage{natbib}
\bibpunct{(}{)}{;}{a}{}{,} 
\usepackage{amssymb}
\usepackage{float}
\usepackage{color}
\usepackage{epstopdf}
\usepackage{ctable}
\usepackage{multirow}
\usepackage[font=small]{caption}
\usepackage{multirow}
\usepackage{comment}
\usepackage{txfonts}
\usepackage{longtable}
\usepackage{esvect}

\usepackage{xcolor}

\newcommand{\kms}{${\rm km~s}^{-1}$}

\newcommand{\target}{A3528 cluster complex}

\begin{document}

\title{Cosmic dance in the Shapley Concentration Core}
\subtitle{II. The uGMRT-MeerKAT view of filaments in the brightest cluster galaxies and tailed radio galaxies in the A3528 cluster complex}

\author{
G.~Di Gennaro\inst{\ref{inst:hamb},\ref{inst:ira}}
\and
T.~Venturi\inst{\ref{inst:ira},\ref{inst:rhodes}}
\and
S.~Giacintucci\inst{\ref{inst:nlr}}
\and
M. Br\"uggen\inst{\ref{inst:hamb}}
\and
E. Bulbul\inst{\ref{inst:mpe}}
\and
J. Sanders\inst{\ref{inst:mpe}}
\and
 A. Liu\inst{\ref{inst:mpe}}
\and
X. Zhang\inst{\ref{inst:mpe}}
\and
K. Trehaeven\inst{\ref{inst:rhodes},\ref{inst:ira}}
\and
D. Dallacasa\inst{\ref{inst:oab},\ref{inst:ira}}
\and
P. Merluzzi\inst{\ref{inst:nap}}
\and
T. Pasini\inst{\ref{inst:ira}}
\and
S. Bardelli\inst{\ref{inst:oab}}
\and
G. Bernardi\inst{\ref{inst:ira},\ref{inst:rhodes},\ref{inst:SA}}
\and
O. Smirnov\inst{\ref{inst:rhodes},\ref{inst:SA},\ref{inst:ira}}
}

\institute{
{Hamburger Sternwarte, Universit\"at Hamburg, Gojenbergsweg 112, 21029 Hamburg, Germany}\label{inst:hamb}
\and
{INAF-Istituto di Radioastronomia, via Gobetti 101, I-40129 Bologna, Italy}\label{inst:ira}
\and
{Department of Physics and Electronics, Rhodes University, PO Box 94, Makhanda, 6140, South Africa.}\label{inst:rhodes}
\and
{Naval Research Laboratory, 4555 Overlook Avenue SW, Code 7213, Washington, DC 20375, USA}\label{inst:nlr}
\and
{MaxPlanck Institute for Extraterrestrial Physics, Giessenbachstrasse 1, 85748 Garching, Germany}\label{inst:mpe}
\and
{INAF-Osservatorio Astronomico di Capodimonte, salita Moiariello 16, I-80131 Napoli, Italy}\label{inst:nap}
\and
{Dipartimento  di  Fisica  e  Astronomia,  Università  di  Bologna,  via Gobetti 93/2, 40129 Bologna, Italy}\label{inst:oab}
\and
{South African Radio Astronomy Observatory, 2 Fir Street, Black River Park, Observatory, Cape Town 7925, South Africa}\label{inst:SA}
}

\date{Received 11 June 2024; Accepted 26 August 2024}

\abstract
{Superclusters represent the largest-scale environments where a number of galaxy clusters interact with each other through minor/major mergers and grow via accretion along cosmic filaments. We focus on the \target\ in the core of the Shapley Supercluster ($z\sim0.05$). This chain includes three clusters, A3528 (which is itself composed by two sub-clusters, namely A3528N and A3528S), A3532 and A3530, which present a mildly active dynamical state.}
{We study how minor mergers affect the evolution of radio galaxies and whether they are able to re-accelerate relativistic electrons in the intracluster medium (ICM). To unveil these energetic processes, extremely sensitive radio observations are required.}
{We used observations from the upgraded Giant Metrewave Radio Telescope (uGMRT) in Band 3 (250--500 MHz), Band 4 (550--900 MHz) and Band 5 (1000--1460 MHz) and MeerKAT L-band (900--1670 MHz) to obtain images and spectral index maps over a wide frequency band and spatial resolutions of the \target. We reach noise levels of $\sim\rm 30-50~\mu Jy\,beam^{-1}$, $\sim\rm 10-20~\mu Jy\,beam^{-1}$, $\sim\rm 20~\mu Jy\,beam^{-1}$ and $\sim\rm 10~\mu Jy\,beam^{-1}$, for uGMRT Band 3, Band 4, Band 5 and MeerKAT L-band, respectively. 
For a comparison with the thermal ICM emission, we also use data from the Spectrum Roentgen Gamma (SRG)/eROSITA X-ray telescope.}
{We detect faint diffuse radio emission associated with the radio galaxies in the \target. Particularly, the brightest cluster galaxies (BCGs) in A3528S and A3532 show filaments of diffuse radio emission which extend for $\sim200-400$ kpc out of the radio galaxy. The spectral index of these filamentary structures is extremely steep and almost constant ($\alpha\sim-2,-2.5$).
{Contrary to the radio tails in A3528N, the spectral properties of these radio filaments are not consistent with standard models of plasma ageing. We also detect roundish diffuse radio emission around the BCG in A3528S which could be classified as a radio mini-halo. The radio tail in this cluster appears longer  that in earlier detections, being $\sim 300$ kpc long at all frequencies.}}
{We linked the presence of extended radio emission in the form of filaments and threads in the \target\ with the effect of minor mergers. This is also reinforced by the increasing X-ray fluctuations in correspondence with the radio extended emission in A3528S. Despite the less energy involved, our findings support the hypothesis that these events can re-energise plasma originating from radio galaxies.}

\keywords{radio continuum: galaxies -- galaxies: clusters: general -- galaxies: clusters: individual: A3528 - galaxies: clusters: individual: A3532}

\titlerunning{The uGMRT-MeerKAT view of the A3528 cluster complex}
\authorrunning{G. Di Gennaro et al.}
\maketitle

%

\section{Introduction}

Galaxy clusters form during the last stages of the evolutionary ladder, by accretion of gas from the cosmic web and/or by collision with other clusters/galaxy groups \citep{press+schecter74,springel+06}. Overdense regions such as {\it superclusters} \citep[e.g.][]{liu+24} represent unique places where to study systems at different evolutionary stages, from groups of galaxies, fairly massive clusters with ongoing accretion activity, and smaller systems located in filaments in the regions between the main clusters. All this makes such environments the perfect test bed where to investigate the effects of mildly interacting clusters on the cluster galaxy population.

Although most energy is released during major mergers (i.e., those involving at least two massive clusters) which then can result in Mpc-scale diffuse radio emission such as  giant radio halos and radio relics \citep[][]{brunetti+jones14,vanweeren+19}, the most common interactions between large-scale structures involve groups, where less energy is involved \citep{cassano+11}, or collisions with a large impact parameter, often causing gas sloshing in the central core \citep{markevitch+vikhlinin07,zuhone+15}. These events are usually referred as minor mergers. 
Major and minor mergers are thought to play an important role in shaping the morphology of cluster radio galaxies. Compared to field radio galaxies, cluster radio galaxies can display asymmetric structures, caused by the bending of the jets by ram pressure. These are then classified as head tail (HT), narrow-angle tail (NAT) and wide-angle tail (WAT) based on the opening angle between the two jets \citep{miley80,o'dea+85}. The direction and the opening angle of the tails can be used to retrieve information on the cluster dynamics and local density, although projection effects can play an important role and are often not easy to assess.
Normally, the radio emission of the radio galaxy lobes fades away in a time-scale of a few tens of millions year, as the plasma loses energy due to synchrotron and inverse Compton (IC) emission \citep{jaffe+perola73}. The rate of energy losses mostly depends on the magnetic field ($B$) and on the energy of the relativistic particles ($E\propto\gamma_L$, with $\gamma_L$ the Lorentz factor). This results in steep spectra ($\alpha\lesssim-1$, being $S_\nu\propto\nu^\alpha$), and sometimes these sources become undetectable at $\sim$GHz frequencies. This can change if relativistic electrons are re-supplied by the active galactic nucleus (AGN) or if the relativistic electrons are re-energiesed by shocks, turbulence or adiabatic compression \citep[e.g.][]{ensslin+gopal-krishna01,ensslin+bruggen02,markevitch+vikhlinin07}. In these cases, some boost in surface brightness and flattening of spectral index is expected \citep[e.g.][]{degasperin+17,vanweeren+21}.

As the relativistic electrons are ejected from the inner AGN, either in single or multiple activity events, bubbles of plasma can rise and be displaced in the intracluster medium  \citep[ICM; e.g.][]{zuhone+13,zuhone+21,vazza+21}.
Recent observations by \cite{brienza+21} have shown that radio bubbles of past AGN activity can be ejected in the ICM up to the virial radius, therefore providing relativistic electron seeds for the formation of the large-scale diffuse emission. Additionally, sources of steep-spectrum fossil plasma on the few tens kpc scales \citep[i.e. radio phoenices with $\alpha\lesssim-2$, see][]{mandal+20} are also thought to be related with past AGN radio activities \citep[e.g.][]{slee+01}. Finally, filaments of diffuse radio emissions in the surroundings of radio galaxies are also visible \citep{giacintucci+22,rudnick+22}. These radio sources are usually extremely collimated and show signs of polarisation, which indicates coherent magnetic field structure \citep[e.g.][]{condon+21}.

Minor mergers are also thought to be able to re-accelerate particles on scales of hundreds of kiloparsec \citep{brunetti+jones14}. Several observational studies have shown that clusters in this dynamical stage host diffuse radio emission in the form of mini radio halos \citep[e.g.][and references therein]{gitti+07,savini+19,vanweeren+19}. These have sizes between 50 kpc and $0.2R_{500}$ \citep{giacintucci+17}, where $R_{500}$ is the radius within which the cluster mean total density is 500 times the critical density of the universe. Mini radio halos are characterised by steep spectra (i.e. $\alpha\lesssim-1$). 
Although not directly associated with the emission of the inner brightest cluster galaxies \citep[BCGs; see][]{richard-laferriere+20}, it is thought that these could contribute to the powering of mini-halos, either by providing seed cosmic-ray electrons (CRe) that are re-accelerated by turbulence in the wake of sloshing  \citep{mazzotta+giacintucci08,zuhone+15} or by providing cosmic-ray protons (CRp) which then hadronically collide with ICM protons \citep{pfommer+ensslin04}.

In this context, the new generation of radio telescopes - such as the LOw Frequency ARray \citep[LOFAR;][]{vanhaarlem+13}, upgraded Giant Metrewave RadioTelescope \citep[uGMRT;][]{gupta+17}, Australia SKA Pathfinder \citep[ASKAP;][]{hotan+21} and MeerKAT \citep{jonas16} - have been crucial in revealing an increasing number of clusters hosting these sources with low-surface brightness radio emission, thanks to their impressive high sensitivity and high resolution. It has become clear that a hidden population of (mildly and highly) relativistic electrons  exists in the ICM.

\section{The Shapley Concentration Core and the \target}
\label{sec:clusters}
The Shapley Supercluster \citep{shapley30} hosts a variety of systems in the early stages of their assembly and/or are already merged systems. It is  identified as the richest and most massive system in the eRASS1 supercluster catalogue \citep{liu+24}, with an estimated total mass of $2.6\pm0.5 \times 10^{16} M_{\odot}$ and 45 member clusters.
Particularly, the inner part of the supercluster, i.e. the Shapley Concentration Core, is dominated by two main chains of Abell clusters: one at its centre, namely the A3558 complex, which includes two additional clusters (A3556 and A3562) and two galaxy groups (SC1327-312 and SC1329-313); and one to the north-west of the centre, namely the A3828 complex, which includes also the A3530 and A3532 clusters. They both have a similar projected linear size of about 7.5 Mpc to the east-west and in the north-south \citep{jones+09}, at the mean redshift of 0.048 and 0.054, respectively \citep{bardelli+98b,bardelli+01}. 

\begin{table}
\caption{Clusters information.}
\vspace{-5mm}
\begin{center}
\resizebox{0.5\textwidth}{!}{
\begin{tabular}{lccc}
\hline
\hline
Cluster name & Abell 3528\,N & Abell 3528\,S & Abell 3532 \\
Redshift ($z$) & 0.0542 & 0.0544 & 0.0554 \\
Right Ascension (RA) & $\rm 12^h54^m23.5^s$ & $\rm 12^h54^m41.4^s$ & $\rm 12^h57^m16.9^s$ \\
Declination (Dec) & $-29^\circ01^\prime22''$ & $-29^\circ13^\prime24''$ & $-30^\circ22^\prime37''$  \\
Mass ($M_{500}~[\times10^{14}~{\rm M_\odot}$]) & 1.5 & 2.1 & 2.3\\
Size ($R_{500}~[\rm Mpc]$) & 0.8 & 0.9 & 0.9 \\ 
Luminosity ($L_{500}~[\times 10^{44}~\rm erg\,s^{-1}] $) & 0.7 & 1.1 & 1.1 \\
Cosmological scale ($\rm kpc/''$) & 1.054 & 1.058 & 1.076 \\
\hline
\end{tabular}
}
\end{center}
\vspace{-5mm}
\tablefoot{Clusters mass, radius and luminosity are taken from \cite{piffaretti+11}.}
\label{tab:cluster}
\end{table}

While multi-wavelength (from X-ray to radio) observations over the past decades have revealed the dynamically disturbed nature of the A3558 complex \citep[and especially the region between A3558 and A3562, see][]{markevitch+vikhlinin97,bardelli+98a,bardelli+98b,ettori+00,venturi+00, venturi+03,finoguenov+04,giacintucci+04, giacintucci+05,rossetti+07,ghizzardi+10,merluzzi+15, venturi+17,haines+18,higuchi+20,giacintucci+22} with the recent detection of a diffuse radio bridge between A3562 and SC1329-313 and other pieces of diffuse emission \citep{venturi+22}, the individual clusters in the \target\ are in an overall relaxed state \citep{bardelli+01}. XMM-Newton X-ray observations of A3528 show the presence of two sub-cores, namely A3528N and A3528S \citep{gastaldello+03}, both characterised by temperature and abundance profiles typical of cool-core clusters. In-between these two sub-cores, \cite{gastaldello+03} also detected a bridge of thermal emission which, together with the asymmetric thermal distribution of the two sub-clusters, pointed to an off-axis post-merger scenario in A3528, where the closest core encounter happened about 1–2 Gyrs ago. On the other side, {\it Chandra} observations revealed the non cool-core nature, and the possible presence of cavities in the distribution of the thermal gas \citep{lakhchaura+13} of A3532 which is thought to be in the pre-merger phase with A3530.
Finally, no diffuse radio emission on cluster scales (e.g. radio halos and radio relics) was found in previous radio observations \citep{venturi+01,digennaro+18b}.

\begin{table*}[h!]
\caption{Radio observation details.}
\vspace{-5mm}
\begin{center}
\resizebox{0.8\textwidth}{!}{
\begin{tabular}{lcccc}
\hline
\hline
Array & \multicolumn{3}{c}{uGMRT} & MeerKAT\\
\cmidrule(lr){2-4}\cmidrule(lr){5-5} 
Project & 32\_024 & 43\_012 & 32\_024 &  AO1 \\
Band & Band 3 & Band 4 & Band 5 & L-band \\
Frequency range ($\Delta\nu$ [MHz]) & 250--500 & 550--900 & 1000--1460 & 900--1670 \\
Central Frequency ($\nu$ [MHz]) & 410 & 650/700 & 1260  & 1284 \\
No. Channels & 4096 & 2048 & 4096 & 4096 \\
Observation Length [h] & 8 & 24 & 8 & 8 \\ 
Time on each target [h] & 3 & 6 & 2 & 2 \\
Observation Date [dd-mm-yyyy] & 08-05-2017  & 16-12-2022 & 10-05-2017 & 15-07-2019 \\ 
                        &  & 19-12-2022 \\
                        &  & 30-03-2023 \\
Minimum {\it uv} coverage [$\lambda$] & 112 & 200 & 200 & -  \\
Largest angular scale [$'$] & 10 & 23 & 5 & 27.5 \\
Primary beam [$'$] & 75 & 38 & 23 & 68 \\
\hline
\end{tabular}
}
\end{center}
\vspace{-5mm}
\tablefoot{The observing time is equally split per cluster, and includes the time on the calibrators. The pointing centres for the uGMRT Band 4 and 5 and MeerKAT L-band observations are the same as the cluster centres listed in Tab. \ref{tab:cluster}, as well as Band 3 observations on A3532. Observations of A3528 with the uGMRT in Band 3 were pointed to the mid-point between the two sub-cores, i.e. $\rm RA_{J2000}=12^h54^m29.87^s, ~DEC_{J2000}=-29^\circ06^\prime59.58''$. The MeerKAT observations are obtained by the authors in response to the January 2019 MeerKAT Announcement of Opportunity (AO1).
}\label{tab:obs}
\end{table*}

In the \target\ the radio galaxy population appears to be substantially different from A3558. In particular, the BCGs in the former chain is dynamically younger, more powerful than the latter, and with possible hints of re-started activity \citep[i.e. the BCGs in A3528S and A3532, see][]{digennaro+18b}. 
Finally, the presence of several tailed radio galaxies in the two sub-cores in A3528, despite their cool-cores, suggests some form of interaction between A3528\,N and A3528\,S \citep{digennaro+18b}.

All these pieces of evidence point to a more complex dynamical state of the \target\ that needs to be better explored. The ultra-relativistic electrons associated with the radio galaxies -- both tails and BCGs -- possibly interact with the surrounding environment through sloshing and feedback. In this paper, we focus on the total intensity emission of the three radio-loud clusters of the \target, namely A3528\,N, A3528\,S and A3532 (Table \ref{tab:cluster}), to shed light on such interactions and on the duty cycle of the BCGs, as well as the dynamics within A3528. 
We combined radio observations with uGMRT Band 3 (250--500 MHz), Band 4 (550--900 MHz), Band 5 (1000--1460 MHz) and MeerKAT L-band (900--1670 MHz). Moreover, we compared the radio footprint of the three clusters with the thermal ICM emission taken from the eROSITA Western All-Sky Survey \citep[eRASS;][Sanders et al. in prep.]{bulbul+24,merloni+24,kluge+24}.

The paper is organised as follows: in Sect. \ref{sec:obs} we describe our observations and the radio data calibration. In Sects. \ref{sec:results} and \ref{sec:disc} we describe and discuss the results of our analysis. We conclude with a summary in Sect. \ref{sec:concl}.
Throughout the paper, we assume a standard $\Lambda$CDM cosmology, with $H_0 = 70$ km s$^{-1}$ Mpc$^{-1}$, $\Omega_m = 0.3$ and $\Omega_\Lambda = 0.7$. This translates to a luminosity distance of $D_{\rm L}=240.8$ Mpc, and a scale of 1.051 kpc/\arcsec at the mean redshift of the cluster complex, i.e. $z = 0.054$.

\section{Observations and data reduction}
\label{sec:obs}
In this section, we describe the both the radio (Tab.~\ref{tab:obs}) and the eROSITA X-ray observations and data reduction strategy for the \target. 

\subsection{Radio data}
The radio observations cover a frequency range between 250 and 1670 MHz and were analysed separately. 

\subsubsection{uGMRT}

The \target\ was observed with the upgraded Giant Metrewave Radio Telescope (uGMRT) in Band 3 (250--500 MHz; project 32\_024, PI: T. Venturi), Band 4 (550--900 MHz; project 43\_012, PI: G. Di Gennaro) and Band 5 (1000-1460 MHz; project 32\_024, PI: T. Venturi). In each band, the single cluster was observed for a total of 8 hours, including the time on the calibrators. For observations in Band 3, given the large field of view, only one pointing was chosen to cover A3528 ($\rm RA_{J2000}=12^h54^m29.87^s, ~DEC_{J2000}=-29^\circ06^\prime59.58''$). For observations in Band 4 and 5, the pointing centres are the same as the cluster coordinates listed in Tab.~\ref{tab:cluster}.
Data were recorded in 2048 (Band 4) and 4096 (Band 3 and 5) frequency channels,  with an integration time of 5.3 (Band 4) and 4 (Band 3 and 5) seconds in full Stokes mode. For all the observations, we used 3C286 as primary calibrator.

We processed the data in a similar way for the three targets. First, we run the Source Peeling and Atmospheric Modeling \citep[\texttt{SPAM}\footnote{\url{http://www.intema.nl/doku.php?id=huibintema:spam:pipeline}},][]{intema+09} pipeline on the narrow-band (i.e. {\it gsb}) data. These provide the models for the wide-band (i.e. {\it gwb}) phase calibration. For the Band 4 observation of A3532, we also included the sky model from Band 3 in the {\it gsb} data, as there were no sufficient {\it peeled} sources to perform the direction-dependent calibration and modelling of ionospheric effects\footnote{\url{https://www.intema.nl/doku.php?id=huibintema:spam:faq}}.
For the same reason and due to the small field of view (i.e. $23^\prime$), the same Band 3 model was applied to all the targets also in Band 5, together with the option \texttt{allow\_selfcal\_skip=True}. The latter allows to continue the calibration in \texttt{SPAM} in case there are too few peeling sources. It does not affect our final calibration as the effect of the ionosphere is negligible at these high frequencies. The wide-band data were then split into four/six sub-bands which were then processed separately. 

Finally, for each target the sub-bands were imaged together at the same central frequency with \texttt{WSClean~v3.4} \citep{offringa+14}, using \texttt{Briggs} \citep{briggs95} weighting and \texttt{robust=-0.5} (Band 3 and 4) and \texttt{robust=0} (Band 5), and \texttt{multiscale} deconvolution with scales of $[1,4,8,16]\times{\rm pixelscale}$ (being pixelscale the size of the pixel in arcseconds (i.e. 2\arcsec, 1\arcsec\ and 0.5\arcsec, for Band 3, 4 and 5 respectively). Additionally, we created the highest-resolution Band 4 images using {\tt robust=-2}.
For Band 3 and 4, we produced images at different resolutions, tapering the {\it uv} plane at 10\arcsec, 15\arcsec\ and 30\arcsec, with the same {\tt multiscale} deconvolution setting (and pixelscale of $2''$). To emphasise the possible presence of diffuse emission, we removed the contribution of compact sources: first we created a clean model only including compact sources, by excluding from the visibilities all the emission with linear sizes $\geq100$ kpc at the clusters redshift (i.e. $\sim2000\lambda$); then we subtracted this model and re-imaged the data, at the same resolution as mentioned above.
For all the images, the final reference frequencies are 410 MHz for Band 3, 700 (650) MHz for Band 4 (A3528\,N) and 1260 MHz for Band 5 (see Table \ref{tab:obs}). All the images were corrected for the primary beam attenuation \citep[i.e. using {\tt EveryBeam} in {\tt WSClean;}][]{offringa+14}, which were used for the flux density estimation.
The systematic uncertainties due to residual amplitude errors are set to 8\% for the observations in Band 3 and 5\% for the observations in Band 4 and 5 \citep{chandra+04}.

\subsubsection{MeerKAT L-band}

MeerKAT observations of the \target\ were performed in L-band (900--1670 MHz) on 15 July 2019 for a total of 8 hours, including the calibrators (e.g. J1311-222 as phase calibrator, and PKS1934-63 and PKS0408-65 as flux and bandpass calibrators). Data were recorded in 4096 frequency channels, with a width of $\sim210$ kHz each. The three clusters in the \target\ were imaged in the same observation, with the pointing centre on A3528S at $\rm RA_{J2000}=12^h55^m00.0^s, ~DEC_{J2000}=-29^\circ40^\prime00.0''$.

For each target, standard data reduction was performed using the Containerized Automated Radio Astronomy Calibration (\texttt{CARACal}) pipeline\footnote{\url{https://github.com/caracal-pipeline/caracal}} \citep{jozsa+20a, jozsa+20b}, which includes different flagging steps \citep[e.g. \texttt{tricolour}, see][for radio frequency interferences, and auto-correlations, shadowed antennas and channels/{\it spw} in \texttt{CASA}]{hugo+22} and delay, bandpass and gain calibrations on PKS1934-63. Solutions were then transferred to the target, channel-averaged by a factor of four, and followed by a second round of flagging. Several rounds of phase and amplitude+phase self-calibrations were done by means of \texttt{WSClean}, using \texttt{Briggs} \citep{briggs95} weighting and \texttt{robust=0}, \texttt{join-channel} deconvolution and 4th-order polynomial fitting options, and using the \texttt{CubiCal} package \citep{kenyon+18}. 

As for the uGMRT observations, final images were done with {\tt WSClean~v3.4} \citep{offringa+14}, using \texttt{Briggs} \citep{briggs95} weighting and \texttt{robust=-0.5} and \texttt{multiscale} deconvolution with scales of $[1,4,8,16]\times{\rm pixelscale}$, where the pixel scale equal to 1\arcsec. Similarly to the uGMRT imaging, low-resolution images at 10\arcsec, 15\arcsec\ and 30\arcsec\ were also produced, as well as those without compact sources to emphasise the diffuse emission. Primary beam correction was also applied, and those corrected images were used for the flux density estimation.
The final uncertainty is set to 5\% \citep{knowles+22}.

\subsection{eROSITA}

The Shapley Supercluster region -- and therefore the \target\ -- has been scanned by the Spectrum Roentgen Gamma (SRG)/eROSITA X-ray telescope \citep{predehl+21}, over the five cycles of the All-Sky Survey (eRASS:5), between December 2019 to February 2022. Data was processed with eROSITA Science Analysis Software System \citep[eSASS,][]{brunner+22} pipeline version 020, which has the improved boresight correction, detector noise suppression, and subpixel position computation over the c010 version which is used for the first All-Sky Survey release \citep{merloni+24}. The data from telescope modules (TMs) 1, 2, 3, 4, 5, 6, 7 \citep{predehl+21} are co-added to generate the image of both clusters. The images were created using the eSASS tool {\tt evtool},  from event files, which were corrected for good time intervals, dead times, and bad pixels.

The final images (Fig. \ref{fig:xray}) were created in the 0.2--2.3 keV energy band. Moreover, to emphasise the dynamics of the three clusters, we also produce Gaussian Gradient Magnitude (GGM) filtered images \citep{sanders+16} with $\sigma=10$ pixels.

\begin{table}
\caption{Radio imaging details.}
\vspace{-5mm}
\begin{center}
\resizebox{0.5\textwidth}{!}{
\begin{tabular}{ccccc}
\hline
\hline
Central frequency & Resolution & $uv$-taper & Robust & Map noise \\
$\nu$ [MHz] & $\Theta$ [$\arcsec\times\arcsec,^\circ$] & [\arcsec] & & $\rm \sigma_{rms}~[\mu Jy~beam^{-1}]$ \\
\hline
\multicolumn{5}{c}{A3528} \\
410  & $7.2\times4.7$, 7 & None & $-0.5$ & 40.9 \\
    & $11.5\times9.7$, 22 & 10 & $-0.5$ & 72.7 \\
    & $17.3\times14.0$, 32 & 15 & $-0.5$ & 122.8 \\
    & $36.0\times26.6$, 33 & 30 & $-0.5$ & 316.6 \\
\hline
\multicolumn{5}{c}{A3528N} \\
650 & $4.3\times2.5$, 12 & None & $-2$ & 13.6 \\
    & $4.7\times2.9$, 8 & None & $-0.5$ & 11.1 \\
    & $11.2\times 9.0$, 26 & 10 & $-0.5$ & 33.7 \\
    & $17.3\times13.0$, 29 & 15 & $-0.5$ & 59.7 \\
    & $45.7\times24.8$, 22 & 30 & $-0.5$ & 148.1 \\
1260  & $2.9\times2.2$, 5 & None & $0$ & 17.8 \\
1284 & $5.0\times4.7$, 166  & None & $-0.5$ & 8.9  \\
     & $10.1\times10.1$, 131 & 10 & $-0.5$ & 7.4 \\
     & $15.1\times15.1$, 130 & 15 & $-0.5$ & 12.7  \\
     & $29.9\times29.9$, 126 & 30 & $-0.5$ & 35.5 \\
\hline
\multicolumn{5}{c}{A3528S} \\
700 & $4.0\times2.2$, 12 & None & $-2$ & 52.7 \\
    & $4.3\times2.9$, 1 & None & $-0.5$ & 17.9 \\
    & $11.5\times9.4$, 32 & 10 & $-0.5$ & 37.1 \\
    & $18.7\times13.3$, 32 & 15 & $-0.5$ & 67.8 \\
    & $38.9\times27.0$, 19 & 30 & $-0.5$ & 170.0 \\
1260 & $2.9\times2.2$, 18 & None & $0$ & 16.9 \\
1284 & $5.0\times4.7$, 166  & None & $-0.5$ & 8.9  \\
     & $10.1\times10.1$, 131 & 10 & $-0.5$ & 7.4 \\
     & $15.1\times15.1$, 130 & 15 & $-0.5$ & 12.7  \\
     & $29.9\times29.9$, 126 & 30 & $-0.5$ & 35.5 \\
\hline
\multicolumn{5}{c}{A3532} \\
410  & $7.6\times4.7$, 9 & None & $-0.5$ & 34.6 \\
    & $12.2\times9.4$, 21 & 10 & $-0.5$ & 58.2 \\
    & $18.0\times13.7$, 31 & 15 & $-0.5$ & 94.8 \\
    & $37.8\times26.3$, 35 & 30 & $-0.5$ & 241.3 \\
700 & $3.6\times2.2$, 10 & None & $-2$ & 39.1 \\
    & $4.3\times2.9$, 176 & None & $-0.5$ & 15.5 \\
    & $11.9\times9.4$, 15 & 10 & $-0.5$ & 27.3 \\
    & $17.3\times14.0$, 20 & 15 & $-0.5$ & 45.0 \\
    & $36.0\times27.7$, 26 & 30 & $-0.5$ & 108.8 \\
1260    & $2.9\times2.2$, 1 & None & $0$ & 20.7 \\
1284 & $5.0\times4.7$, 164  & None & $-0.5$ & 9.2  \\
     & $10.1\times10.1$, 140 & 10 & $-0.5$ & 6.8 \\
     & $15.1\times15.1$, 133 & 15 & $-0.5$ & 11.2  \\
     & $29.9\times29.9$, 128 & 30 & $-0.5$ & 30.1 \\
\hline
\end{tabular}
}
\end{center}
\vspace{-5mm}
\label{tab:images}
\end{table}

\begin{figure*}
\centering
\includegraphics[width=\textwidth]{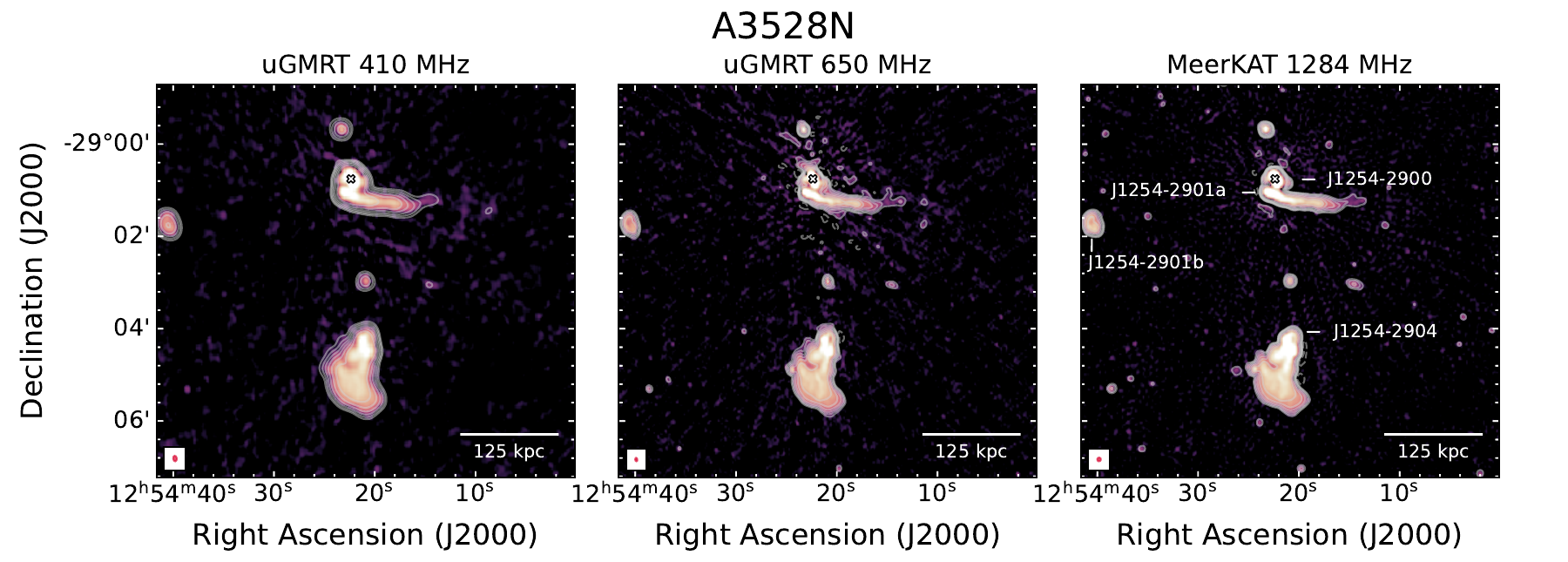}\\
\includegraphics[width=\textwidth]{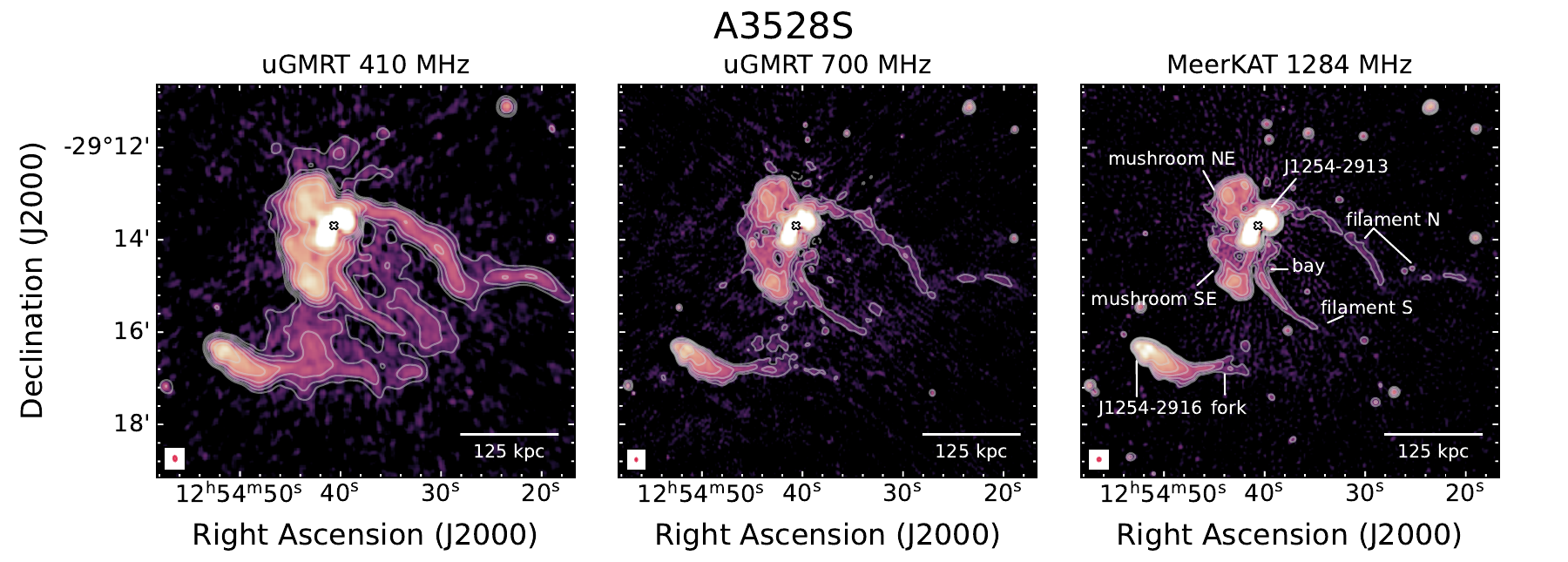}\\
\includegraphics[width=\textwidth]{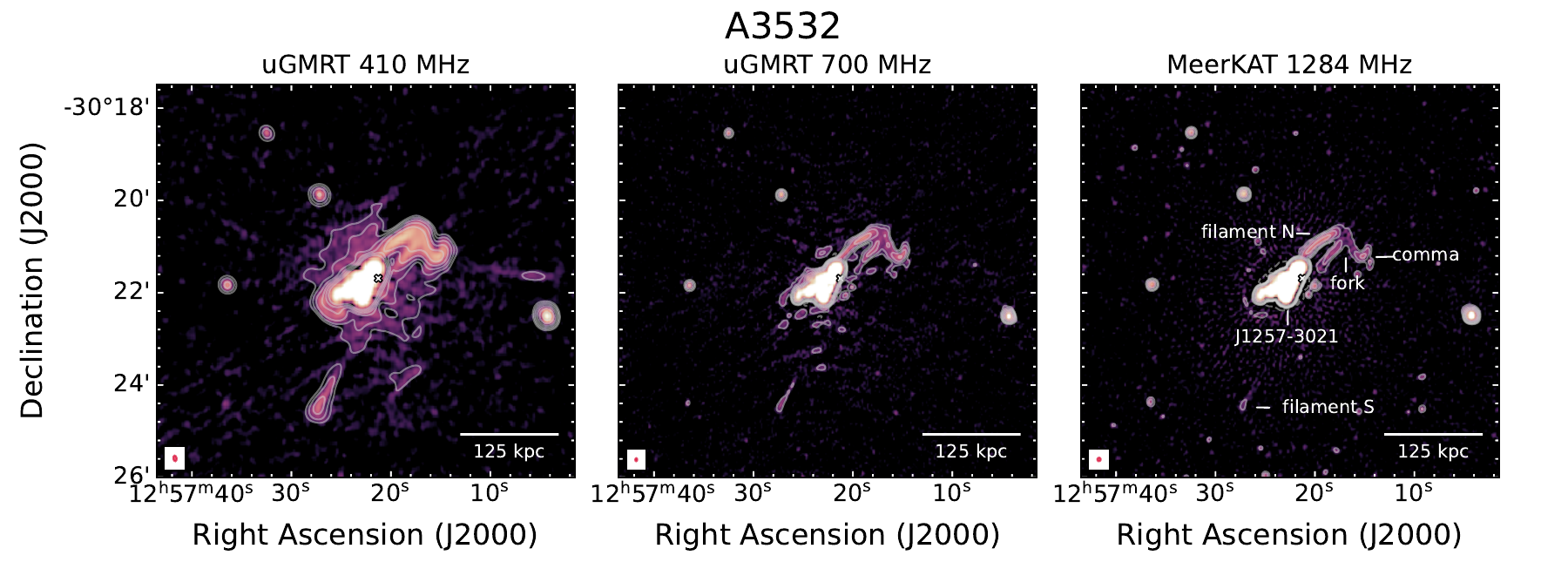}
\caption{Full-resolution images of the central region of A3528N (top row), A3528S (central row) and A3532 (bottom row). From left to right: uGMRT Band 3 (410 MHz), uGMRT Band 4 (650/700 MHz) and MeerKAT (1284 MHz). Radio contours are displayed at $3\sigma_{\rm rms,\nu}\times[-1,1,2,4,8,16,32]$ levels (when present, negative contours are in dashed line), where $\sigma_{\rm rms,\nu}$ the map noise at the different frequencies (see Tab. \ref{tab:images}). Labels on the radio sources/features are given in the right panel. The beams are displayed in the bottom left corner of each panel (see Tab.~\ref{tab:images}). The white `$\times$' marker shows the location of the X-ray peak.}
\label{fig:images_alltelescope_fullres}
\end{figure*}

\begin{figure*}
\centering
\includegraphics[width=0.48\textwidth]{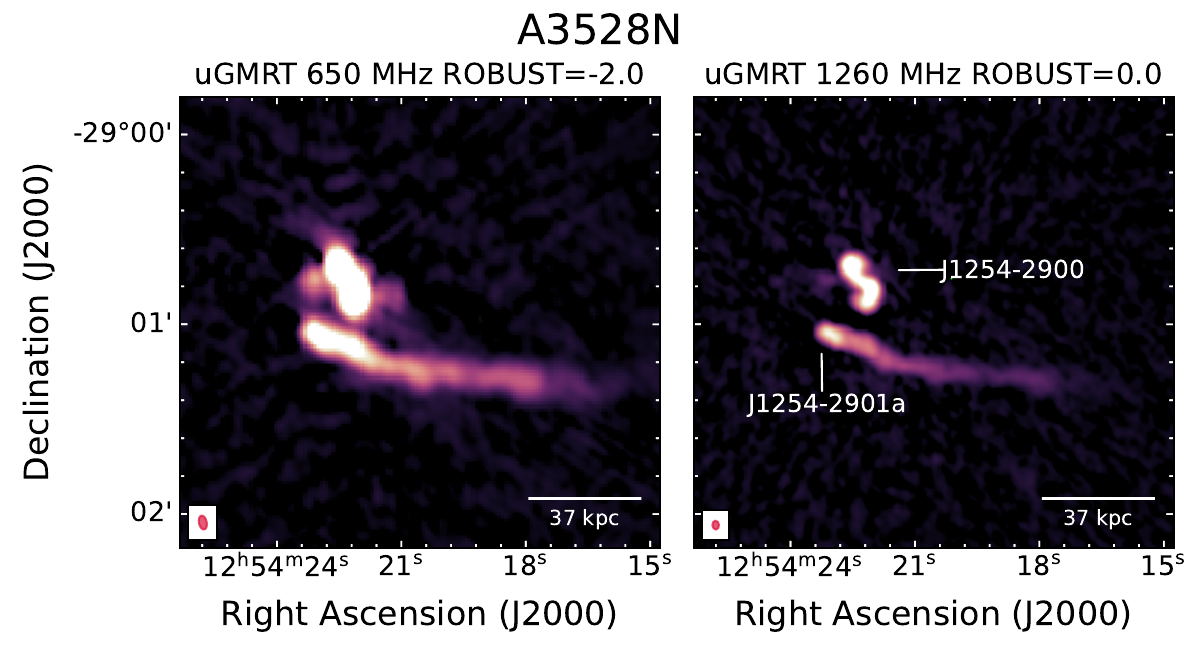}
\includegraphics[width=0.48\textwidth]{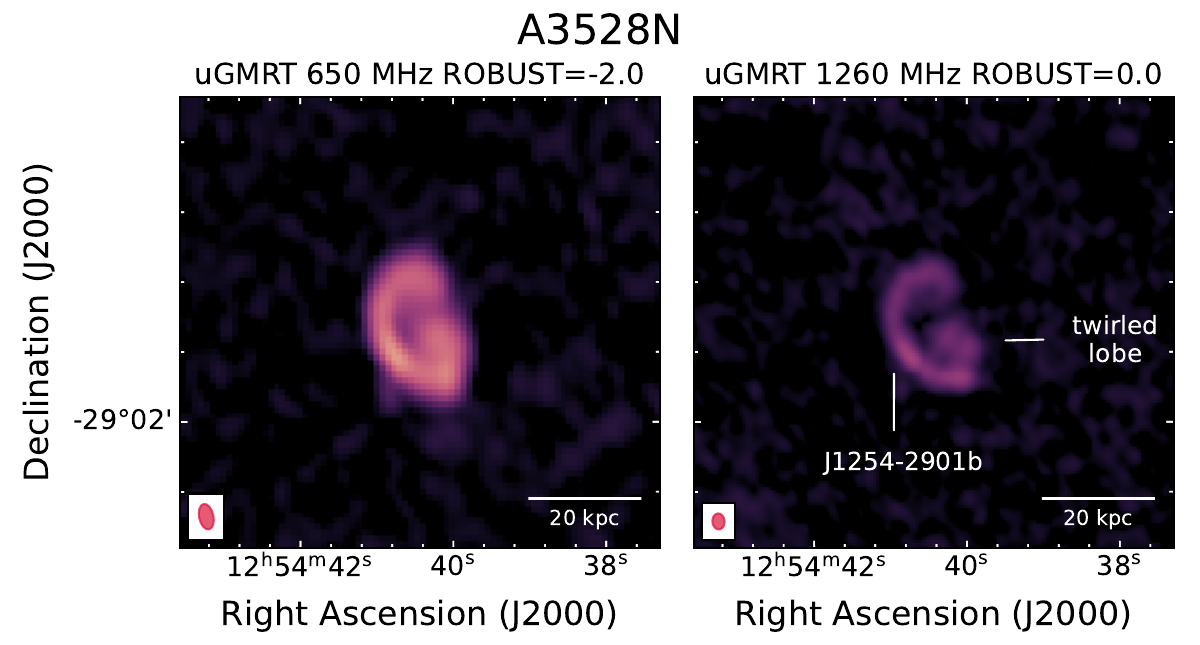}
\includegraphics[width=0.48\textwidth]{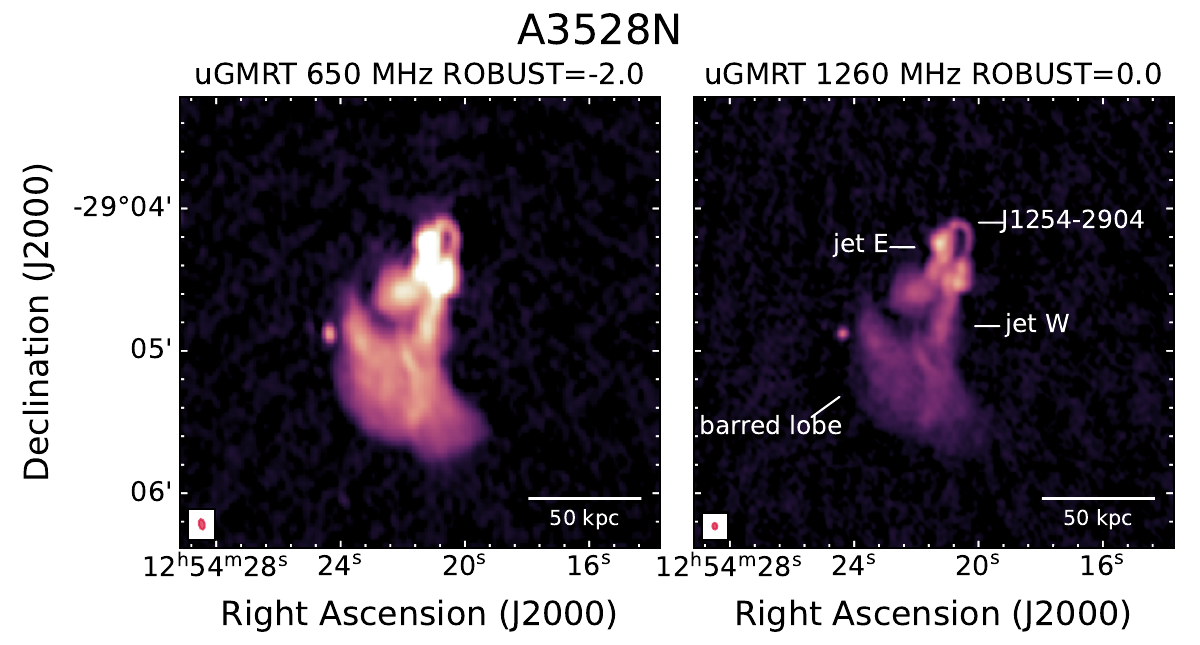}
\includegraphics[width=0.48\textwidth]{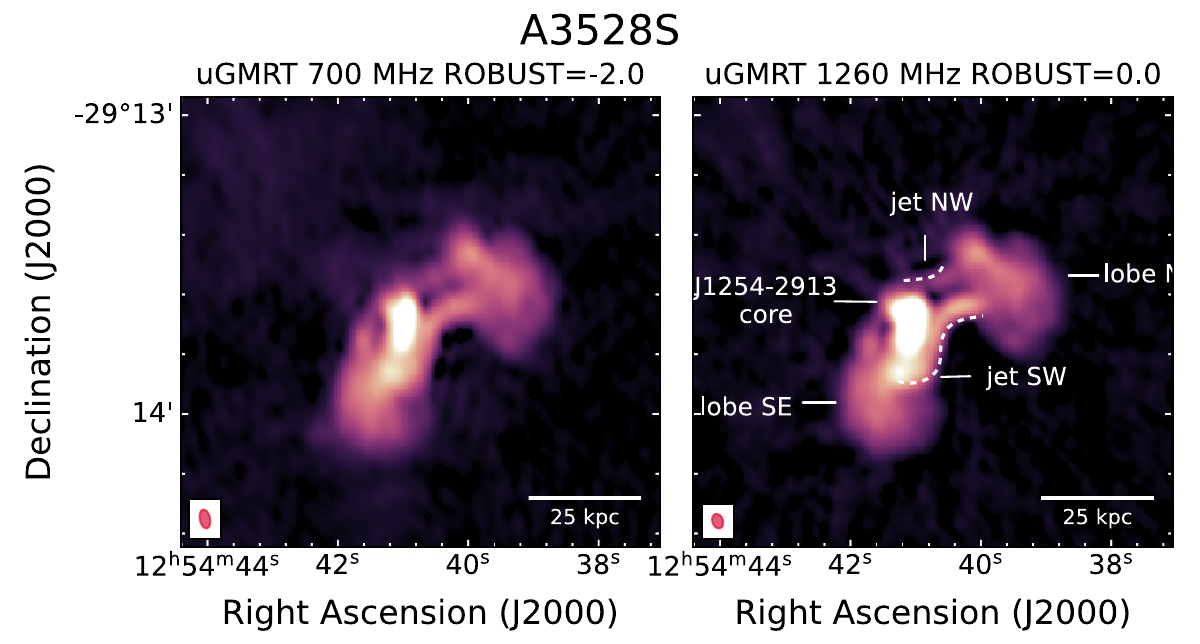}
\includegraphics[width=0.48\textwidth]{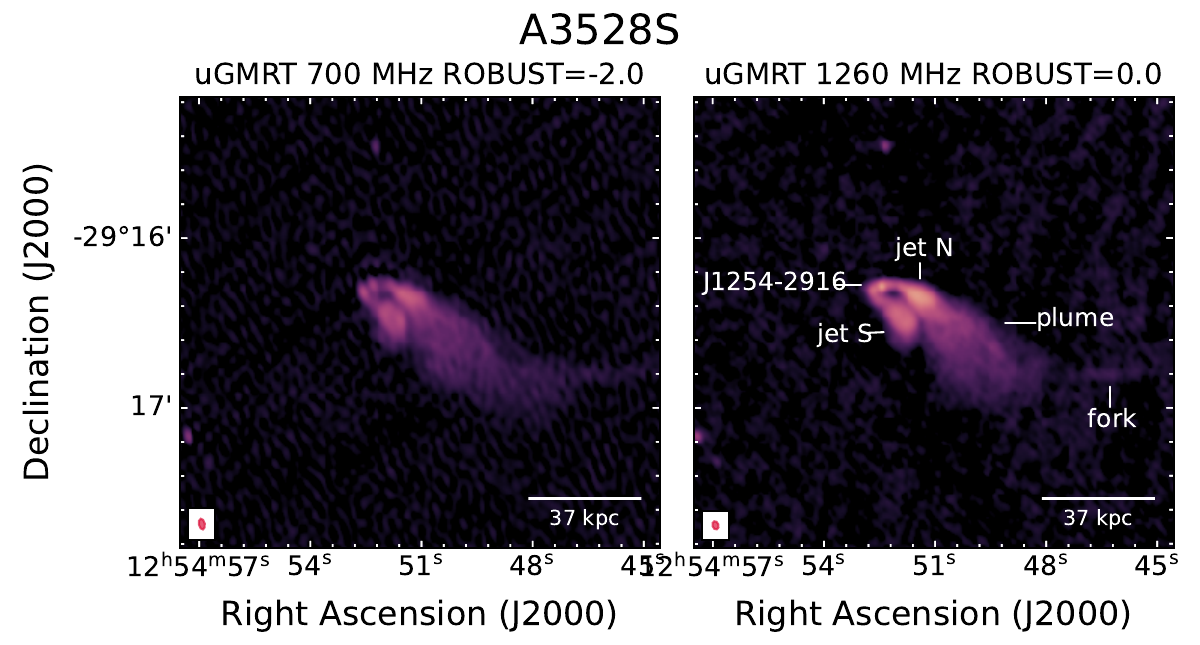}
\includegraphics[width=0.48\textwidth]{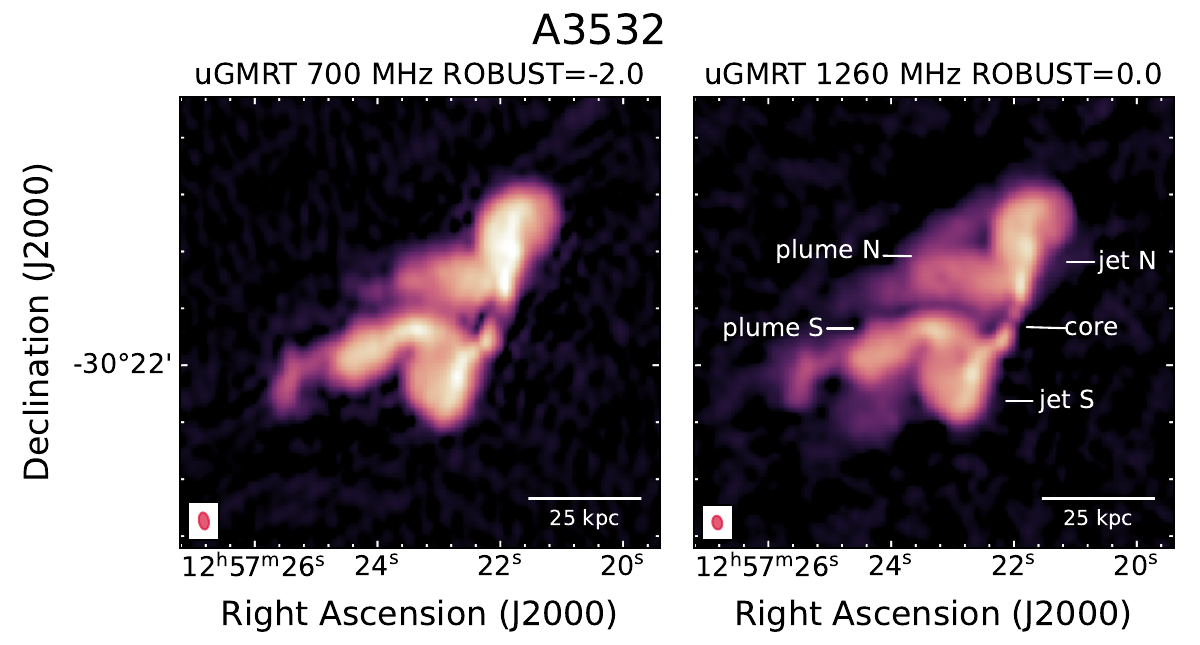}
\caption{High-resolution images of the cluster radio sources. Top left: J1254-2900 and J1254-2901a (A3528N); Top right: J1254-2901b (A3528N); Central left: J1254-2904 (A3528N); Central right: J1254-2913 (A3528S); Bottom left: J1254-2916 (A3528S); Bottom right: J1257-3021 (A3532). For each radio galaxy, the uGMRT Band 4 ({\tt robust=-2}; left) and the uGMRT Band 5 ({\tt robust=0}; right) are displayed. The beam is displayed in the bottom left corner of each panel (see Tab. \ref{tab:images}).}
\label{fig:band5}
\end{figure*}

\section{Results}
\label{sec:results}

If compared with the previously-published observations \citep[i.e.][]{venturi+01,digennaro+18b}, the new high-fidelity uGMRT and MeerKAT images (Figures \ref{fig:images_alltelescope_fullres} and \ref{fig:band5}, and Table \ref{tab:images}) reveal a plethora of new diffuse radio sources as well as new details of the radio galaxies in the three clusters of the \target. Here, we describe the radio emission for each cluster and the comparison with the X-ray thermal emission. Finally, we also measure the integrated flux densities of the radio sources.

\subsection{Radio morphology}

\subsubsection{A3528N}

The BCG in A3528N, namely J1254-2900, does not show any additional diffuse radio emission, neither at full nor low resolutions (top panels in Fig.~\ref{fig:images_alltelescope_fullres} and Appendix \ref{apx:lowres}, Fig.~\ref{fig:images_A3528Nalltelescope_allres}, respectively). At the uGMRT Band 5 high resolution ($\Theta_{1260}=2.9''\times2.2''$, see top left panels in Fig. \ref{fig:band5}), J1254-2900 presents the same `S' shape that was detected at 8.4 GHz \citep{digennaro+18b}. This source is also totally embedded in the inner core of the ICM, where the gas density is higher (left panel in Fig.~\ref{fig:xray}).

South of the BCG, we find two tailed radio galaxies, i.e. the head tail (HT) J1254-2901a and the narrow-angle tail (NAT) J1254-2904 (see top panels in Fig. \ref{fig:images_alltelescope_fullres}). The head and the core of J1254-2901a remain unresolved also at the uGMRT Band 5 high resolutions (top left panels in Fig.~\ref{fig:band5}). The length of the tail reaches $\sim130$ kpc in the 410 MHz low resolution image. On the other hand, J1254-2904 shows an asymmetric length of the two jets, with the western one ending in a sharp, barred lobe that extends in the north-east/south-west direction for $\sim80$ kpc. At the highest resolution (see central left panel in Fig.~\ref{fig:band5}), we distinguish the two unresolved jets which expand first into two small bubbles and then turning into a plume (eastern jet) and the barred lobe (western jet). Not accounting for projection, we measure a length of $\sim45$ kpc and $\sim95$ kpc for the eastern and western (including the sharp lobe) jet, respectively. Interestingly, the barred lobe of J1254-2094 does not coincide with any particular thermal discontinuity, also in the GGM-filtered X-ray contours (left panel in Fig. \ref{fig:xray}).
Finally, \cite{digennaro+18b} noticed an additional WAT radio galaxy, J1254-2901b, east of the cluster centre, at the edge of the sloshed core (left panel in Fig.~\ref{fig:xray}). For the first time at low frequencies, we clearly detect the `C' shape tail that was only visible at 8.4 GHz, with the southern jet turning into a twirled lobe (see top right panel in Fig.~\ref{fig:band5}).

\begin{figure*}
\centering
\includegraphics[width=\textwidth]{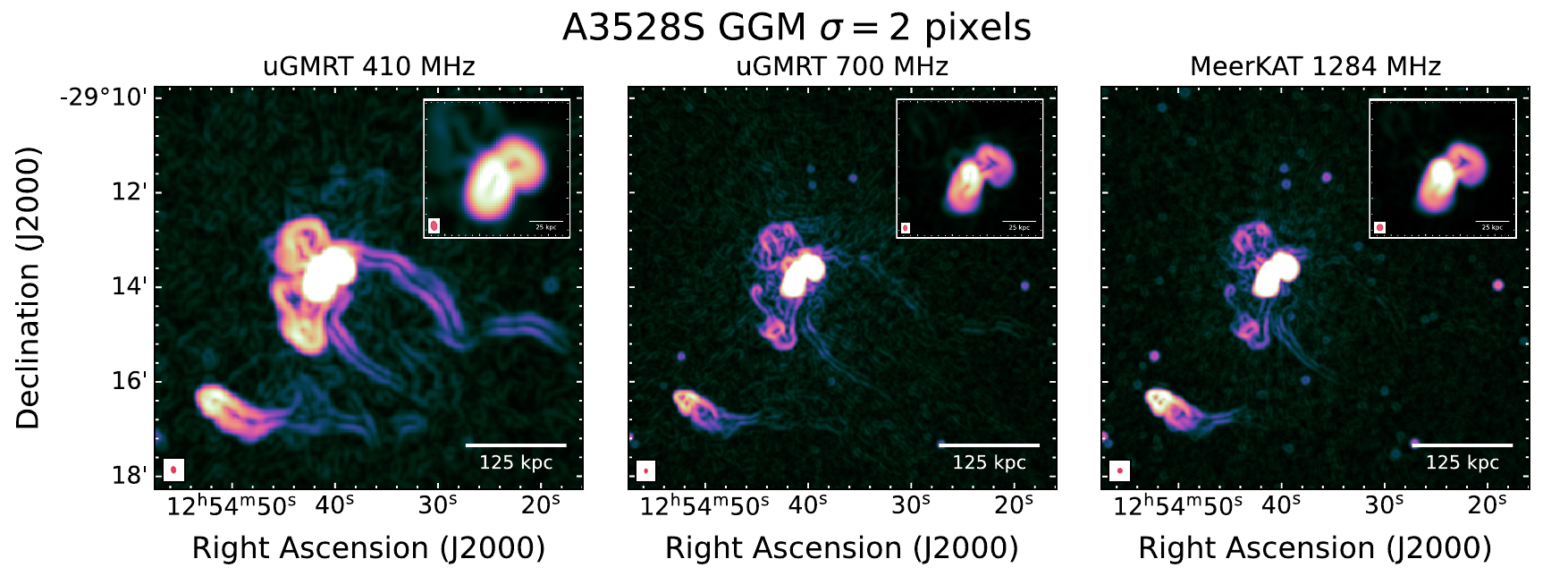}
\caption{Unsharped Gaussian Gradient Magnitude \citep[GGM;][]{sanders+16} filtered images with $\sigma=2$ pixels of the central region of A3528S. From left to right: uGMRT Band 3 (410 MHz), uGMRT Band 4 (700 MHz) and MeerKAT L-band (1284 MHz), at full resolution (see Tab. \ref{tab:images}). On the top right corner of each panel, a zoom-in on the WAT, at the same resolution and frequency. }\label{fig:a3528s_ggm}
\end{figure*}

\subsubsection{A3528S}
The most remarkable diffuse radio source in the \target\ is associated with the wide-angle tail (WAT) BCG in A3528S, i.e. J1254-2913 (middle panels in Figs.~\ref{fig:images_alltelescope_fullres} and \ref{fig:images_A3528Salltelescope_allres}). We confirm the detection of two ``{\it mushroom}''-shaped radio sources, north-east and south-east of the radio galaxy, which were previously marginally detected only at the lowest frequency \citep[i.e. 235 MHz,][]{digennaro+18b}. The new detailed radio images reveal not only the mushroom-shaped sources at all three frequencies (i.e. up to 1.28 GHz), but also their filamentary structure (Fig. \ref{fig:a3528s_ggm}).  This emission resemble, on smaller scales, that detected in the Virgo cluster \citep{owen+00}.
Moreover, we detect two thin radio filaments pointing towards south-west (see central panels in Fig.~\ref{fig:images_alltelescope_fullres}). The northern filament extends further west than the southern filament, breaking into a `V' shape which is mostly visible at 410 MHz (e.g., uGMRT Band 3) for a total length of $\sim390$ kpc. On the other hand, the southern filament seems to start directly from the core of the WAT, bending of $90^\circ$ in the south-east direction (``bay'', see right middle panel Fig.~\ref{fig:images_alltelescope_fullres}) and then turning again westward for $\sim200$ kpc. Between the two radio filaments, additional diffuse radio emission is  present at low resolutions (see Appendix \ref{apx:lowres}, Fig.~\ref{fig:images_A3528Salltelescope_allres}). 
The high-resolution, high-frequency image (i.e. uGMRT Band 5, see central right panels in Fig.~\ref{fig:band5}) of the BCG reveals a more complex morphology than the lower-frequency observations: two unresolved jets are ejected from the north and south of the inner core, with the southern one showing a `S' shape towards the west (see dashed line in the central right panels in Fig.~\ref{fig:band5}), where both turn into a lobe.

South of the complex, diffuse radio emission of J1254-2913, we find the NAT radio galaxy J1254-2916. The extent of the tail is substantially longer than found in previous observations \citep[$\sim80$ kpc][]{digennaro+18b}, at all frequencies. In particular, we clearly see the tail changing direction from south-west to west, $\sim80$ kpc from the core. It then bifurcates at a distance of $\sim160$ kpc of the core and reaches the terminal point of the southern filament in J1254-2913, for a total projected length of $\sim280$ kpc. At the high resolution of the uGMRT Band 5 observations (bottom left panels in Fig.~\ref{fig:band5}), it appears that this long tail is connected only to the northern jet of J1254-2916.

\subsubsection{A3532}

Similarly to the BCG in A3528S, that in A3532, i.e. J1257-3021, shows a filament in the north at all frequencies. At $\sim80$ kpc from its start, the filament bifurcates, generating two additional filaments, the western BCG turning into a comma (bottom panels in Fig.~\ref{fig:images_alltelescope_fullres}). Interestingly, the 410 MHz full-resolution image reveals an additional piece of diffuse emission south of the BCG with no clear optical counterpart. This becomes more visible in the low-resolution images, also at higher frequencies, where the two sources connect at the $3\sigma_{\rm rms}$ level (see Appendix \ref{apx:lowres}, Fig.~\ref{fig:images_A3532alltelescope_allres}). If  associated with J1257-3021, this kind of emission could be caused by an earlier episode of activity of the BCG. Moreover, hints of presence of diffuse radio emission around the BCG (e.g. mini radio halo) are visible at 410 MHz (at both full and low resolutions), although there is no clear detection at higher frequencies.

The uGMRT Band 5 image (bottom right panels in Fig.~\ref{fig:band5}) shows impressive details of the BCG. Two jets are ejected in the north-west/south-east direction, turning into two plumes that get distorted and dragged towards the east. This suggests a motion of the radio galaxies in the ICM towards the west, in the direction of the radio-quiet member of the \target\ (i.e. A3530), as also suggested in \cite{digennaro+18b}.

\begin{figure*}
\centering
\includegraphics[width=0.33\textwidth]{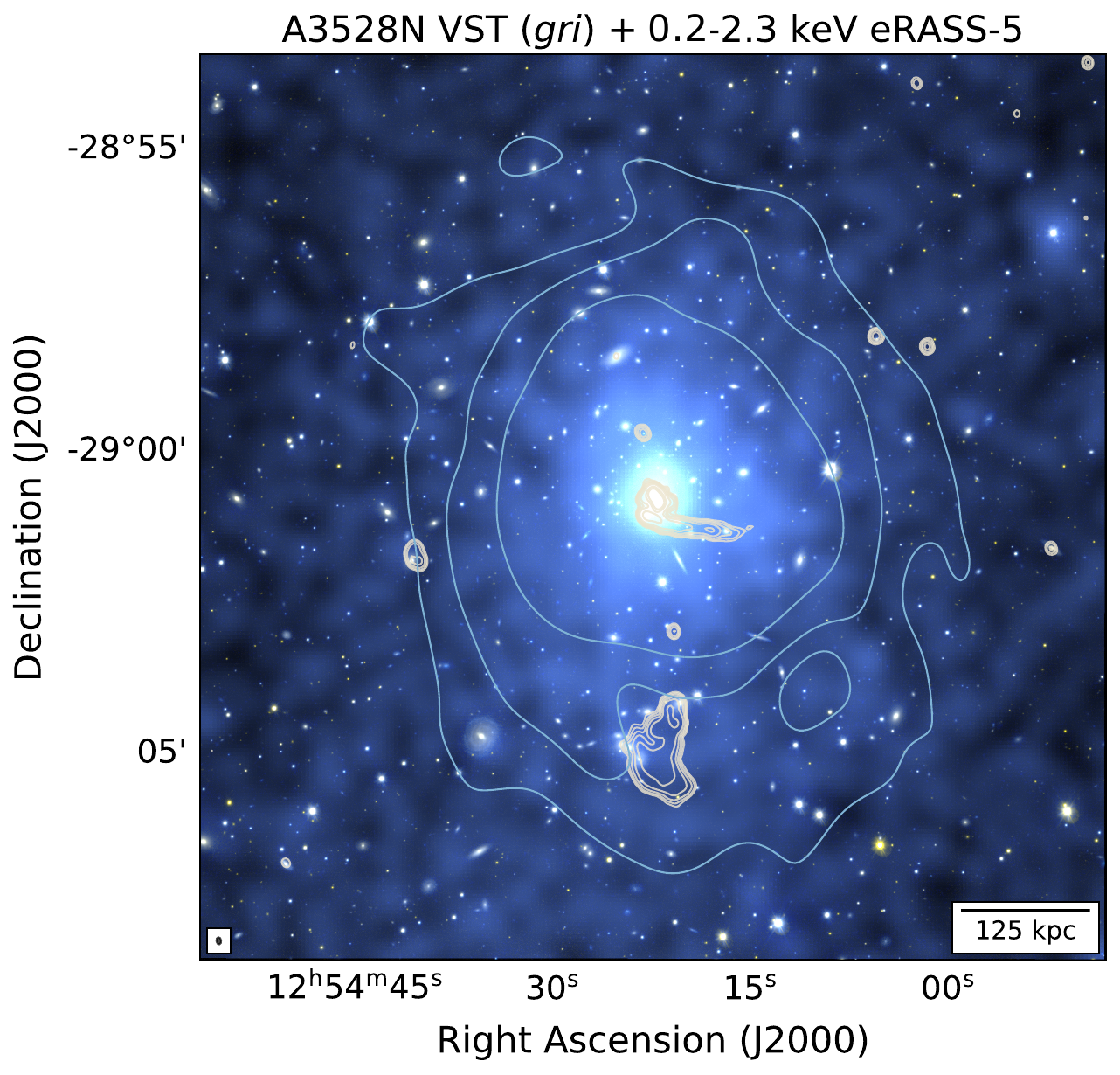}
\includegraphics[width=0.33\textwidth]{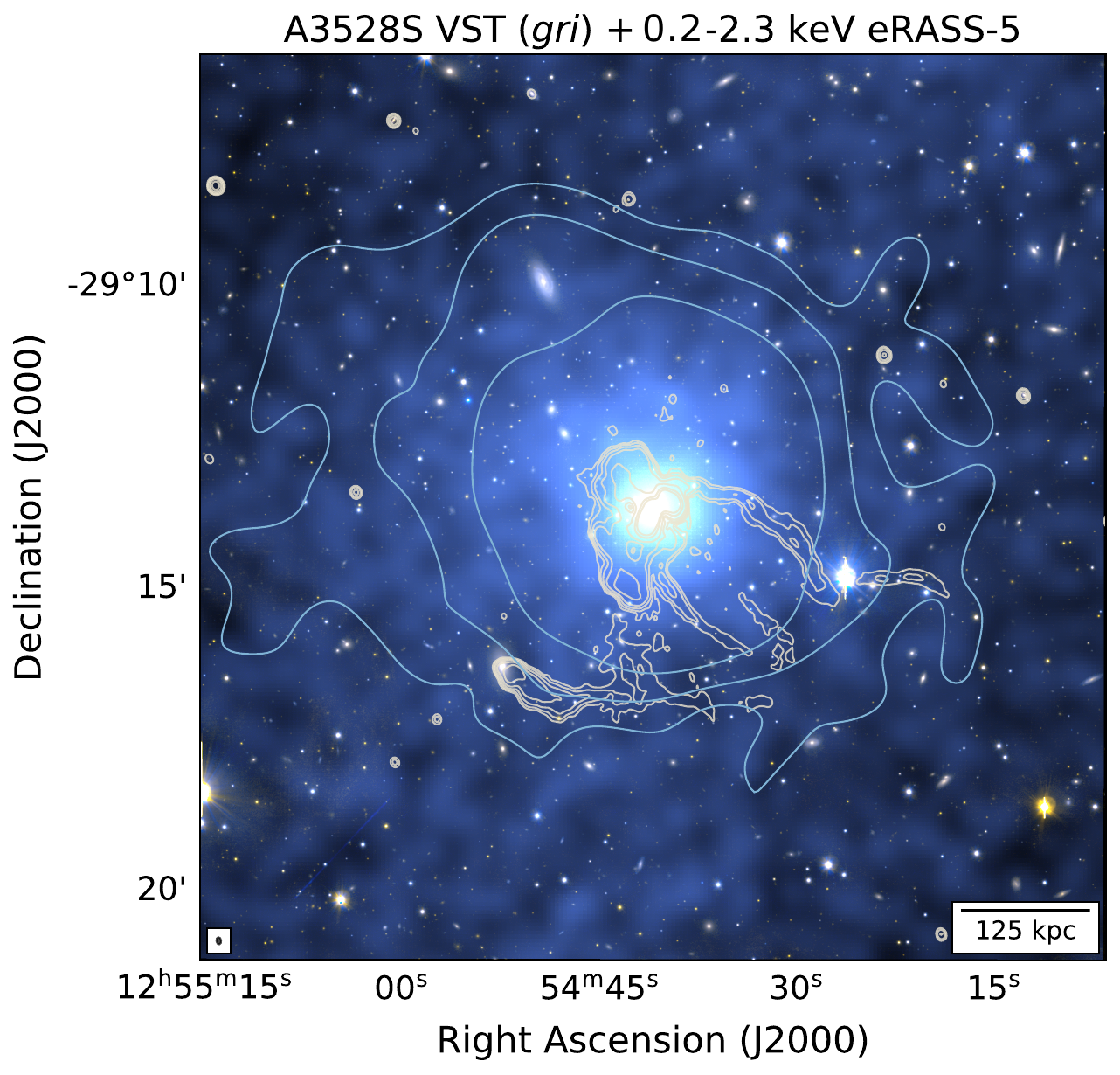}
\includegraphics[width=0.33\textwidth]{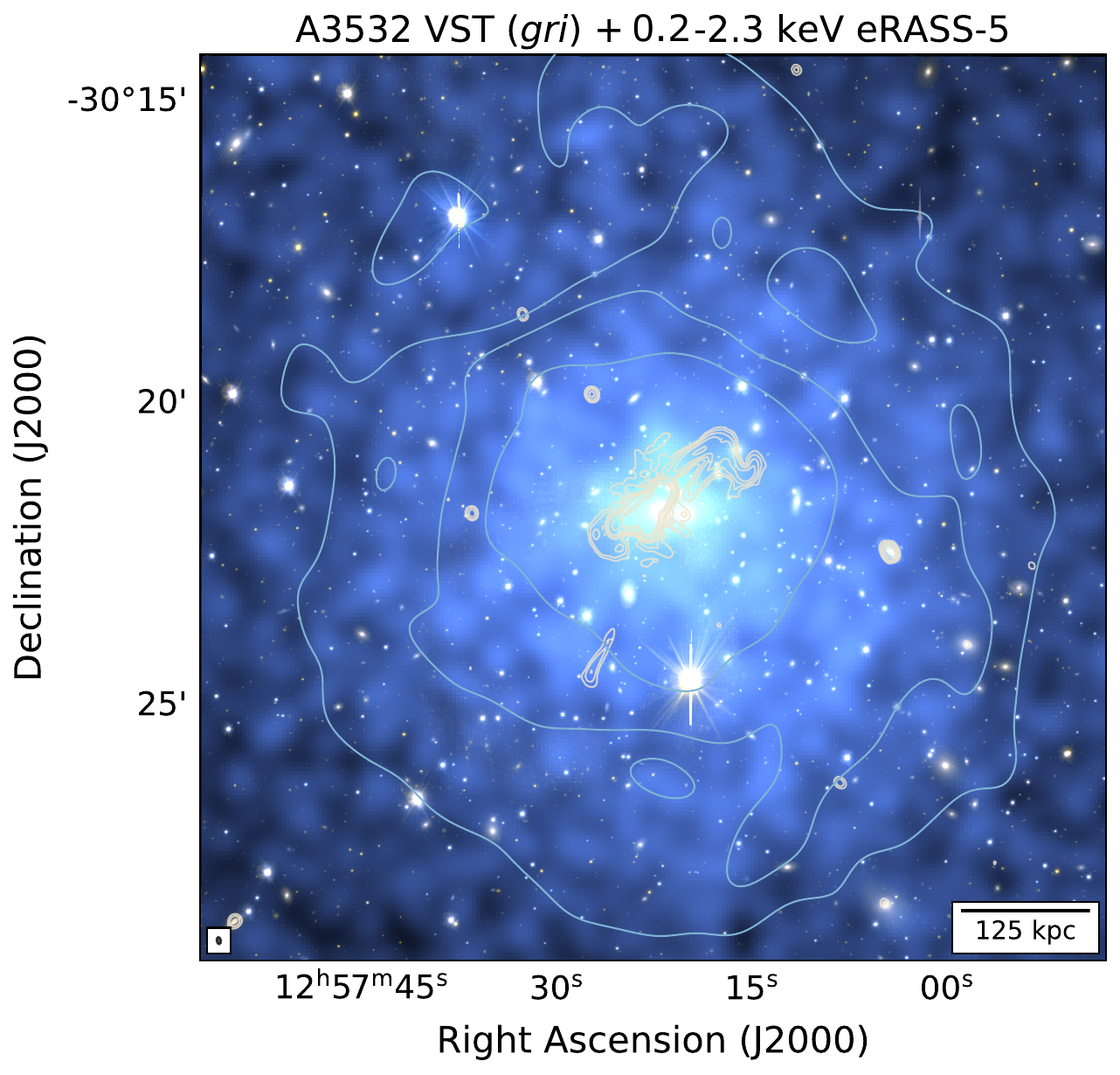}\\
\includegraphics[width=0.33\textwidth]{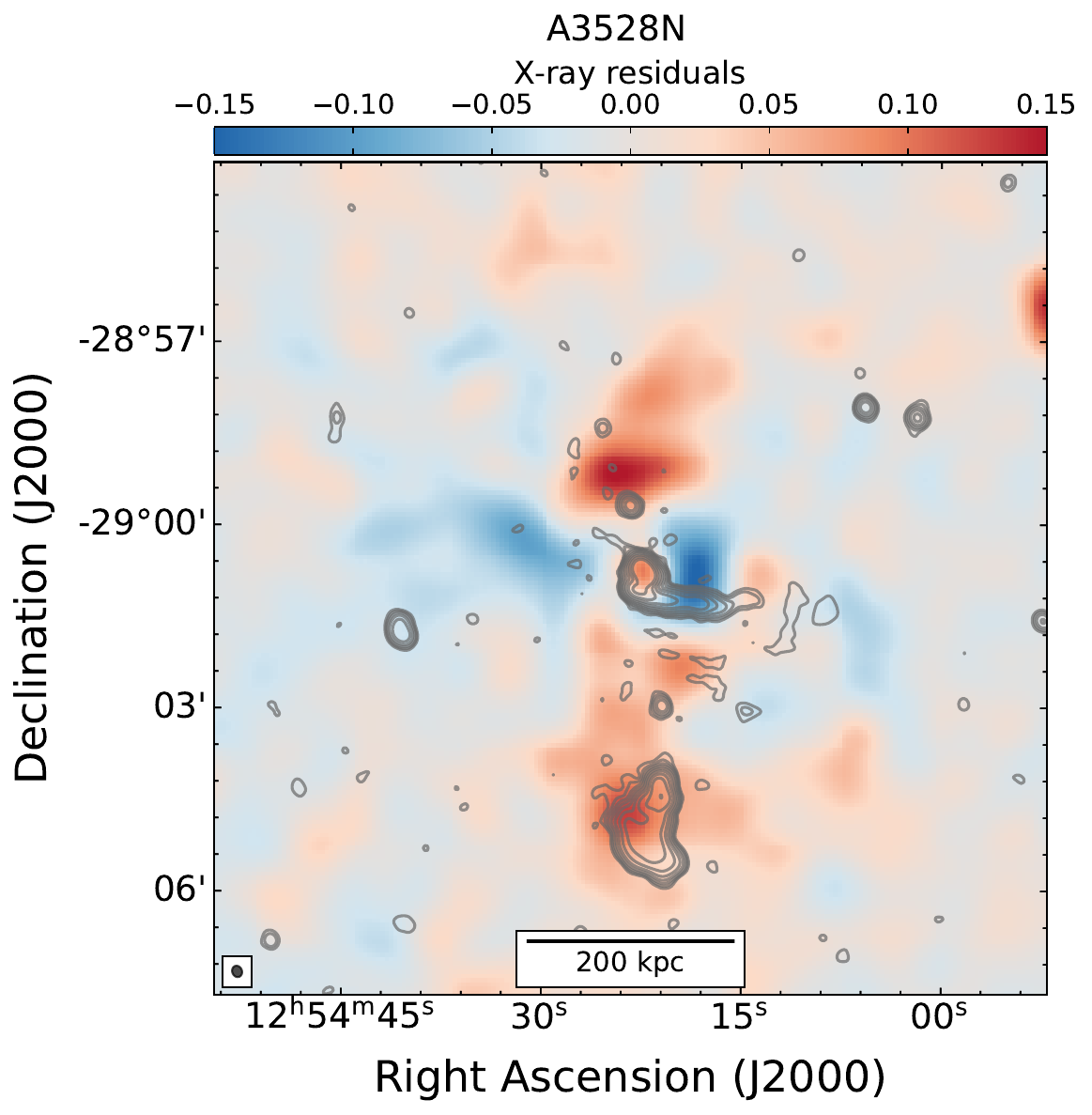}
\includegraphics[width=0.33\textwidth]{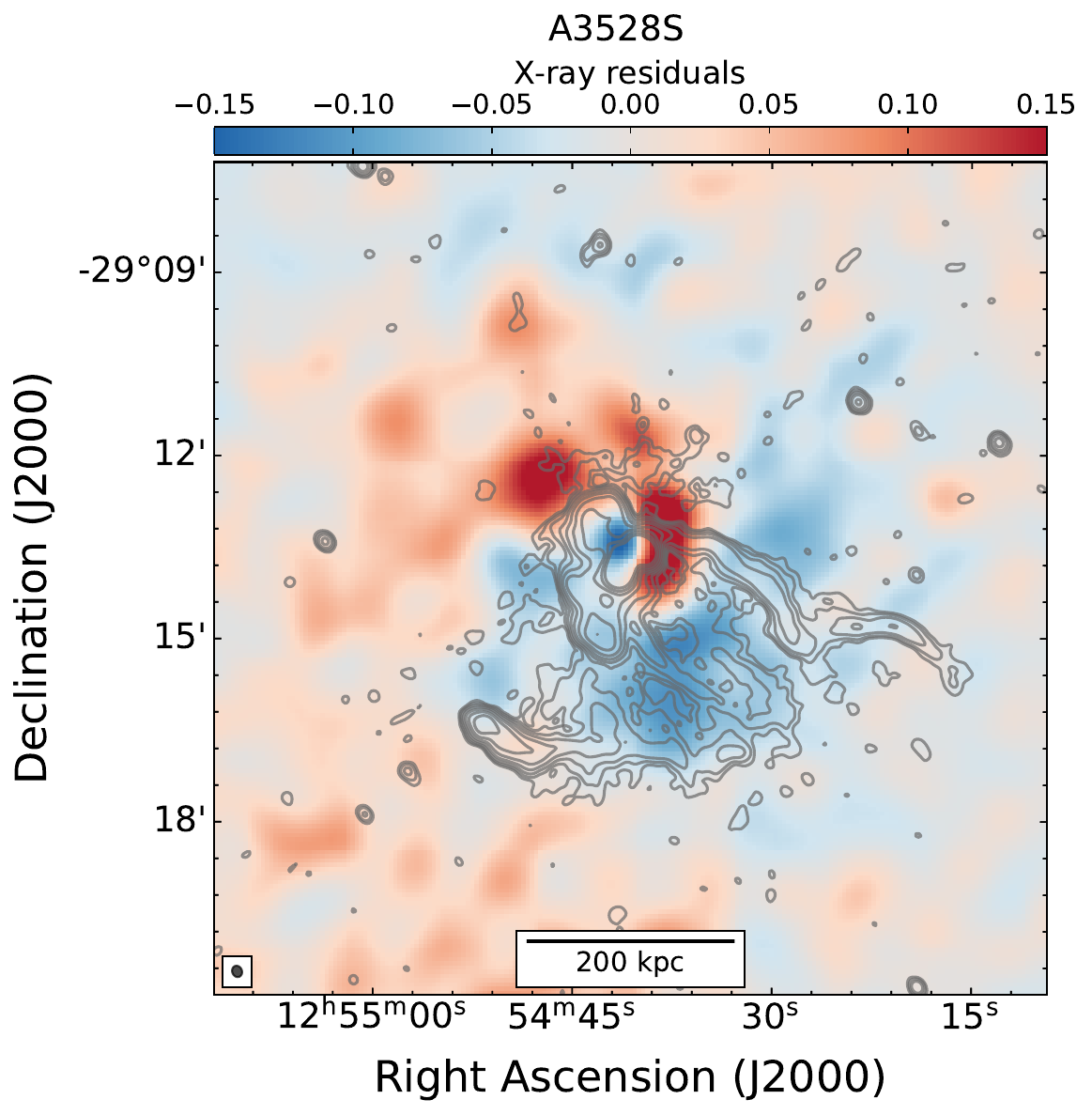}
\includegraphics[width=0.33\textwidth]{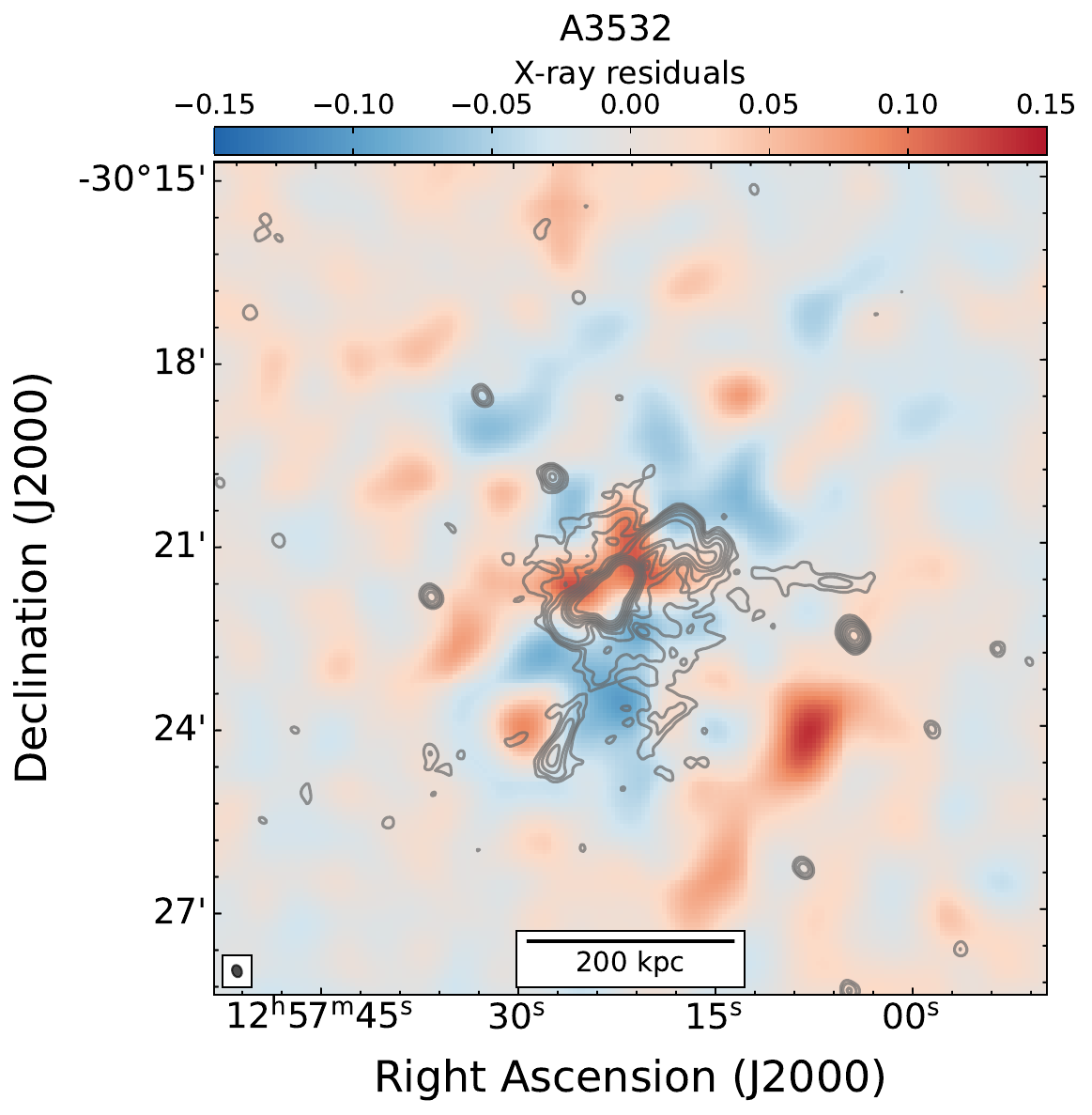}
\caption{Radio/X-ray/optical comparison. Top panels: Optical \citep[VLT Survey Telescope {\it gri};][]{merluzzi+15} and X-ray \citep[0.2--2.3 keV eROSITA All-Sky Survey (eRASS), blue;][Sanders et. al, in prep.]{bulbul+24,kluge+24} composite images of A3528N (left), A3528S (middle) and A3532 (right) with uGMRT Band 3 radio contours (white) drawn at $4\sigma_{\rm rms}\times[2,4,8,16,36,128,256]$. To emphasise the X-ray features, we also overplot the contours of the unsharped Gaussian Gradient Magnitude (GGM) filtered image \citep{sanders+16} at $\rm [3,6,20]\times10^{-6}~photons\,cm^{-2}\,s^{-1}$ levels (light blue). Bottom panels: Residual X-ray images after subtracting a standard $\beta$-model \citep{cavaliere+fusco-femiano76}. Radio contours are from the uGMRT Band 3 {\tt taper=10\arcsec} image (beam in the bottom left corners), at the $4\sigma_{\rm rms}\times[1,2,4,8,16,36,128,256]$ levels.}
\label{fig:xray}
\end{figure*}

\subsection{Morphology of the ICM}
We used the data from the first eROSITA Western All-Sky Survey \citep[eRASS1;][Sanders et al. in prep.]{bulbul+24,merloni+24,kluge+24} to estimate the dynamical state of the three clusters in the \target. From this catalogue, we obtained the concentration parameter ($c$), the centroid shift ($w$) and the power ratio ($P_3/P_0$), which indicate how strong is the X-ray peak compared with the rest of the cluster surface brightness, how much the X-ray peak is shifted from the cluster centroid, and how many X-ray peaks are present in the cluster, respectively. These parameters are commonly used in the literature to investigate the dynamical state of a cluster \citep[e.g.][]{santos+08,lovisari+17,andrade-santos+17,rossetti+17}. Relaxed clusters are usually found in the parameter space bound by $c>0.2$, $w<0.012$ and  $P_3/P_0<1.2\times10^{-7}$ \citep[e.g.][]{cassano+10,cassano+13,cuciti+21b}. We find $c=0.31$, $w=0.006$ and $P_3/P_0=1.2\times10^{-7}$ for A3528N, $c=0.34$, $w=0.004$ and $P_3/P_0=3.7\times10^{-7}$ for A3528S, and $c=0.75$, $w=0.05$ and $P_3/P_0=2.7\times10^{-7}$ for A3532. These values are indicating a complex morphology of the clusters, with A3528N and A3528S being in line with undisturbed systems according to the $c$-$w$ plot, but not with the $c$-$P_3/P_0$ and $w$-$P_3/P_0$ ones. On the other hand, A3532 agrees with the ``disturbed'' morphological classification according to the $c$-$w$ and $c$-$P_3/P_0$ plots, but not with the $w$-$P_3/P_0$ one. These results point to a more complex dynamical state than previously studied.

We therefore produced the residual images of the three clusters, after subtracting an isothermal $\beta$-model (see Appendix \ref{apx:betamodel}) which usually describes a relaxed system \citep{cavaliere+fusco-femiano76}. 
A3528N appears to have positive residuals along the north-south direction (bottom left panel in Fig. \ref{fig:xray}), which reflects the elongated morphology also highlighted by the GGM contours (top left panel in Fig. \ref{fig:xray}). A3528S presents the typical traits of a sloshing core, with a spiral shape of negative and positive residuals (bottom central panel in Fig. \ref{fig:xray}), as also suggested by previous studies \citep{gastaldello+03}. The extended radio emission associated with the BCG J1254-2913 lays in the region of negative residual region, with the northern filament of J1254-2913 and the long tail in J1254-2916 partially following the northern and southern edges of the region, respectively. Similarly, the two filaments in A3532 fill the two regions of negative residuals (bottom right panel in Fig. \ref{fig:xray}), with the ``comma'' radio feature that is probably produced by hitting the excess of surface brightness of the ICM. 

Finally, we also give the eRASS1 masses ($M_{500}$), averaged temperature ($T_{500}$) and bolometric luminosities ($L_{500}$) at $R_{500}$ \citep{bulbul+24}: 
$M_{500}=3.4^{+0.3}_{-0.3}\times10^{14}~\rm M_\odot$, $T_{500}<4.9$ keV and $L_{500}=2.0^{+0.7}_{-0.4}\times10^{44}~{\rm erg\,s^{-1}}$ for A3528N (with $R_{500}=1.0$ Mpc); $M_{500}=4.2^{+0.3}_{-0.4}\times10^{14}~\rm M_\odot$, $T_{500}=3.3^{+1.0}_{-0.8}$ keV and $L_{500}=1.8^{+0.3}_{-0.2}\times10^{44}~{\rm erg\,s^{-1}}$ for A3528S (with $R_{500}=1.1$ Mpc); $M_{500}=4.1^{+0.4}_{-0.3}\times10^{14}~\rm M_\odot$, $T_{500}=3.5^{+1.0}_{-0.9}$ keV and $L_{500}=2.0^{+0.3}_{-0.2}\times10^{44}~{\rm erg\,s^{-1}}$ for A3532 (with $R_{500}=1.1$ Mpc). These values are slightly above those reported from literature \citep[e.g.][]{piffaretti+11,gastaldello+03}, as also reported by \cite{liu+23} and \cite{migkas+24}.

\begin{figure*}
\centering
\includegraphics[width=\textwidth]{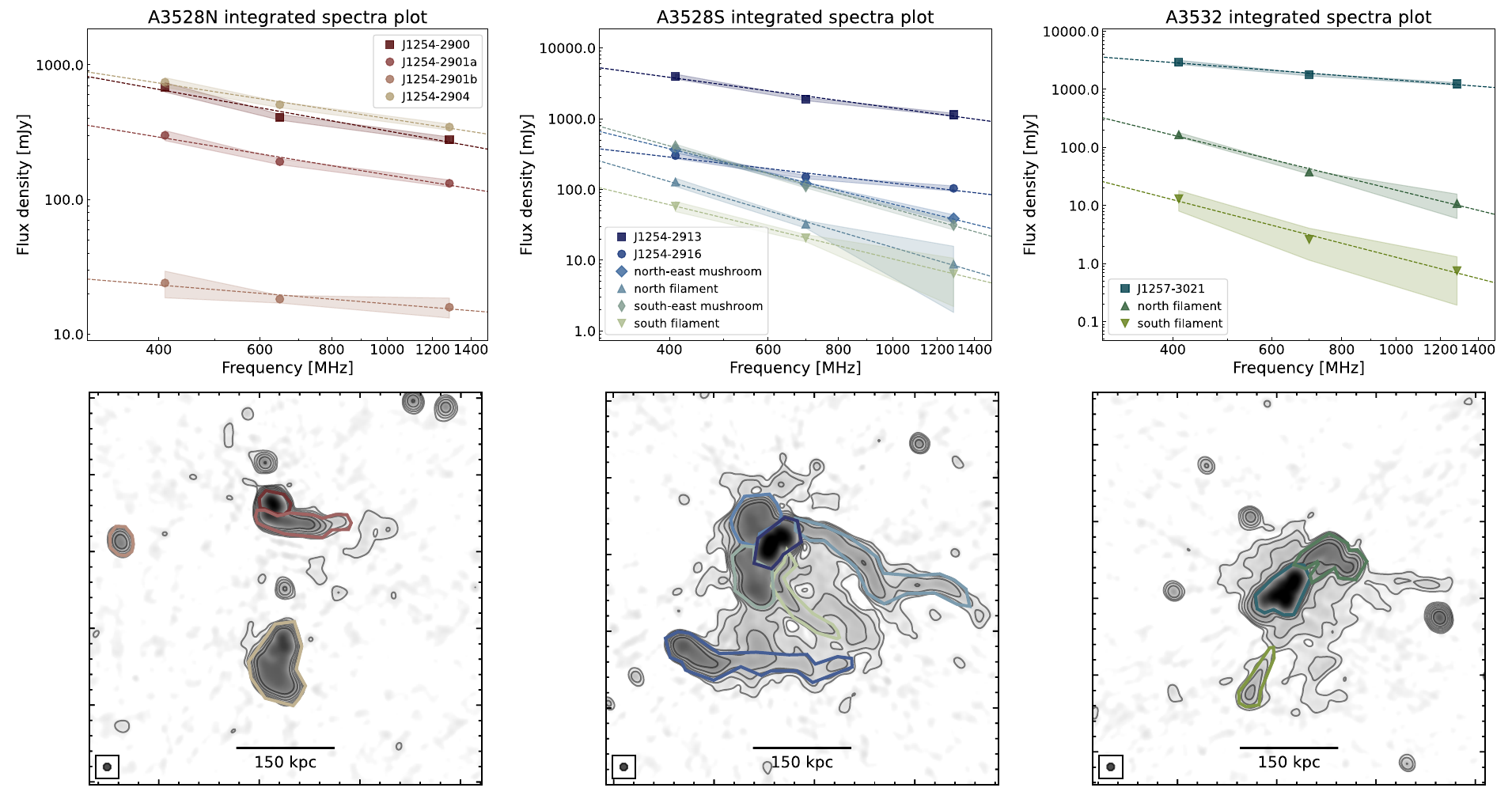}
\vspace{-5mm}
\caption{Integrated spectra of the radio sources in the \target. Top row: Flux densities were measured after convolving all the frequencies at the same resolution. Different dashed lines in each panel are colour-coded based on the region where the flux density was extracted (see bottom panels), and describe the fitted $\alpha$ (see Column 6 in Tab. \ref{tab:fluxes}). Bottom row: uGMRT Band 3 images of the three clusters (from left to right: A3528N, A3528S and A3532) at $\sim12''$ (see beam in the bottom-left corner of each panel). Radio contours start at the $3\sigma_{\rm rms}$ level, which was used as reference for drawing the regions.}\label{fig:intspectra}
\end{figure*}

\begin{table*}
\caption{Integrated radio fluxes and radio power at  1.28 GHz obtained from the regions drawn in the bottom panels in Fig.~\ref{fig:intspectra}, at a common resolution of 12\arcsec. The radio powers have been calculated using the integrated spectral index ($\alpha$) obtained by fitting the radio flux densities with a power-law in logarithm space ($\log S = \beta + \alpha\log\nu$) and listed in Column 6.}
\vspace{-5mm}
\begin{center}
\resizebox{0.85\textwidth}{!}{
\begin{tabular}{ccccccc}
\hline\hline
Region & $S_{410}$ & $S_{650/700}$ & $S_{1260}$ & $S_{1284}$ & $\alpha$ & $P_{1284}$ \\ 
& [mJy] & [mJy] & [mJy] & [mJy] & & $[{\rm W\,Hz^{-1}}]$ \\ 
\hline
\multicolumn{7}{c}{A3528N} \\
J1254-2900 & $683.4\pm54.9$ & $408.6\pm20.4$ & $201.6\pm10.1$ & $277.4\pm4.1$ & $-0.77\pm0.07$ & $(1.6\pm0.1)\times10^{24}$ \\ 
J1254-2901a & $299.1\pm25.5$ & $190.9\pm9.6$ & $100.1\pm5.2$ & $131.8\pm8.0$ & $-0.70\pm0.08$ & $(7.6\pm0.4)\times10^{23}$\\ 
J1254-2901b & $24.0\pm5.4$ & $18.2\pm1.2 $ & $9.7\pm1.0$ & $15.8\pm2.7$ & $-0.35\pm0.22$ & $(9.0\pm0.2)\times10^{22}$\\ 
J1254-2904 & $742.4\pm60.6$ & $505.9\pm25.4$ & $264.5\pm13.6$ & $345.3\pm18.4$ & $-0.66\pm0.08$ & $(2.0\pm0.1)\times10^{24}$\\ 
\hline
\multicolumn{7}{c}{A3528S} \\
J1254-2913 & $3949.6\pm316.1$ & $1892.3\pm94.6$ & $785.1\pm39.3$ & $1136.8\pm60.0$ & $-1.09\pm0.08$ & $(6.7\pm0.3)\times10^{24}$\\ 
mushroom NE & $366.1\pm30.6$ & $120.2\pm6.5$ & N/A & $38.8\pm5.0$ & $-1.96\pm0.13$ & $(2.4\pm0.3)\times10^{23}$\\ 
mushroom SE & $400.5\pm33.2$ & $110.0\pm6.0$ & N/A & $31.7\pm4.7$ & $-2.23\pm0.13$ &  $(2.0\pm0.3)\times10^{23}$\\ 
filament N & $130.6\pm17.4$ & $33.0\pm4.0$ & N/A & $9.0\pm7.2$ & $-2.34\pm0.28$ & $(6.4\pm0.4)\times10^{22}$\\ 
filament S & $61.7\pm9.8$ & $21.1\pm2.5$ & N/A & $6.6\pm4.3$ & $-1.93\pm0.32$ & $(4.4\pm2.4)\times10^{22}$\\ 
J1254-2916 & $296.3\pm27.3$ & $148.9\pm8.3$ & $71.7\pm3.6$ & $103.2\pm8.7$ & $-0.93\pm0.11$ & $(6.0\pm0.5)\times10^{23}$\\  
\hline
\multicolumn{7}{c}{A3532} \\
12547-3021 & $2934.7\pm234.9$ & $1795.2\pm89.8$ & $885.6\pm44.3$ & $1242.2\pm62.3$ & $-0.75\pm0.08$ & $(7.2\pm0.4)\times10^{23}$\\ 
filament N & $161.9\pm14.6$ & $36.8\pm2.7$ & N/A & $10.8\pm4.6$ & $-2.38\pm0.21$ & $(6.8\pm2.8)\times10^{22}$\\ 
filament S & $13.0\pm0.1$ & $2.6\pm0.1$ & N/A & $0.75\pm0.05$ & $-2.49\pm0.81$ & $(5.6\pm2.1)\times10^{21}$\\ 
\hline
`\end{tabular}}
\end{center}
\vspace{-5mm}
\tablefoot{For the uGMRT Band 5 at 1260 MHz we measured the flux densities only for the radio galaxies (i.e. BCGs, HTs, NATs and WATs) at full resolution (i.e. $\Theta_{1260}=2.9''\times2.2''$). 
}
\label{tab:fluxes}
\end{table*}

\begin{figure*}
\centering
\includegraphics[height=0.32\textwidth]{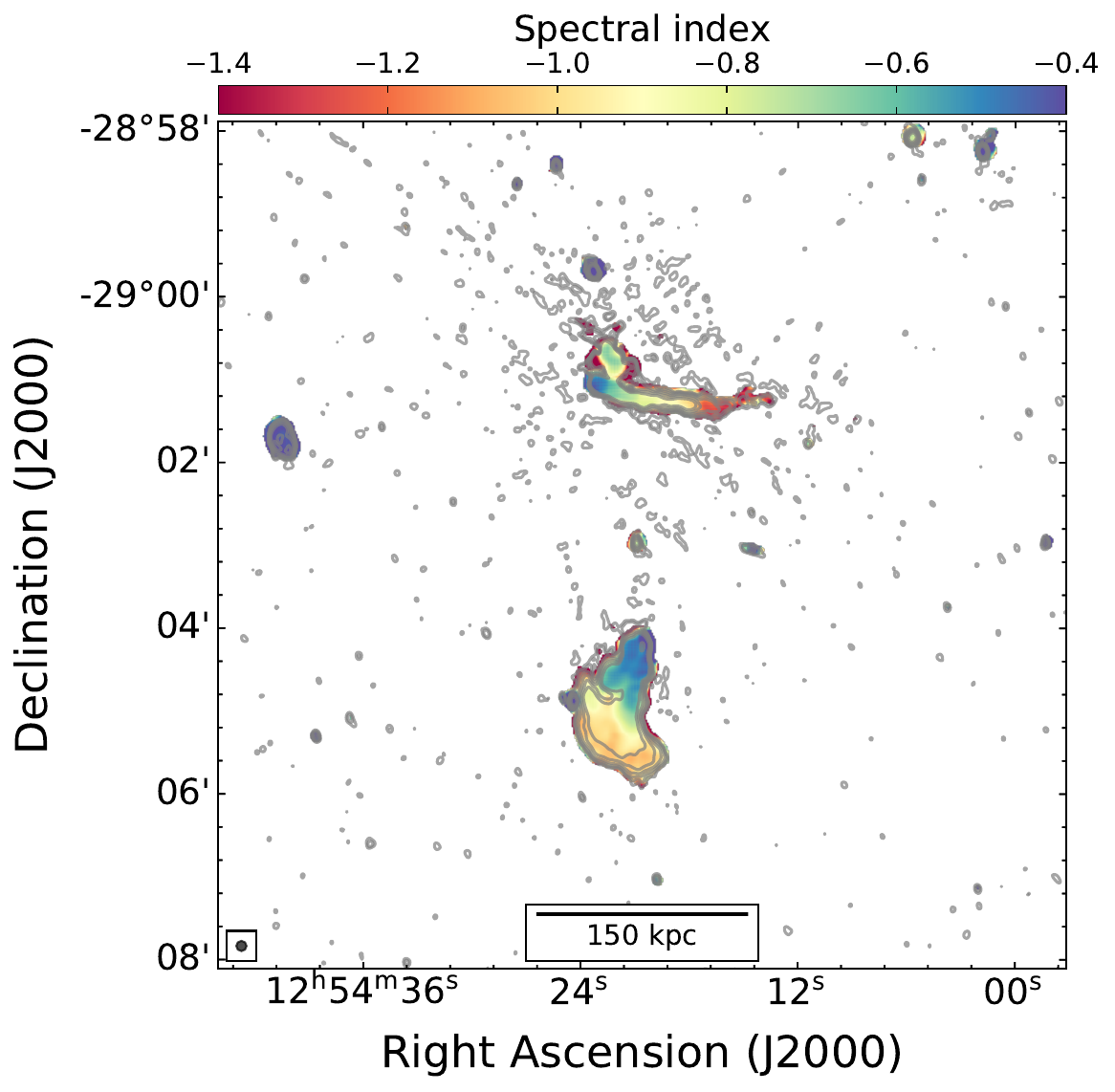}
\includegraphics[height=0.32\textwidth]{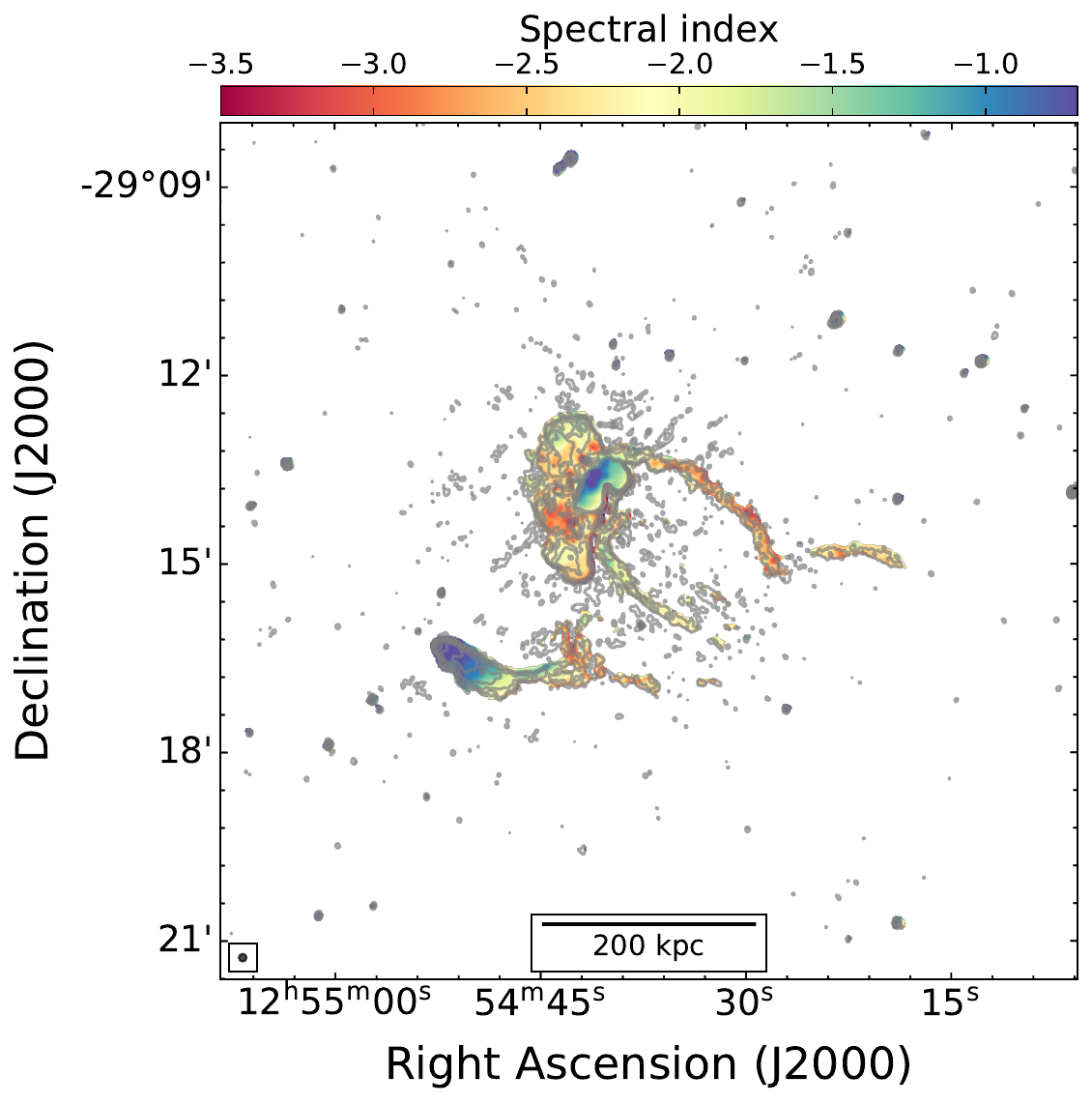}
\includegraphics[height=0.32\textwidth]{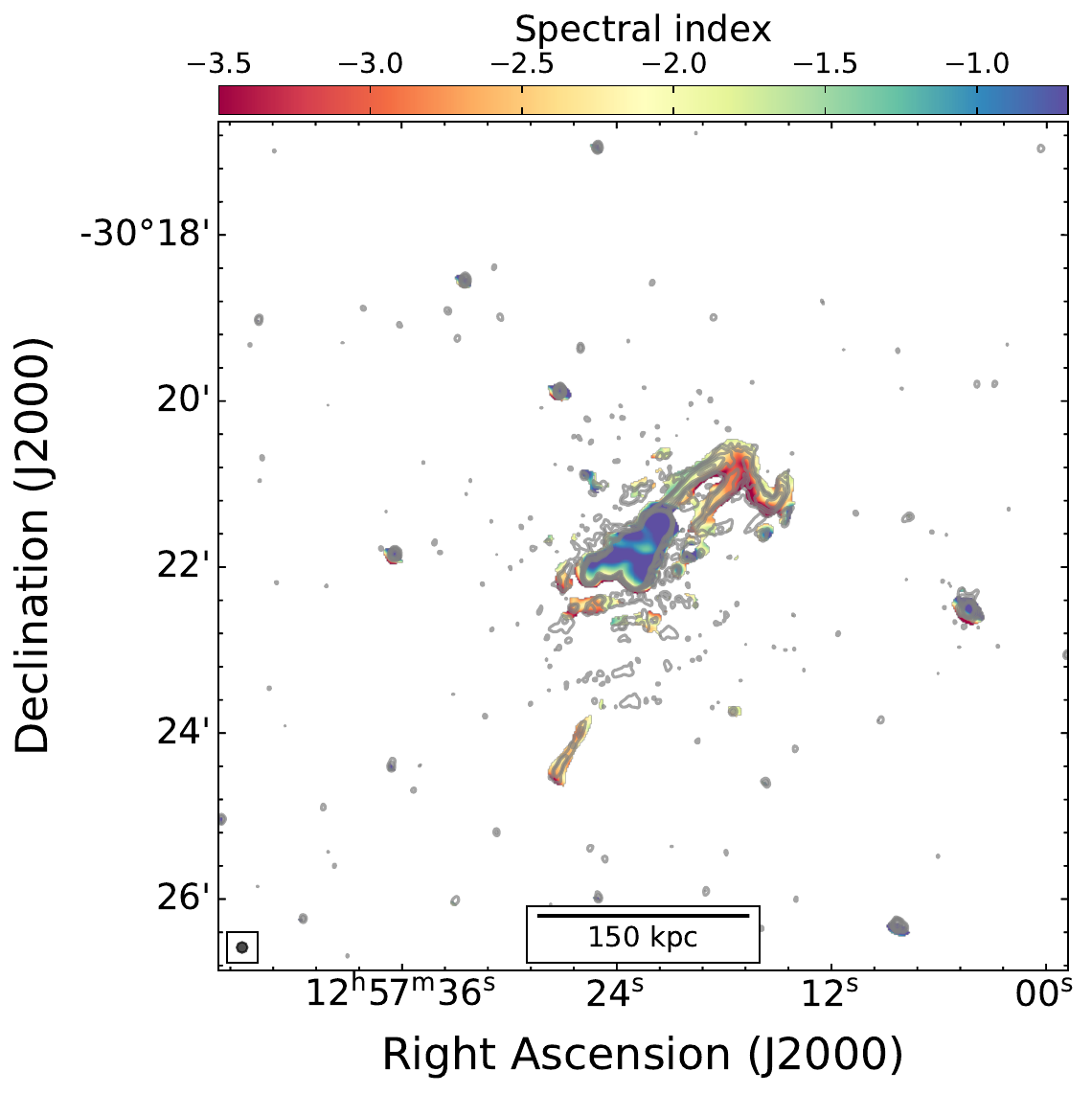} \\
\includegraphics[height=0.32\textwidth]{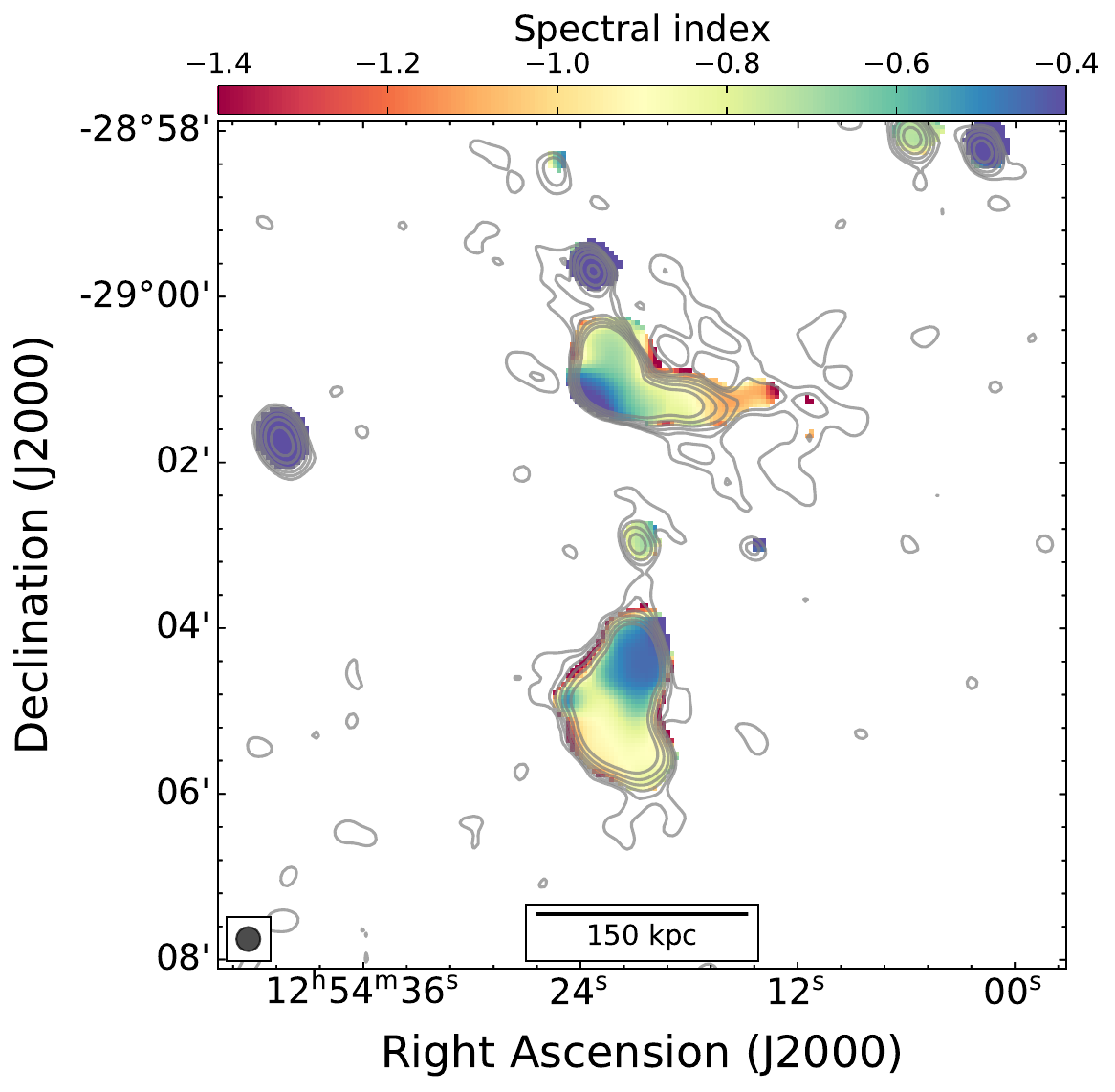}
\includegraphics[height=0.32\textwidth]{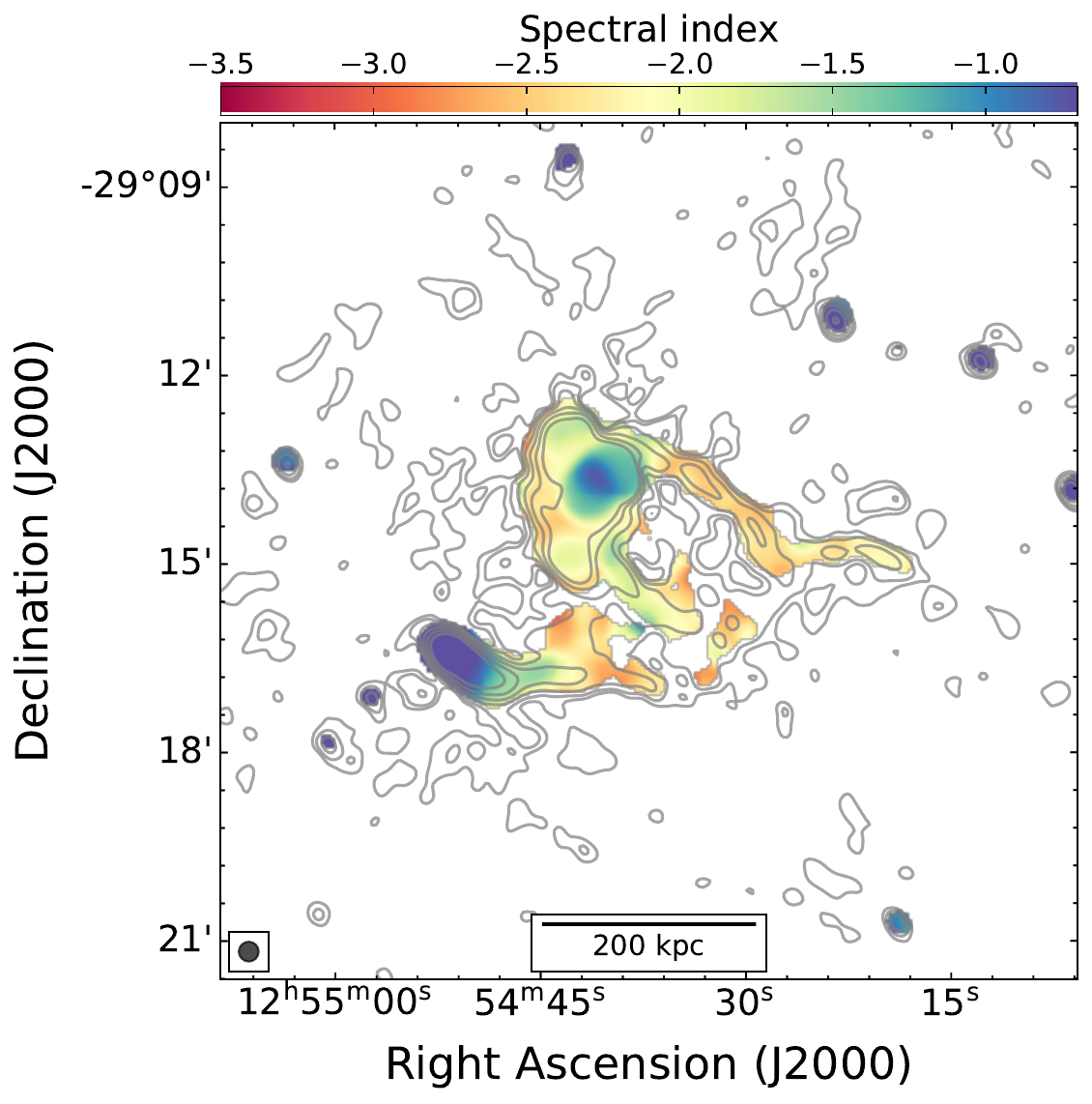}
\includegraphics[height=0.32\textwidth]{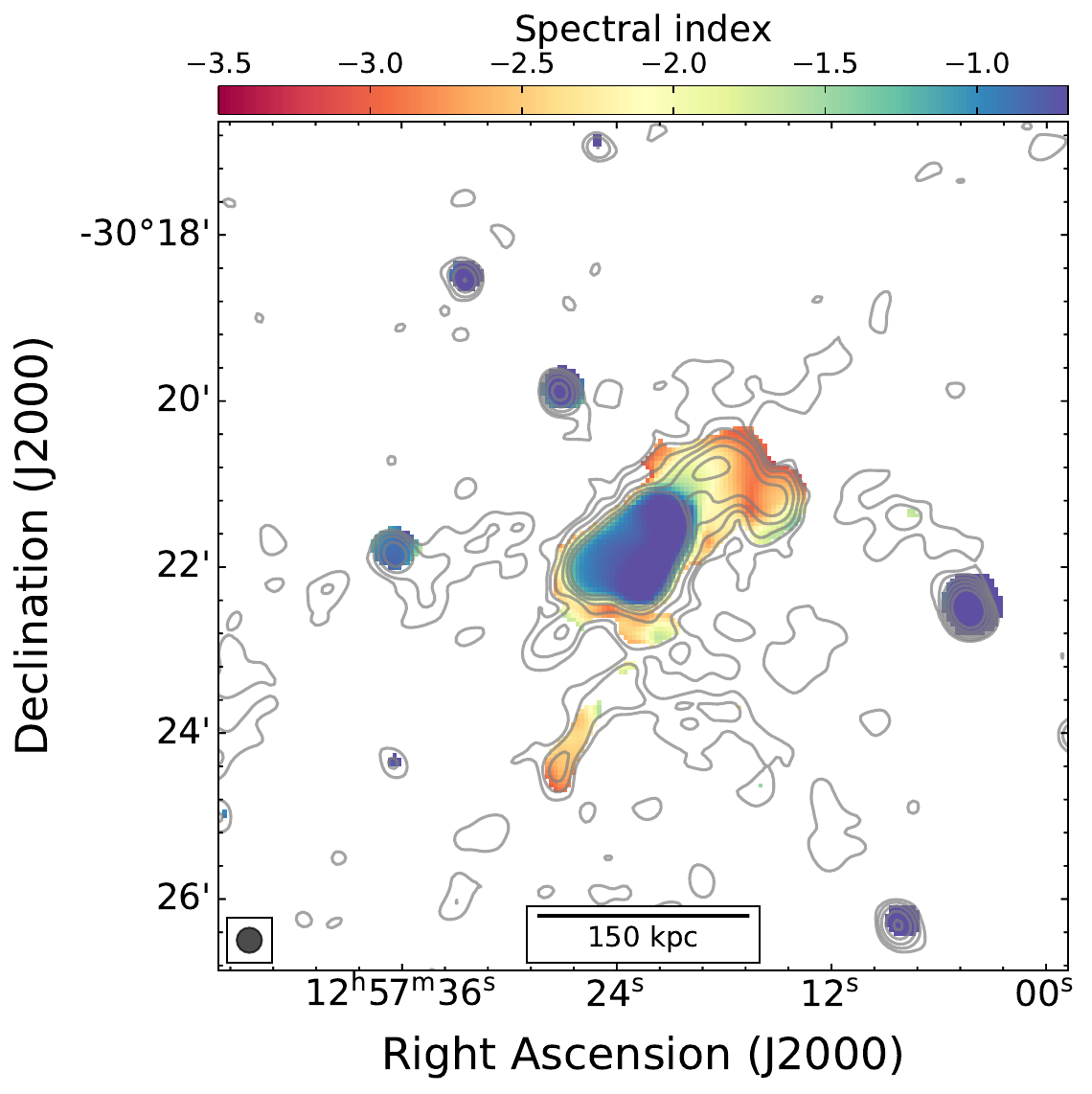}
\caption{Spectral index maps of the inner region of the three clusters in the \target. From left to right: A3528N, A3528S and A3532. Top rows: high resolution ($\Theta\sim7''$); bottom rows: low resolution ($\Theta\sim19''$, i.e. {\tt taper=15\arcsec}). The correspondent uncertainty error maps are shown in Appendix \ref{apx:spixerr}, Fig.~\ref{fig:spixerr} Radio contours are drawn at $3\sigma_{\rm rms}\times[1,2,4,8,16,32]$ of the uGMRT Band 4 observations at the same resolution (not primary beam corrected. See Tab.~\ref{tab:images} for the noise level).}\label{fig:spix}
\end{figure*}

\begin{figure*}
\centering
\includegraphics[height=0.32\textwidth]{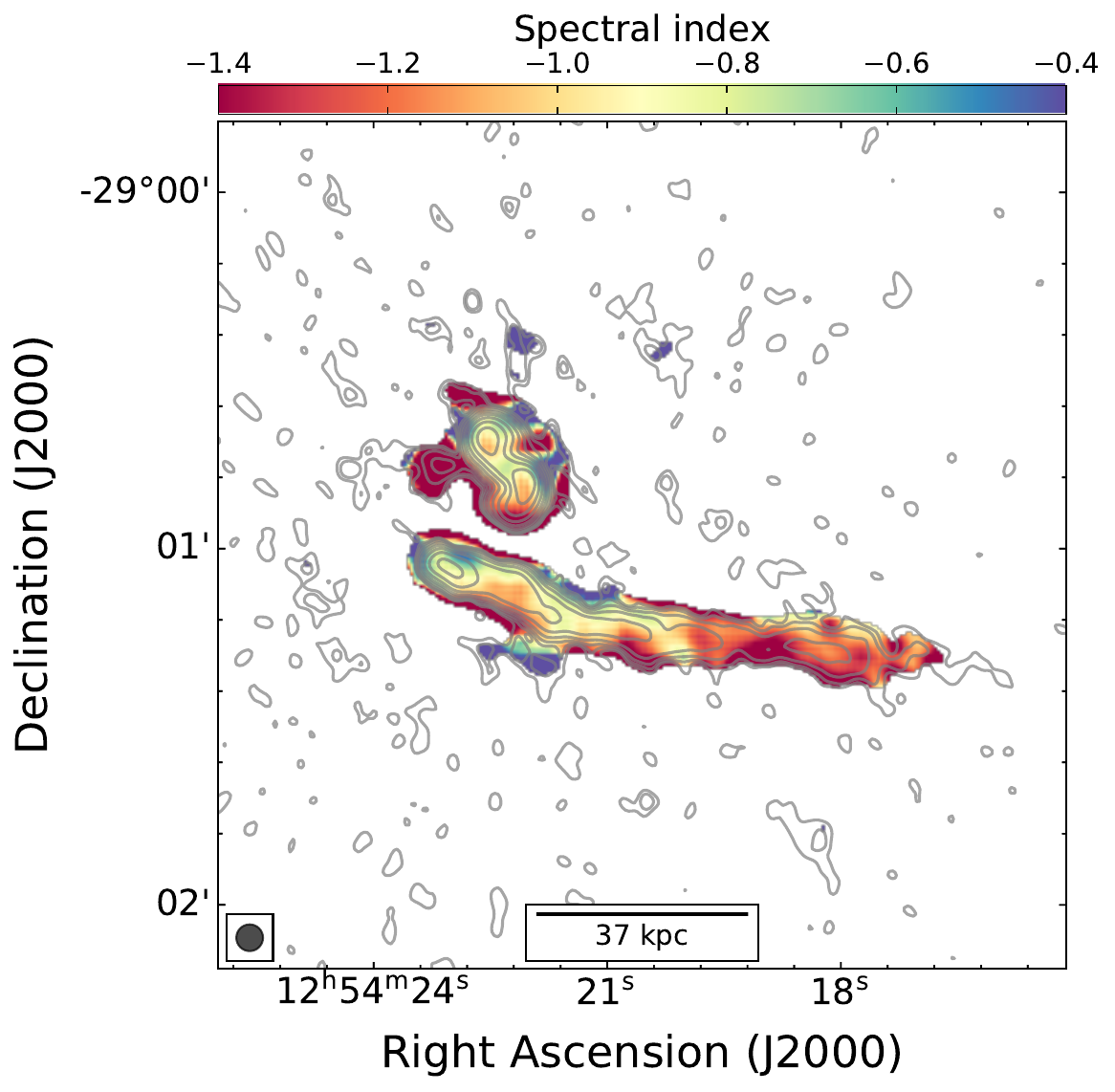}
\includegraphics[height=0.32\textwidth]{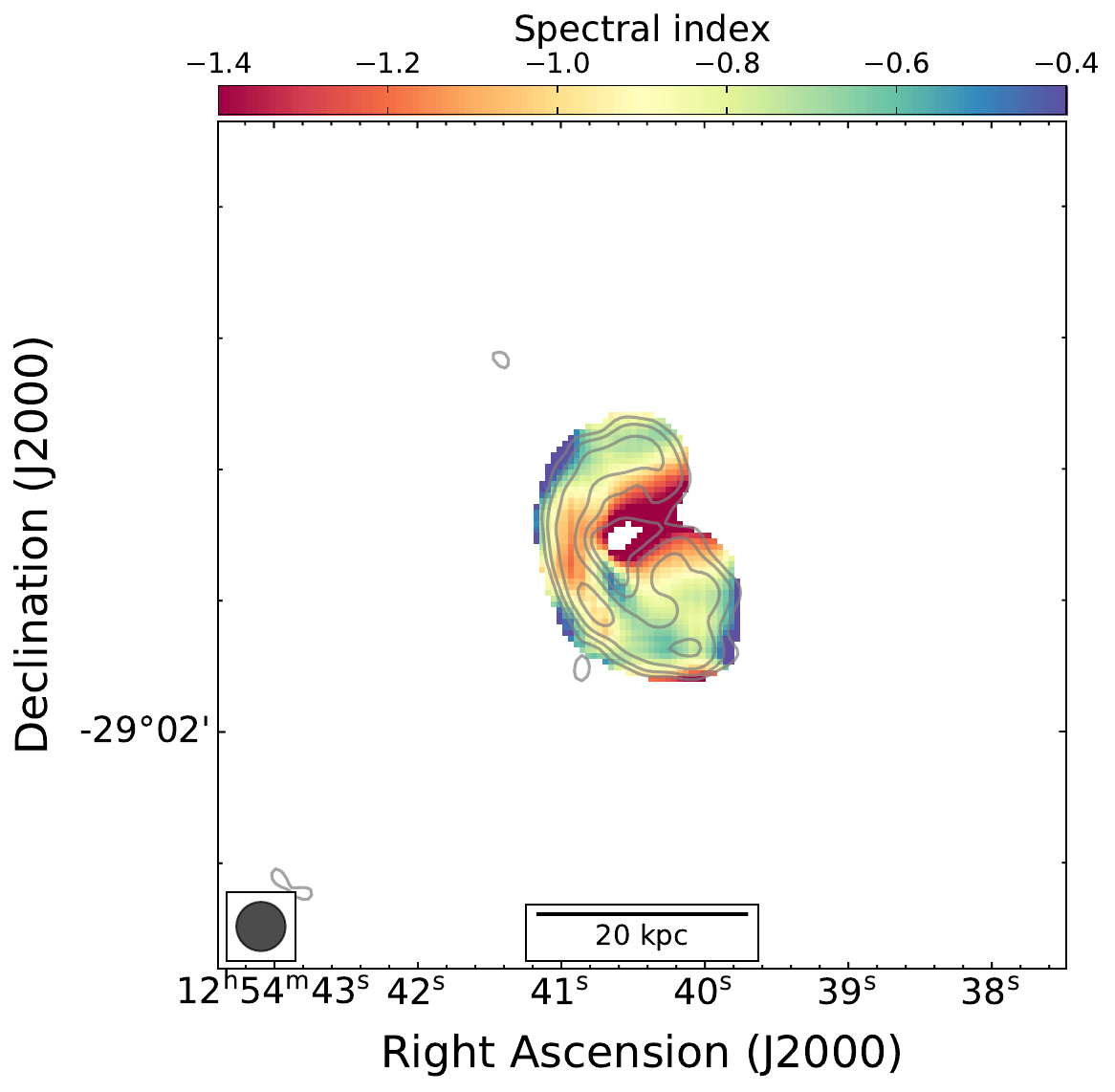}
\includegraphics[height=0.32\textwidth]{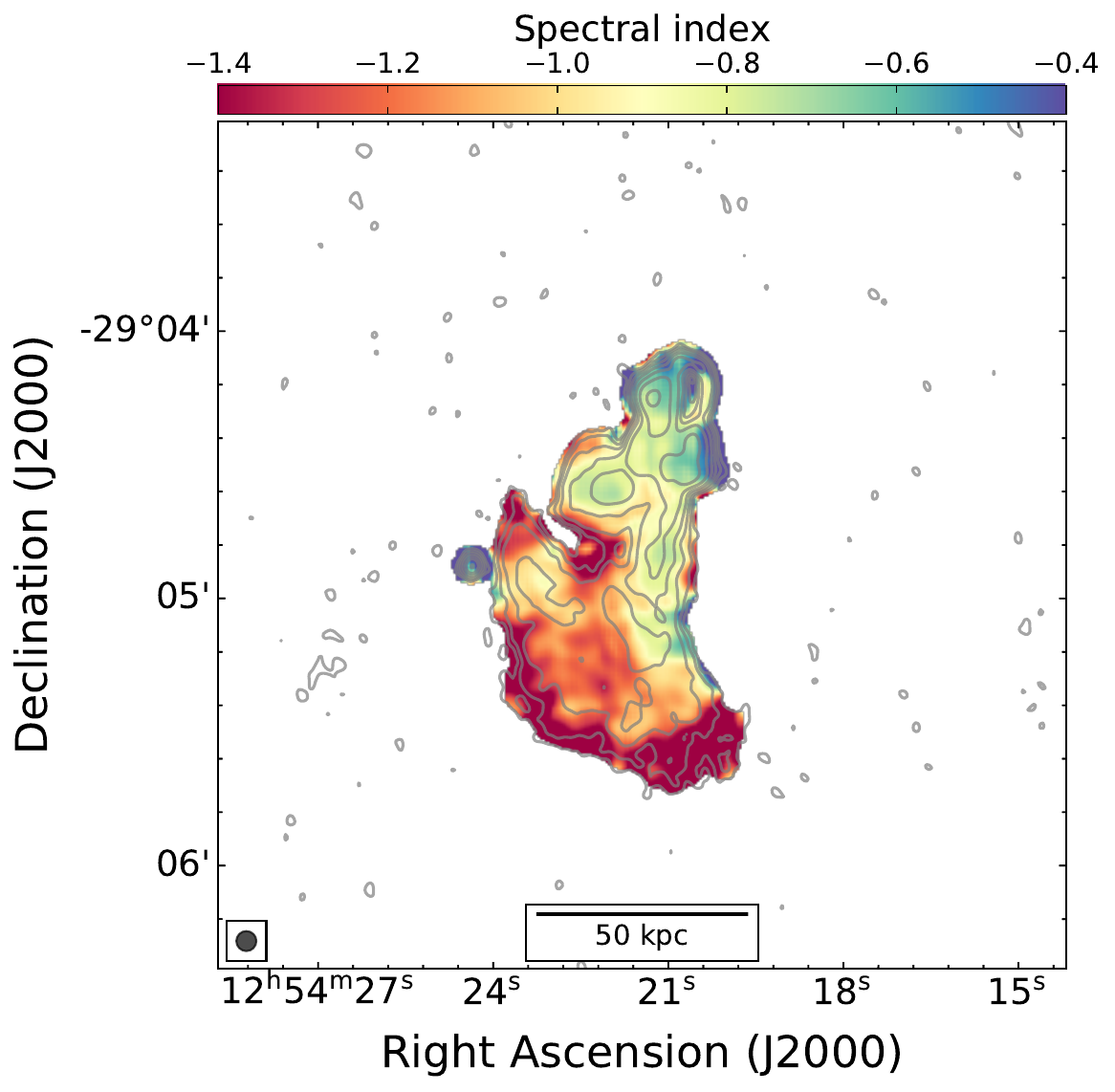} \\
\includegraphics[height=0.32\textwidth]{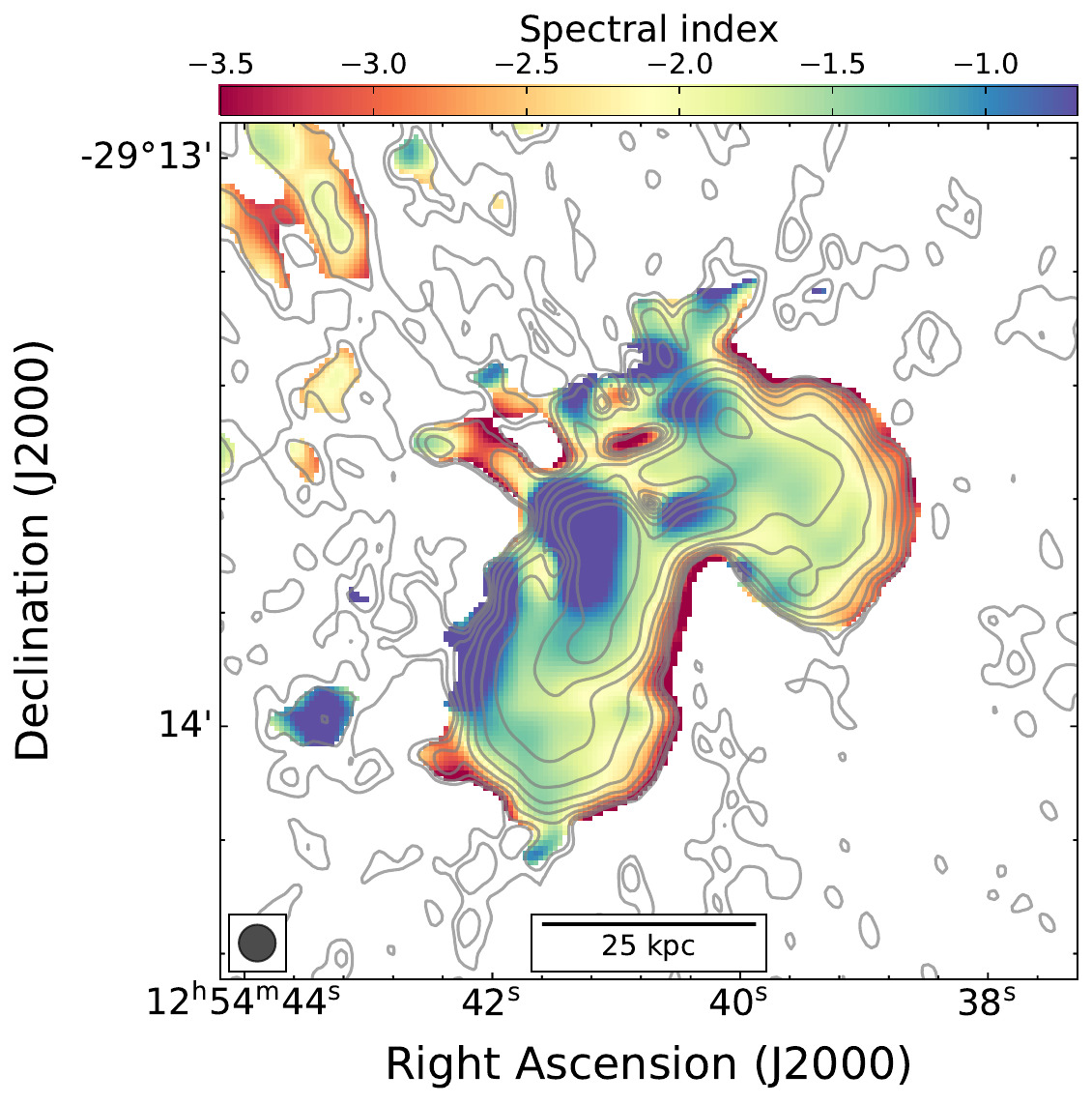}
\includegraphics[height=0.32\textwidth]{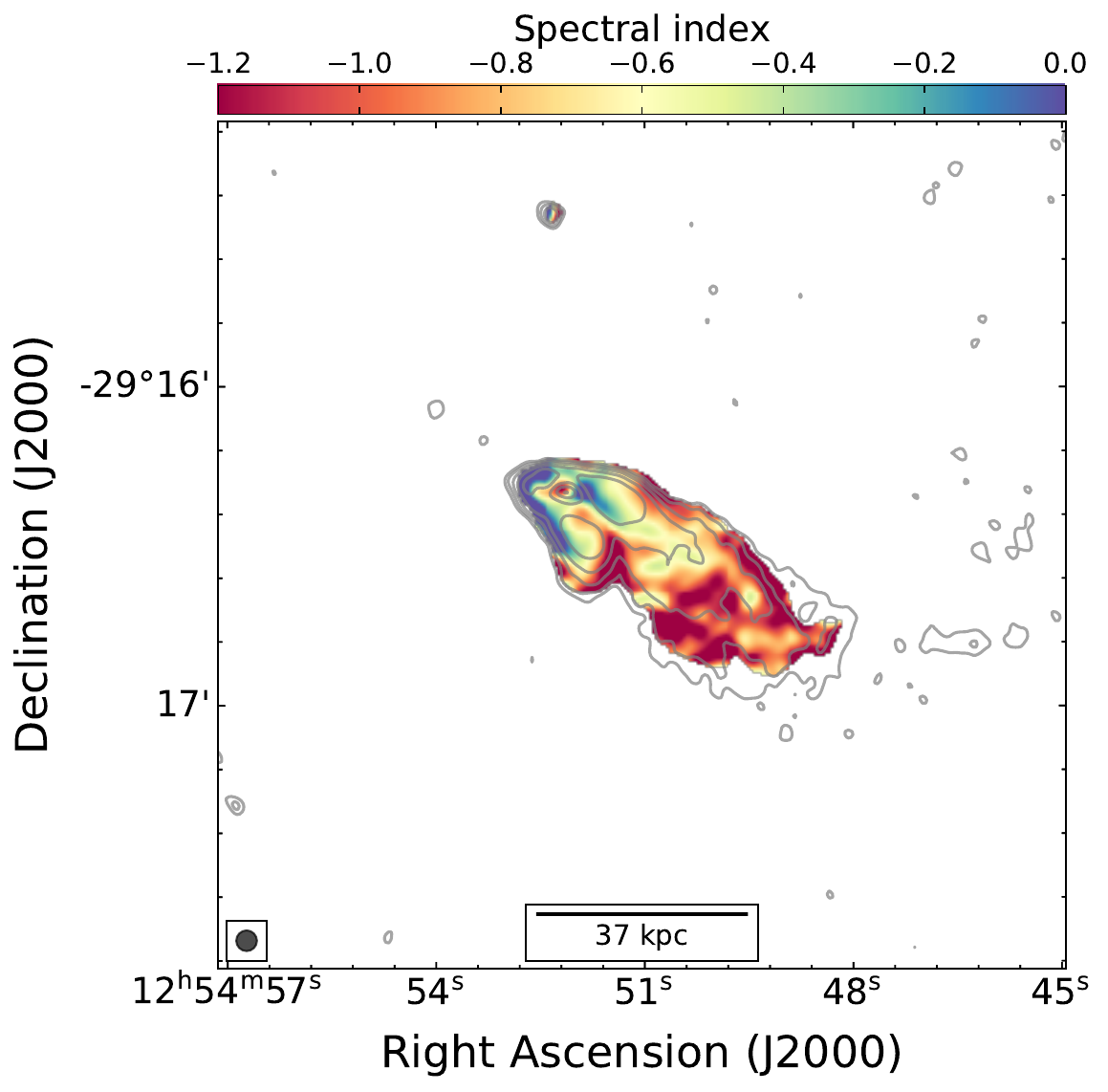}
\includegraphics[height=0.32\textwidth]{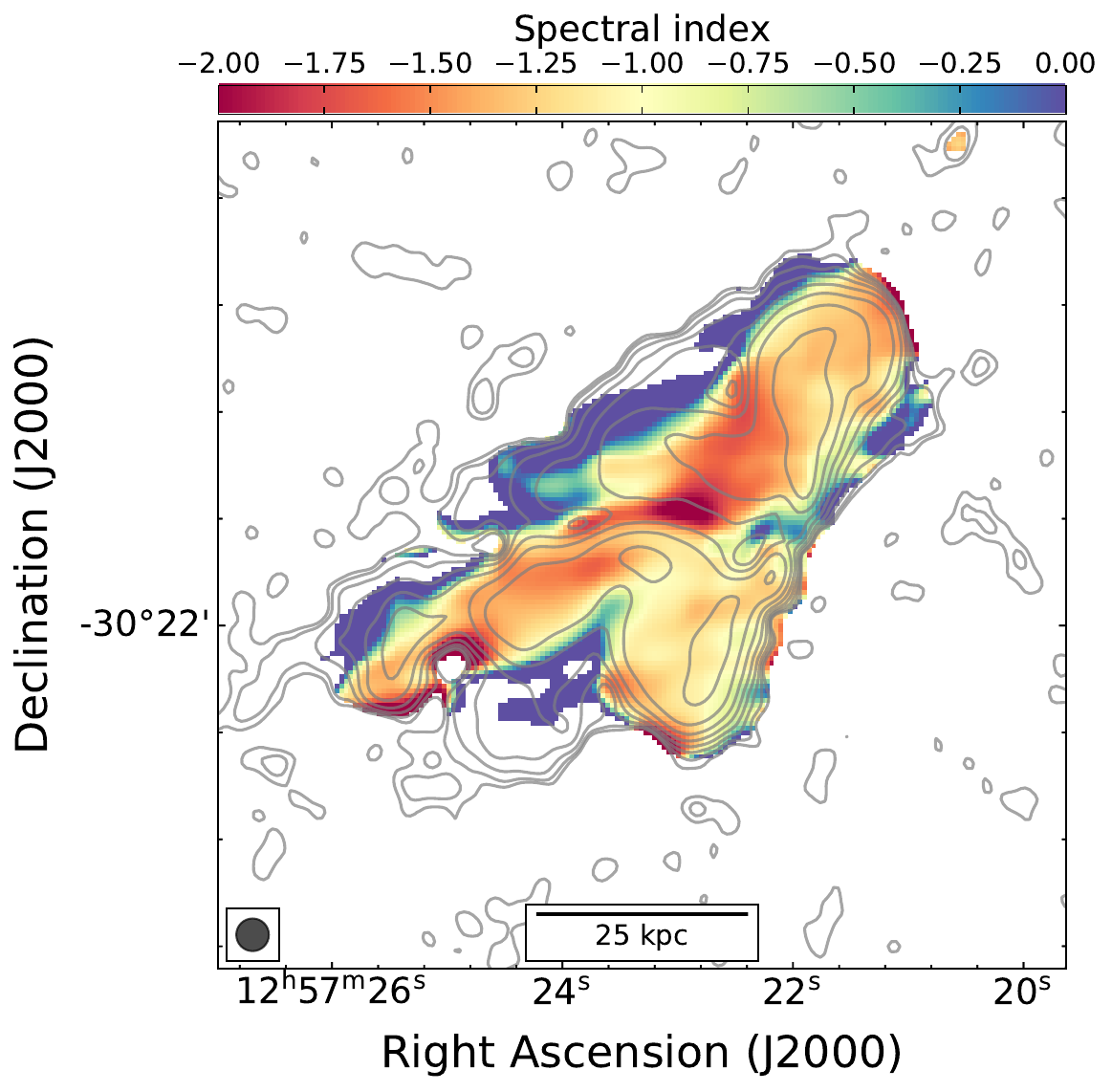}
\caption{Spectral index maps between uGMRT Band 4 ({\tt robust=-2.0}) and Band 5 ({\tt robust=0.0}) of the radio galaxies in the three clusters in the \target, namely, J1254-2900 and J1254-2901a (top left panel), J1254-2901b (top central panel), J1254-2904 (top right panel), J1254-2913 (bottom left panel), J1254-2916 (bottom central panel) and J1257-3021 (bottom right panel). The resolutions of the maps are $4.5''$ (A3528N), $3.9''$ (A3528S) and $3.7''$ (A3532). The corresponding uncertainty error maps are shown in Appendix \ref{apx:spixerr}, Fig.~\ref{fig:spixmaperrRG}. Radio contours are drawn at $3\sigma_{\rm rms}\times[1,2,4,8,16,32,\dots]$ of the uGMRT Band 5 observations at the same resolution (not primary beam corrected. See Tab.~\ref{tab:images} for the noise level).}\label{fig:spixRG}
\end{figure*}

\subsection{Integrated flux densities and spectra}
In order to study the radio spectral properties of all the radio galaxies and radio features in the \target, we measured the total flux densities at 410 MHz (uGMRT Band 3), 650/700 MHz (uGMRT Band 4) and 1284 MHz (MeerKAT L-band). To this aim, we used images corrected for the primary beam, at the common intermediate resolution of $\Theta\sim12''$ (i.e. {\tt taper=10\arcsec}). This resolution was chosen as it provides a good compromise between separating the individual sources (i.e. J1254-2900 and J1254-2901a) and recovering the largest extension of the diffuse radio emission (i.e. the radio filaments in A3528S and A3532, and the long tail in J1254-2916). We also applied an inner $uv$-cut at $200\lambda$, which represents the shortest baseline of the uGMRT Band 4 observations (see Tab.~\ref{tab:obs}), to assure the recovering of flux density on the same maximum spatial scales.
Regions of flux density measurements were chosen starting from the lowest frequency observation (see bottom panels in Fig.~\ref{fig:intspectra}), and then kept constant for all the observing frequencies. 
Uncertainties of the flux density measurements are given by:

\begin{equation}
\label{eq:fluxerr}
\Delta S_\nu=\sqrt{(f S_\nu)^2 + N_{\rm beam}\sigma_{\rm \nu, rms}^2} \, ,
\end{equation}
where $f$ describes the systematic uncertainties of the observation (i.e. , 8\% for the uGMRT Band 3 observations and 5\% for the uGMRT Band 4 and MeerKAT L-band observations), $N_{\rm beam}$ the number of beams covering the selected region, and $\sigma_{\rm\nu, rms}$ the noise level at each observation. The resulting flux density measurements and their errors are listed in Tab.~\ref{tab:fluxes}. 
We then fit our measurements of the radio spectrum with a first order polynomial in logarithmic space (i.e. $\log S_\nu= \beta +\alpha\log \nu$, with $\beta$ the flux normalisation parameter), in order to obtain the integrated spectrum ($\alpha$) of each source. 

The integrated spectra for the radio galaxies in the three clusters (reds, A3528N; blues, A3528S; greens, A3532) are shown in Fig.~\ref{fig:intspectra}, top panel, where we display BCGs with squares, radio tails with circles, filaments with up- and low-vertices triangles, and mushrooms with thin and large diamonds, colour-coded based on the host cluster.
All these regions are well described by a power-law spectrum. The three BCGs show standard spectral indices, with $\alpha=-0.77\pm0.07$, $\alpha=-1.09\pm0.07$ and $\alpha=-0.75\pm0.07$,  for J1254-2900, J1254-2913 and J1257-3021 respectively. The slightly steeper spectrum for J1254-2913 is probably due to the fact that we integrated over the lobes of the WAT (see darkest blue region in the bottom panel in Fig.~\ref{fig:intspectra}). The emission of this source gets steeper in the two mushrooms-shaped sources ($\alpha=-1.96\pm0.13$ and $\alpha=-2.23\pm0.15$, for the north-east and the south-east one respectively). Interestingly, the two filaments in A3528S have even steeper spectra, i.e. $\alpha=-2.34\pm0.33$ and $\alpha=-1.93\pm0.36$ for the  northern and southern filament. Similarly, the northern filament in A3532 has a spectral index of $\alpha=-2.38\pm0.20$, while the candidate filament in the south of the BCG has a spectral index of $\alpha=-2.49\pm0.71$ (see Tab.~\ref{tab:fluxes}).

We also determined the radio power at 1.28 GHz, using the integrated spectral index listed in Tab. \ref{tab:fluxes}:
\begin{equation}
P_{1284} = \frac{4\pi D_L^2}{(1+z)^{\alpha+1}}\, S_{1284} \quad [{\rm W\,Hz^{-1}}] \, , 
\end{equation}
where $D_L$ is the luminosity distance of the cluster. Uncertainties on the radio power take into account both the uncertainties on the flux density and on the spectral index measurement, and they were obtained through Monte-Carlo simulations.

\section{Discussion}
\label{sec:disc}
In recent years, extended radio emission associated with cluster galaxies have been found, either long radio tails associated with the AGN or filaments and bubbles in the AGN proximity. However, it is still not clear how they form, what their magnetic field properties are and how the CRe can get re-accelerated. In the following sections, we investigate the spectral properties of the diffuse emission associated with the radio galaxies in \target.

\subsection{Spectral index maps}
We obtained spectral index maps using the primary beam corrected images, with common $uv$-cut of $200\lambda$ (e.g. the uGMRT Band 4 observation, see Tab.~\ref{tab:obs}), pixel scale and angular resolution. Before any convolution and regridding, we also checked our images for the astrometry precision, using the FIRST survey as reference \citep{degasperin+23}. 
For each cluster, the first-order polynomial in logarithm space (i.e. $\log S= \beta +\alpha\log \nu$, with $\beta$ the flux normalisation parameter) was fit in those pixels where all images have a signal above the $3\sigma_{\rm rms}$ threshold. Uncertainties on the fit were obtained as standard deviation of 150 Monte-Carlo simulations, assuming the flux uncertainty as the sum in quadrature of the noise map and the systematic flux uncertainties, i.e. $\Delta S_\nu=\sqrt{(f S_\nu)^2 + \sigma_{\rm rms}^2}$. 

Spectral index maps of the inner part of the three clusters in the \target\ at full and low resolutions (i.e., $\sim7''$ and $\sim18''$ respectively) are shown in the top and bottom panels in Fig.~\ref{fig:spix}. The corresponding uncertainties maps are given in Appendix \ref{apx:spixerr} (Fig.~\ref{fig:spixerr}). Additionally, we used the uGMRT Band 4 with {\tt robust=-2.0} and Band 5 images to produce the highest-resolution ($\sim4''$) spectral index maps of the radio galaxies (see Fig.~\ref{fig:spixRG}, and \ref{fig:spixmaperrRG} for the corresponding error maps).

In A3528N, the spectral index distribution of the BCG (J1254-2900) is well resolved only at the highest frequency using uGMRT Band 4 and Band 5 observations (upper left panel in Fig. \ref{fig:spixRG}). 
At the location of the core, we measure a spectral index $\alpha\sim-0.7$ which then steepens ($\alpha\sim-1.0$) in the two S-shaped lobes. The two radio tails (J1254-2901a and J1254-2904) show standard spectral index values and gradient at all resolutions (Fig. \ref{fig:spix}, and top left and right panels in Fig. \ref{fig:spixRG}), with the starting point of the head/narrow-angle tail being $\alpha\sim-0.4$ and then steepening along the tail up to $\alpha\sim-1.2$. On the contrary, J1254-2901b has an overall homogeneous flat spectrum (i.e. $\alpha\sim-0.7$, see top central panel in Fig.~\ref{fig:spixRG}) along the `C'-shaped tails.

\begin{figure*}
\centering
\includegraphics[width=0.7\textwidth]{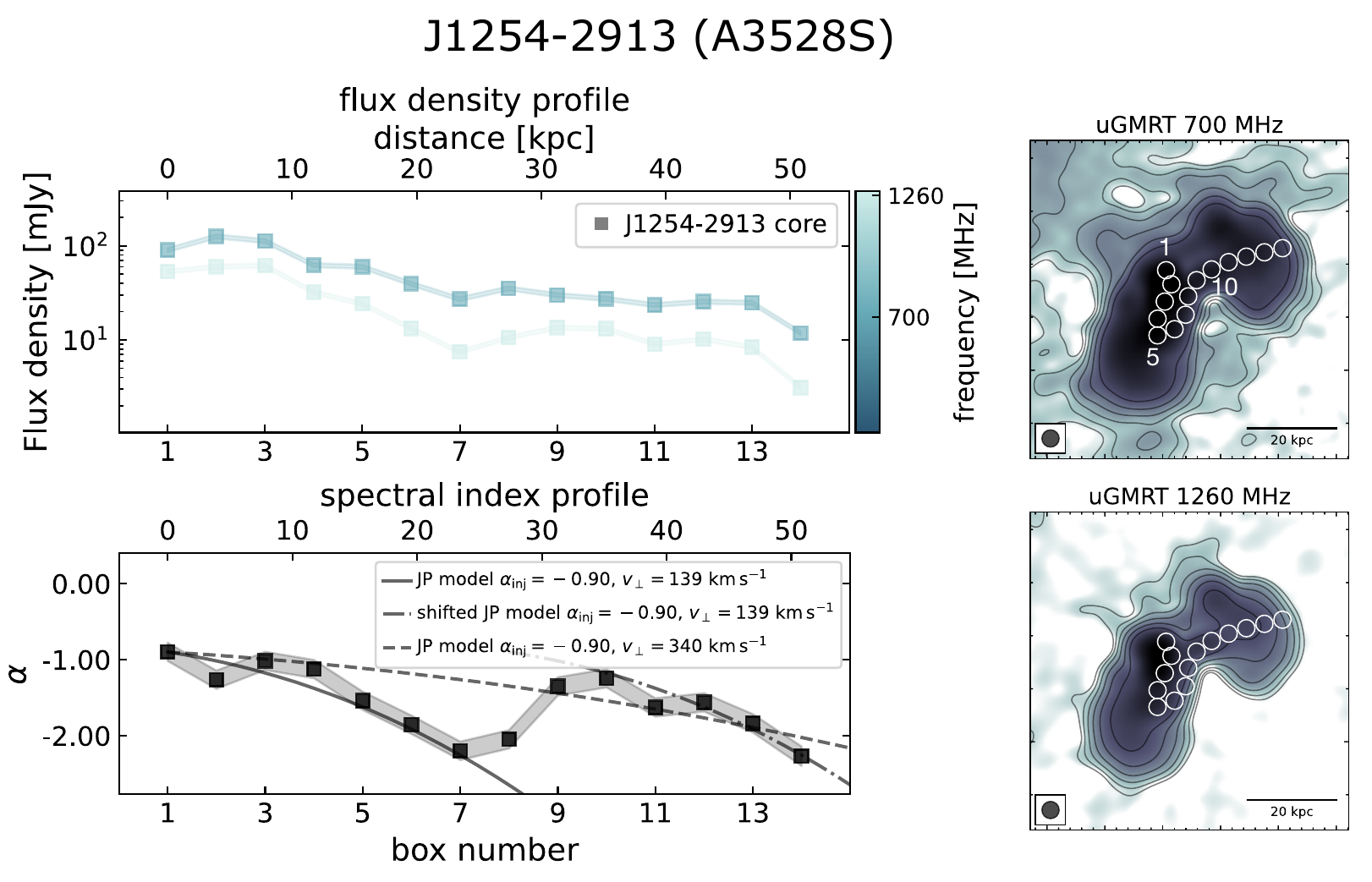}
\caption{Flux density and spectral index  profiles of the WAT J1254-2913 in A3528S. 4\arcsec-resolution images of the uGMRT Band 4 (top right) and uGMRT Band 5 (bottom right) are shown on the right of the profiles.}\label{fig:profiles_J1254-2913}
\end{figure*}

For the BCG WAT in A3528S (J1254-2913), we can clearly distinguish an inner core with flat spectral index ($\alpha\sim-0.7$) from the two lobes with steeper spectra ($\alpha\sim-1.5$), see central column in Fig \ref{fig:spix}. The 4\arcsec-resolution spectral index map (Fig \ref{fig:spixRG}) reveals interesting substructures in the SE and NW lobes of the wide angle tail, with the SW jet being flatter than the NW one (i.e., $\alpha_{\rm jet,SW}\sim-0.7$ and $\alpha_{\rm jet,NW}\sim-1.7$). The diffuse radio emission associated to this radio galaxy is only visible at mid/low resolutions. The two mushroom-shaped sources north-east and south-east the WAT present rather homogeneous spectral index distribution around $\sim-2$ (see central column in Fig. \ref{fig:spix}), with hints of steepening in the SSE ($\alpha\sim-2.5$). Different behaviour is seen for the two filaments: while the northern one is rather constant at values of $\sim-2.5$, the southern one seems to be characterised by a gradient, from $\sim-1.5$ around the bay to values of $\sim-2$. Peculiar spectral index distribution is also seen in J1254-2916. The high resolution own by the combination of the uGMRT Band 4 and Band 5 observations (central bottom panel in Fig. \ref{fig:spixRG}) reveal a standard value of the core ($\alpha\sim-0.4$) and of the two jets ($\alpha\sim-0.7$) up to $\alpha\sim-1.3$ in the inner part of the radio tail. At lower resolutions, this radio tail steepens up to $\alpha\sim-2$ for about 200 kpc.

Finally, the radio jets and lobes of the BCG in A3532 (J1257-3021, right column in Fig. \ref{fig:spix}) have a standard spectra index for a radio sources (i.e. $\alpha\sim-0.8$), with increasing steepening visible at high resolution in the plumes ($\alpha\sim-1.5$, see bottom left panel in  Fig. \ref{fig:spixRG}).  The northern filament presents a gradient from $-1.5$ to $-3$ when approaching the fork and the comma. Similarly, the southern filament has steep spectra, $\sim-2$, with no clear gradients, which possibly exclude the association with a radio galaxy.

\subsection{Spectral index profiles}
\begin{figure*}
\centering
\includegraphics[height=0.38\textwidth]{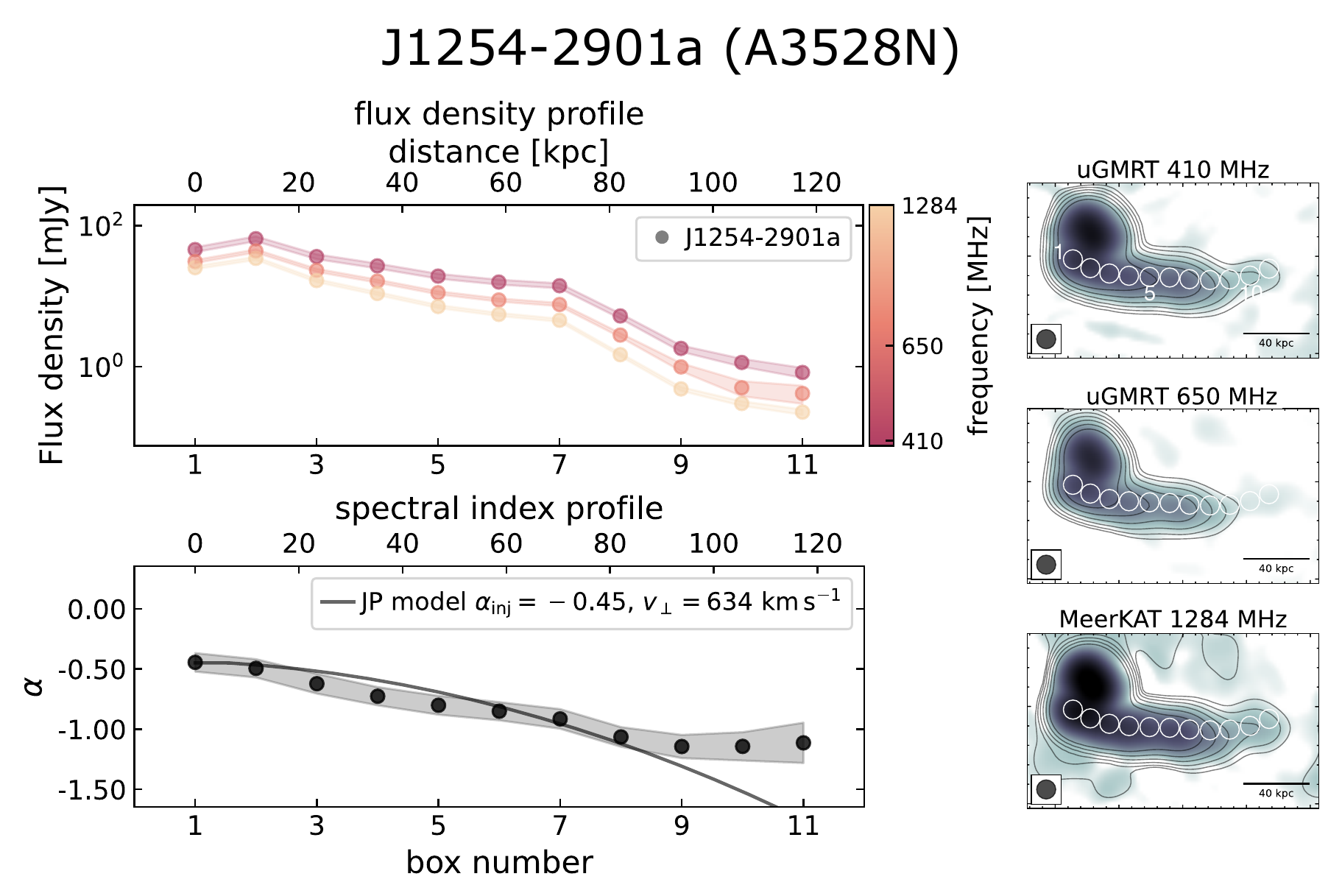}\\
\includegraphics[height=0.38\textwidth]{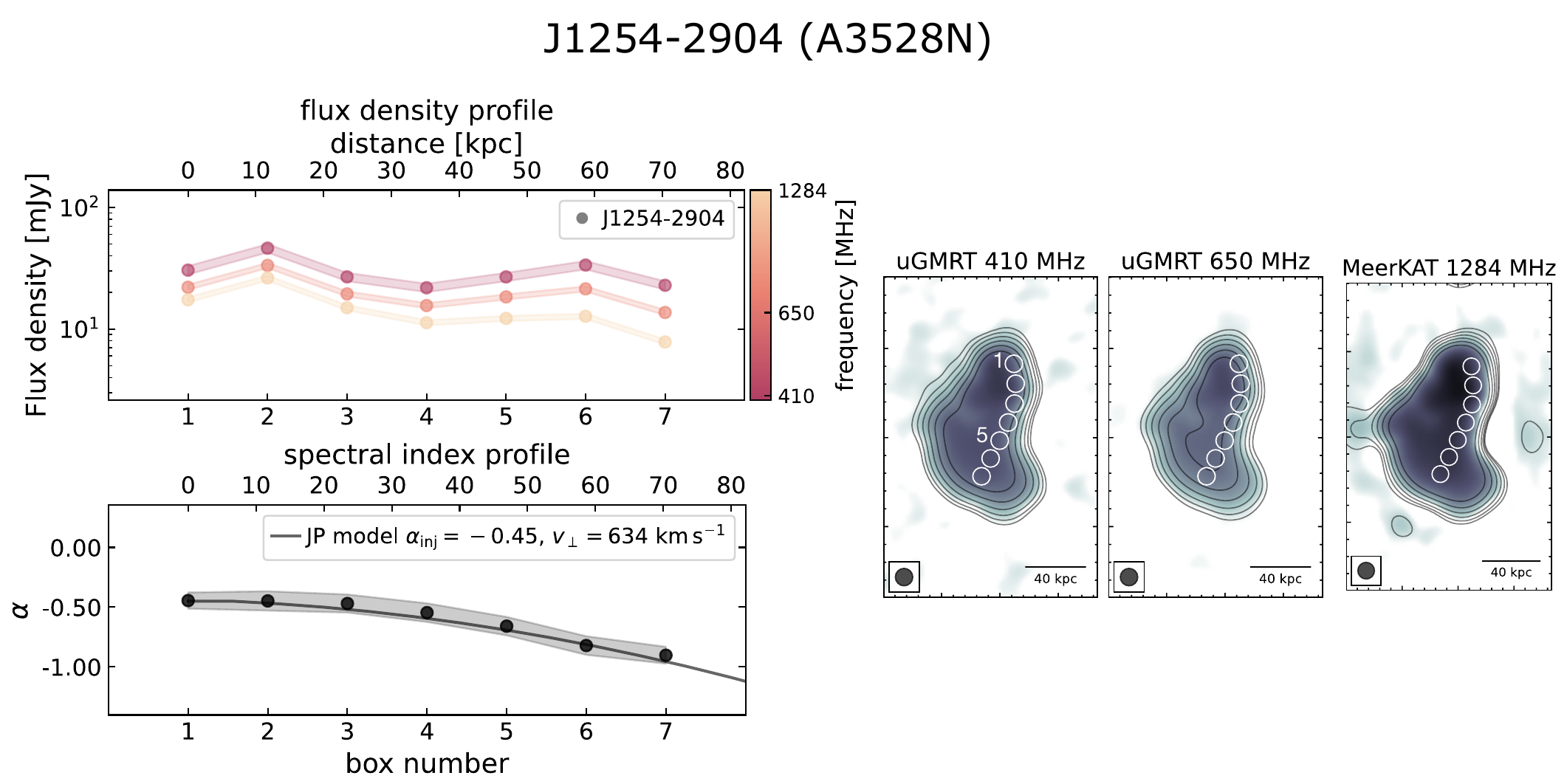}
\caption{Flux density and spectral index profiles of the radio tails in A3528N, i.e. J1254-2901a (top row) and J1254-2904 (bottom row). 12\arcsec-resolution images of the uGMRT Band 3 (top right), uGMRT Band 4 (middle right) and MeerKAT L-band (bottom right) are shown on the right of the profiles.}\label{fig:profiles_J1254-2901a}
\end{figure*}

\begin{figure*}
\centering
\includegraphics[width=0.8\textwidth]{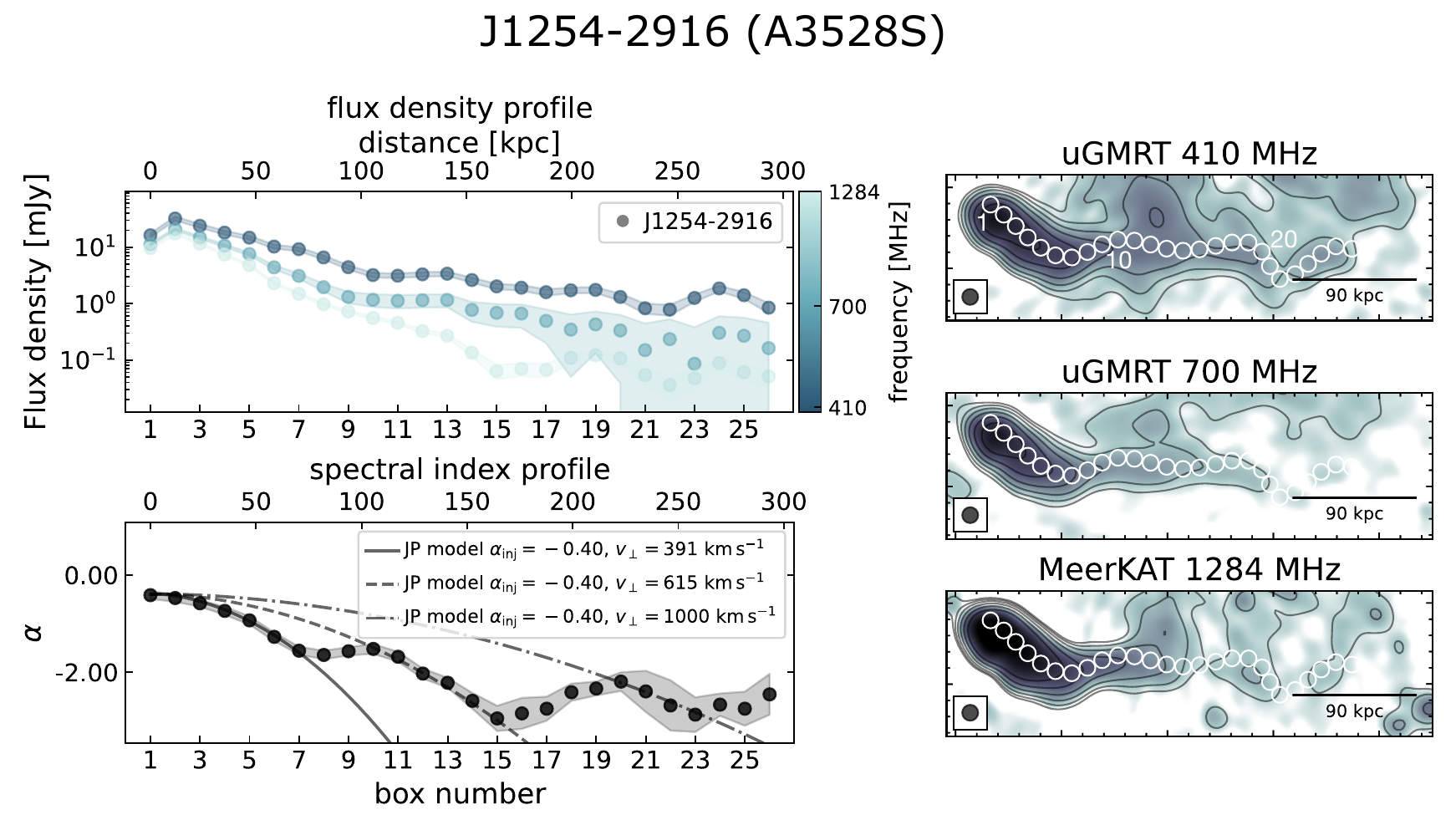} 
\caption{Flux density and spectral index  profiles of J1254-2916 in in A3528S. 12\arcsec-resolution images of the uGMRT Band 3 (top right), uGMRT Band 4 (middle right) and MeerKAT L-band (bottom right) are shown on the right of the profiles. }\label{fig:profiles_J1254-2916}
\end{figure*}

To investigate the spectral properties of the radio emission associated with the radio galaxies in the \target, we extracted the flux densities and spectral index profiles along the radio tails (e.g. J1254-2901a, J1254-2904 and J1254-2916) and from the north and south filaments in J1254-2913 and J1257-3021. These profiles were obtained from the uGMRT Band 3, uGMRT Band 4 and MeerKAT L-band observations, convolved at the same intermediate resolution of  $\sim 12''$ (i.e. {\tt taper=10\arcsec}), from beam-size circles following the brightest edge of the uGMRT Band 3 radio contours \citep{edler+22}. These regions correspond to $\sim12$ kpc each, at the cluster redshift. The profiles are shown in Figs. \ref{fig:profiles_J1254-2913}--\ref{fig:profiles_A3532_filament}, where we also show the distance travelled by the plasma, after the injection, using the conversion kpc/\arcsec\ in Tab. \ref{tab:cluster}. As for the spectral index maps and integrated spectral indices, a single power-law was also assumed in each beam-size circle.

At the time of the injection, the relativistic plasma is described by a single power-law of the type $N(E)\propto E^{-\delta_{\rm inj}}$, where the power-law index ($\rm \delta_{inj}$) is related to the injected spectral index according to $\rm \delta_{inj}=1-2\alpha_{inj}$. The radiative loss time can be then estimated as:
\begin{equation}\label{eq:trad}
t_{\rm rad} = 1.6\times10^3 \, \frac{B^{1/2}}{B^2 +B_{\rm CMB}^2} \, [\nu_{\rm break}(1+z)]^{-1/2} \quad[\rm Myr] \, ,
\end{equation}
where $B$ and $z$ are the magnetic field and redshift of the AGN, respectively, and $B_{\rm CMB}=3.25(1+z)^2~\mu$Gauss the equivalent magnetic fields of the Cosmic Microwave Background (CMB).

In order to investigate the radiative loss time of each source, we derived profiles of the spectral index. We assumed a standard Jaffe-Perola \citep[JP;][]{jaffe+perola73} model, which assumes a single burst of cosmic-ray injection, a radiative time that lasts much longer than the electron isotropisation, and a uniform and non-evolving magnetic field. In particular, in our modelling we use $\alpha_{\rm inj}$ from the spectral index profiles at the location of the core of each radio galaxy, and we assume the minimum condition for magnetic fields to maximise the age of the particles. The latter requirement means that we can estimate the magnetic fields as $B=B_{\rm CMB}/\sqrt{3}~\mu$Gauss, which corresponds to $B\sim2.1~\mu$Gauss at the redshift of our sources. 
We then use the {\tt synage} package \citep{murgia+99} to obtain the evolution of the spectral index, up to 500 Myr after the burst of particles. The theoretical time-evolution of the spectral index and the observed profile were then compared to determine the theoretical spatial evolution of $\alpha$, and to estimate the source projected velocity ($\varv_\perp$) which is assumed to be constant for the entire time. Using the spectral index profile for this kind of analysis is particularly useful, as we remove the contribution of the flux normalisation in the analysis \citep{edler+22}.

\subsubsection{J1254-2913}
\label{sec:watJ1254-2913}
The new high-resolution uGMRT Band 4 and uGMRT Band 5 observations (see central right panel in Fig. \ref{fig:band5}) have shown an interesting and peculiar morphology of the wide-angle tail (WAT) in J1253-2913, with two jets being ejected from the inner core and deviating towards west, ending in the W lobe.

To understand the physical behaviour of this radio galaxy as shown in the high-resolution images (Fig. \ref{fig:band5}), we extracted the flux density and spectral index profile from 4\arcsec-size circles (i.e. the common resolution of the 700 MHz and 1260 MHz observations). We traced the path  along the `S'-shaped SW jet, from the inner region ({\it core}, Fig. \ref{fig:band5}) to the western lobe ( Fig. \ref{fig:profiles_J1254-2913}).  
The spectral profile of the source is well described by a standard JP model with $\alpha_{\rm inj}=-0.9$ and $\varv_\perp=140$ \kms, from the inner core up to the start of the SW jet (i.e. $\alpha\sim-2$, box \#6-7; see solid line in Fig.~\ref{fig:profiles_J1254-2913}). This velocity is typical of the WAT associated with BCGs in clusters' centre \citep[i.e. $\varv_{\rm gal}\sim100$ \kms;][]{o'dea+23}.
At the location where the southern jet deviates towards west (boxes \#8-10), the spectral index flattens back to $\alpha\sim-1$ for $\sim10$ kpc, and then steepens again when the jet turns into the lobe (up to $\alpha\sim-2$). Interestingly, this second steepening could be explained by a JP ageing model with same injected spectral index but with a higher velocity (i.e. $\sim340$ \kms, see dashed line in Fig.~\ref{fig:profiles_J1254-2913}), or with a shifted JP model with same velocity as the inner core (i.e. $\varv_\perp=140$ \kms, see dot-dashed line in Fig.~\ref{fig:profiles_J1254-2913}). This latter explanation would imply multiple events of activity of the AGN. If we consider the projected distance from the core to the location of the spectral flattening (i.e. $\sim10$ kpc) and the projected velocity $\varv_\perp=140$ \kms, we find that time interval between the two activity events is about 70 Myr.
At the location of the spectral flattening, hints of flux flattening are also present. On the other hand, the flux density profile along the SW jet appears to be rather constant up to the edge of the western lobe.

Alternatively, the `S'-shaped jet could be explained as a filament connected the two lobes, as also seen in ESO\,137-G006 \citep{koribalski+24}. However, the thin threads in ESO\,137-G006 have a steep spectral index ($\alpha\sim-2.5$) and no signs of flattening, as we instead measure in boxes \#9-10 in J1254-2913 ($\alpha\sim-1$; Figs. \ref{fig:spixRG} and \ref{fig:profiles_J1254-2913}). Considering the small opening angle of J1254-2913, and the limitation due to the projection on the plane of the sky, it results non-trivial to determine the nature of the thread/jet in this WAT.

\subsubsection{Radio tails}
The two radio tails in A3528N (see Fig. \ref{fig:profiles_J1254-2901a}) have brightness and spectral profiles that are typical of tailed radio galaxies, with a standard spectral index at the position of the AGN ($\sim-0.45$) which then steepens at the end of the tail (up to $\sim-1.1$), i.e. after $\sim 120$ kpc and $\sim70$ kpc for J1254-2901a and J1254-2904, respectively. We note the complex and asymmetric morphology of the two jets in J1254-2904, which probably outlines the combination of the cluster bulk motion and the galaxy's orbit in the cluster potential well. For this source, the profiles were extracted following only the western longer jet, up to the barred lobe.  
The profiles are typical from sources that follow the expected JP ageing model. For both, we assume a distribution of electrons with $\alpha_{\rm inj}=-0.45$, i.e. the spectral index value at the head of the tails, and a velocity $\varv_\perp\sim634$ \kms. Considering the position in the cluster (i.e. the galaxies' redshift of $z=0.0545$), these velocities appear high, although are still in line with the velocity dispersion of the cluster, i.e. $\sim 900$ \kms\ for A3528N  \citep{bardelli+01}.
Under these assumptions, we estimate that it takes about 150 Myr and 100 Myr for the plasma to age from the head to the end of the tail in J1254-2901a and J1254-2904, respectively.

The flux density and spectral index profiles for J1254-2916 in A3528S are shown in Fig. \ref{fig:profiles_J1254-2916}. For the first $\sim70$ kpc, i.e. just before the fork (box \#7), the JP model with $\alpha_{\rm inj}=-0.4$ and $\varv_\perp\sim390$ \kms\ reproduce well the spectral index profile, with values from $-0.4\pm0.1$ to $-1.6\pm0.1$. As in the fork (boxes \#7-10) the spectrum is constant around values of $-1.5\pm0.1$, and then successively steepens again up to $-3.0\pm0.4$, after $\sim 60$ kpc. From box \#15 to \#20 (i.e. $\sim60$ kpc), i.e. where the tail disappears in the uGMRT Band 4 and MeerKAT observations, we estimate an upper limit on the spectral index of $\alpha=-2.4$. Interestingly, we note that the second steepening in the tail (i.e. those from box \#10 to \#15) could be explained by the JP ageing model with a different HT projected velocity, i.e. $\varv_\perp\sim620$ \kms\ (see dashed line in Fig. \ref{fig:profiles_J1254-2916}). Similarly, the third steepening (boxes \#20-23) would agree with a source velocity of $\varv_\perp\sim1000$ \kms\ (see dot-dashed line in Fig. \ref{fig:profiles_J1254-2916}). This change of velocity, which is not included in our simple modelling, could occur during the galaxy motion in the cluster, which would decelerate after leaving the minimum point of the cluster potential well. This velocity would be more than twice the velocity dispersion of the sub-cluster \citep[i.e. $\sim454$ \kms, see][]{bardelli+01}. Moreover, the change of velocity would not explain the ``inverse'' gradient profile of the spectral index from boxes \#16 to \#20 (i.e. 175--220 kpc from the head of the tail). This flattening could be explained by re-acceleration processes triggered by the turbulence generated during the sloshing event between A3528S and A3528N, as also suggested for other literature examples such as Abell 1033 \citep{degasperin+17,edler+22}. In support of that, we note that the long tail of J1254-2916 lays in the proximity of the southern edge in the X-ray, which is highlighted in the Gaussian Gradient Magnitude \citep[GGM;][]{sanders+16} image with $\sigma=10$ pixels and in the comparison with the X-ray residual map (see middle panels in Fig. \ref{fig:xray}).

\subsubsection{Mushrooms in J1254-2913}

\begin{figure*}
\centering
\includegraphics[width=0.8\textwidth]{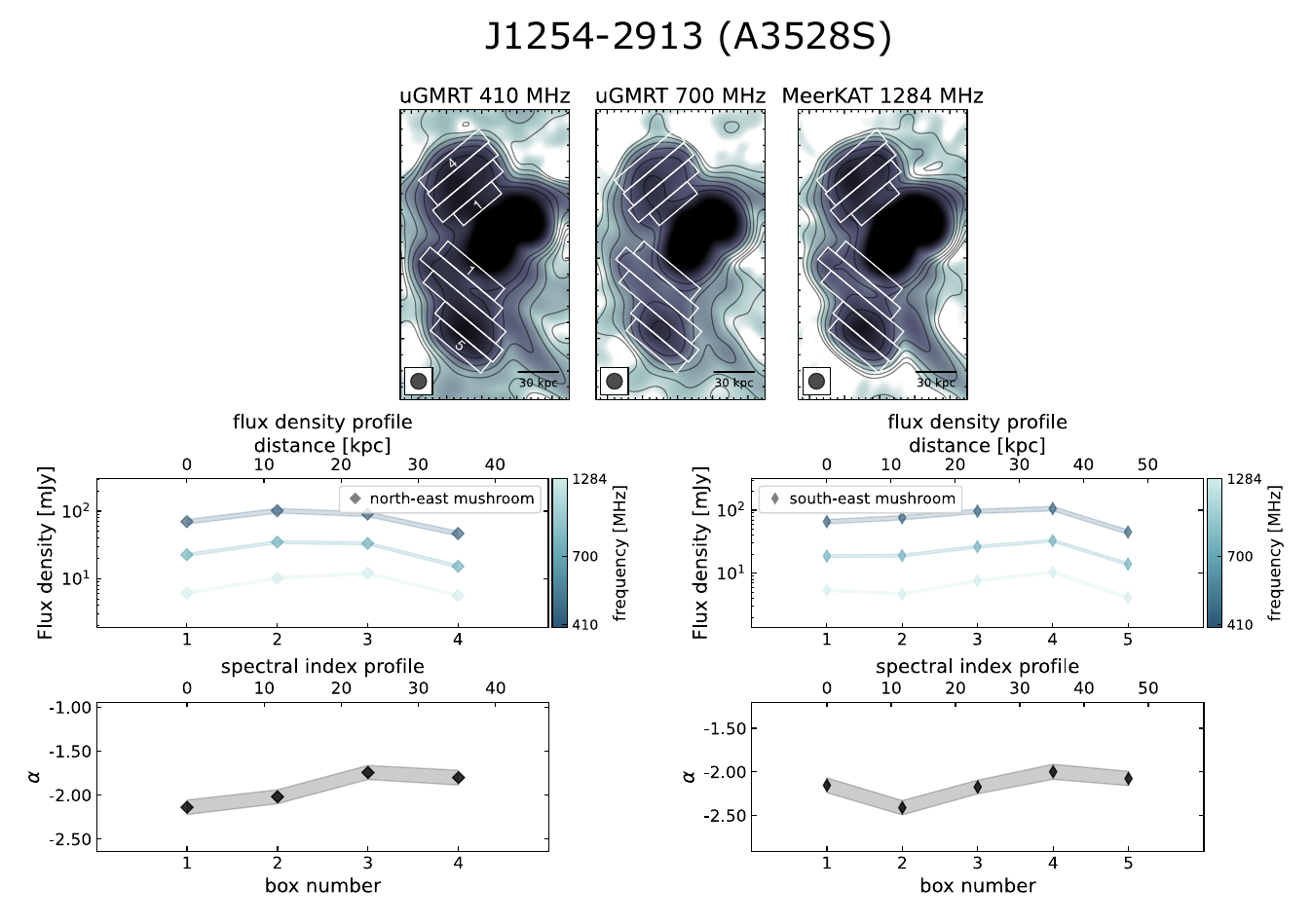}
\caption{Flux density and spectral index  profiles of the two mushrooms in J1254-2913. 12\arcsec-resolution images of the uGMRT Band 3 (top left) uGMRT Band 4 (top middle) and MeerKAT Lband (top right) are shown on the top of the profiles.}\label{fig:profiles_J1254-2913_mushrooms}
\end{figure*}

J1254-2913 is surrounded in the north-east and south-east by two mushrooms of emission. For these radio features, we extracted the flux density and spectral index profiles from rectangular boxes, with a height equal to the beam size (i.e. $\sim12''$) and width to cover the emission up to the $3\sigma_{\rm rsm}$ level, paying attention to avoid, in the SE mushroom, the emission associated with the southern filament (see Fig. \ref{fig:profiles_J1254-2913_mushrooms}).

The flux density profiles appear to be constant for the full extent of the two structures (i.e. $\sim35$ kpc and $\sim50$ kpc, for the NE and SE mushroom, respectively), at all observing frequencies. Interestingly, while the SE mushroom seems to have a quite constant spectral index profile ($\alpha\sim-2.1$), with a hint of a dip in box \#2, for the NE mushroom we detect a flattening towards the edge of the radio emission, from  $\alpha=2.1\pm0.1$ to $\alpha=-1.8\pm0.1$. Hints of cavities - or depressions in the X-ray surface brightness distribution - are visible at the location of the two mushrooms in the X-ray residuals (bottom row in Fig. \ref{fig:xray}), with the radio emission in the NE mushroom being confined by the positive residuals due to the sloshing motion. Assuming that these radio bubbles have been ejected by the AGN in J1254-2913, we can calculate the raising time as $t= H\,/\varv_{\rm mushroom}$, where $H$ is the position of the mushrooms with respect to the the AGN (being 35 kpc and 45 kpc for the NE and SE mushrooms, respectively) and $\varv_{\rm mushroom}$ their raising velocity. This latter parameter can be calculated using buoyancy arguments \citep[e.g.][]{brienza+21}:
\begin{equation}
\varv_{\rm mushroom} = \sqrt{\frac{2gV}{C \Phi}} \, ,
\end{equation}
where  $g=2\sigma^2/H$ \citep[with $\sigma$ the cluster velocity dispersion, being $\sim454$ \kms\ for A3528S, see][]{bardelli+01}
is the gravitational acceleration assuming hydrostatic equilibrium, $V=4/3 \pi ab^2$ and $\Phi=\pi ab$ the  volume and the cross-section of the mushroom, assuming ellipsoidal geometry (with $b=35$ kpc the minor semi-axis of the mushroom\footnote{Here, we assume that the third dimension of the bubble is the same as the minor ellipse semi-axis.}), and $C=0.75$ the drag coefficient \citep{churazov+01,brienza+21}. Using these parameters, we obtain $\varv_{\rm mushroom}\sim1200-1100$ \kms. Here we also assume a volume filling factor of 1, which is probably an overestimation, given the filamentary structure highlighted by the GGM image (Fig. \ref{fig:a3528s_ggm}). Assuming a filling factor of 0.5, the velocity of the {\it mushroom} would decrease to $\sim700$ \kms.

Alternatively, we can assume that the mushrooms are moving at the sounds speed $c_s$:
\begin{equation}
\varv_{\rm mushroom} = c_s = \sqrt{\Gamma \frac{k_BT}{\mu m_p}} \, ,
\end{equation}
where $\Gamma=5/3$ is the adiabatic index, $k_BT$ is the ICM temperature at the location of the mushroom which can be obtained from the X-ray observations, $\mu=0.62$ is the mean molecular weight, and $m_p=1.67\times10^{-24}$ g is the proton mass. Using XMM-Newton observations, \cite{gastaldello+03} measured a temperature of $k_BT=4-4.5$ keV at the mushrooms locations, leading to $\varv_{\rm mushroom}\sim1000$ \kms, in line with the velocity found using buoyancy arguments. With this velocity and assuming that the {\it mushrooms} are in the same plane of the sky as the cluster centre and neglecting the velocity of the radio galaxy, the  raising time of the NE and SE mushroom-shaped sources is $\sim30$ Myr and $\sim40$ Myr, respectively.

\subsubsection{Filaments around the BCG}

\begin{figure*}
\centering
\includegraphics[width=0.9\textwidth]{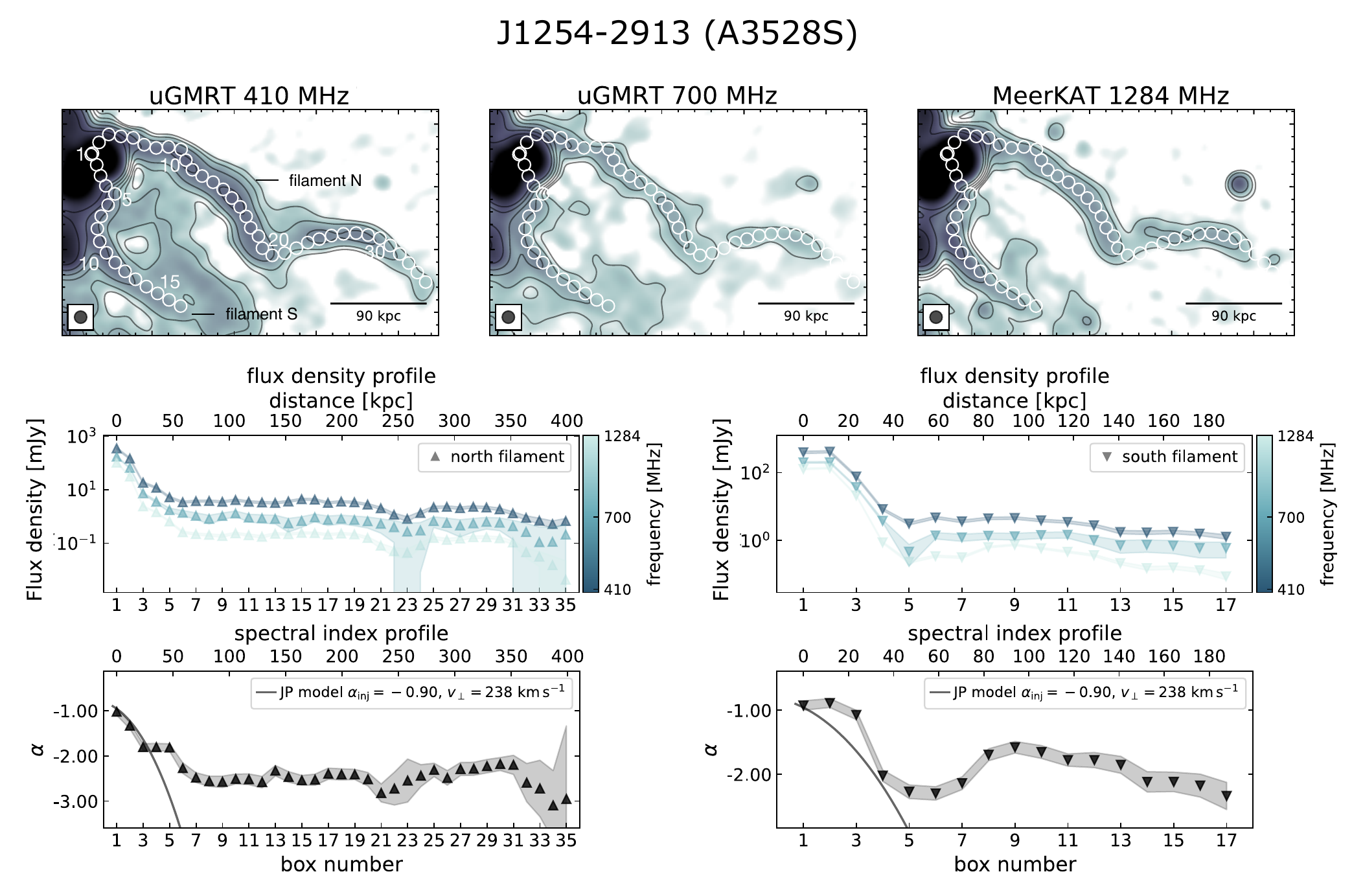}
\caption{Flux density and spectral index profiles of the two filaments in A3528S. Top row: 12\arcsec-resolution images of the uGMRT Band 3 (left), uGMRT Band 4 (central) and MeerKAT L-band (right). Bottom left: profiles for the northern filament; bottom right: profiles for the southern filament.}\label{fig:profiles_A3528S_filament}
\end{figure*}

For both filaments in A3528S, we extracted the flux density and spectral index profiles starting from the location of J1243-2913 core. For the northern filament, we then proceeded within the NW lobe (boxes 2 and 3 in the top panels in Fig. \ref{fig:profiles_A3528S_filament}) and then proceeded for the full extent of the filament ($\sim350$ kpc). The spectral index profile (bottom left panel in Fig. \ref{fig:profiles_A3528S_filament}) is consistent with a JP model with $\alpha_{\rm inj}=-0.9$ for these first three boxes, namely for the WAT source, assuming a projected velocity of $\rm\sim240$~\kms\ (consistent with the finding shown in Sect. \ref{sec:watJ1254-2913}). These values translate to a radiative ageing time of $t_{\rm rad}\sim100$ Myr for the WAT.

The filament does not follow the predicted ageing profile of a JP model, but the spectral index reaches a plateau around $\alpha\sim-2.5$ and remains constant for about 200 kpc, i.e. up to when the filament breaks to the `V'-shape (box \#21). Similarly, the flux density profiles also remain approximately constant along the full extent of the filament, and well above the noise level (see Tab. \ref{tab:images}). At the location of the `V'-shape break, we measure a mild steepening of the spectrum ($\alpha\sim-2.8$), to then get flatter again to values of $\alpha\sim-2$ for additional $\sim 100$ kpc. A steepening occurs again after box \#32 (i.e. after $\sim100$ kpc, with  $\alpha\sim-3$), although here the filament is only clearly detected in the uGMRT Band 3 image. 
On the other hand, the morphology of the southern filament in A3528S is more complex, as presented in Sect.~\ref{sec:results}. In this case, the radio filament does not seem to be ejected directly from the radio galaxy lobe but it proceeds south-west the galaxy, i.e. in the opposite direction of its putative motion. We therefore follow this morphology, then following the bay and the rest of the filament (see top panels in Fig.~\ref{fig:profiles_A3528S_filament}). The same JP ageing model assumed for the northern filament (i.e. $\alpha_{\rm inj}=-0.9$ and $\varv_\perp\sim 240$ \kms) explains well the spectral profile up to the beginning of the `bay' (i.e. box \#5, see bottom right panel in Fig. \ref{fig:profiles_A3528S_filament}). The filament bay (boxes \#5-8) shows a constant spectral index profile, with values of $\sim-2.2$; then the spectral index flattens ($\alpha\sim-1.6$) and gradually steepens again to $\alpha\sim-2.3$ at the end of the filament (i.e. $\sim100$ kpc).

\begin{figure*}
\centering
\includegraphics[width=0.9\textwidth]{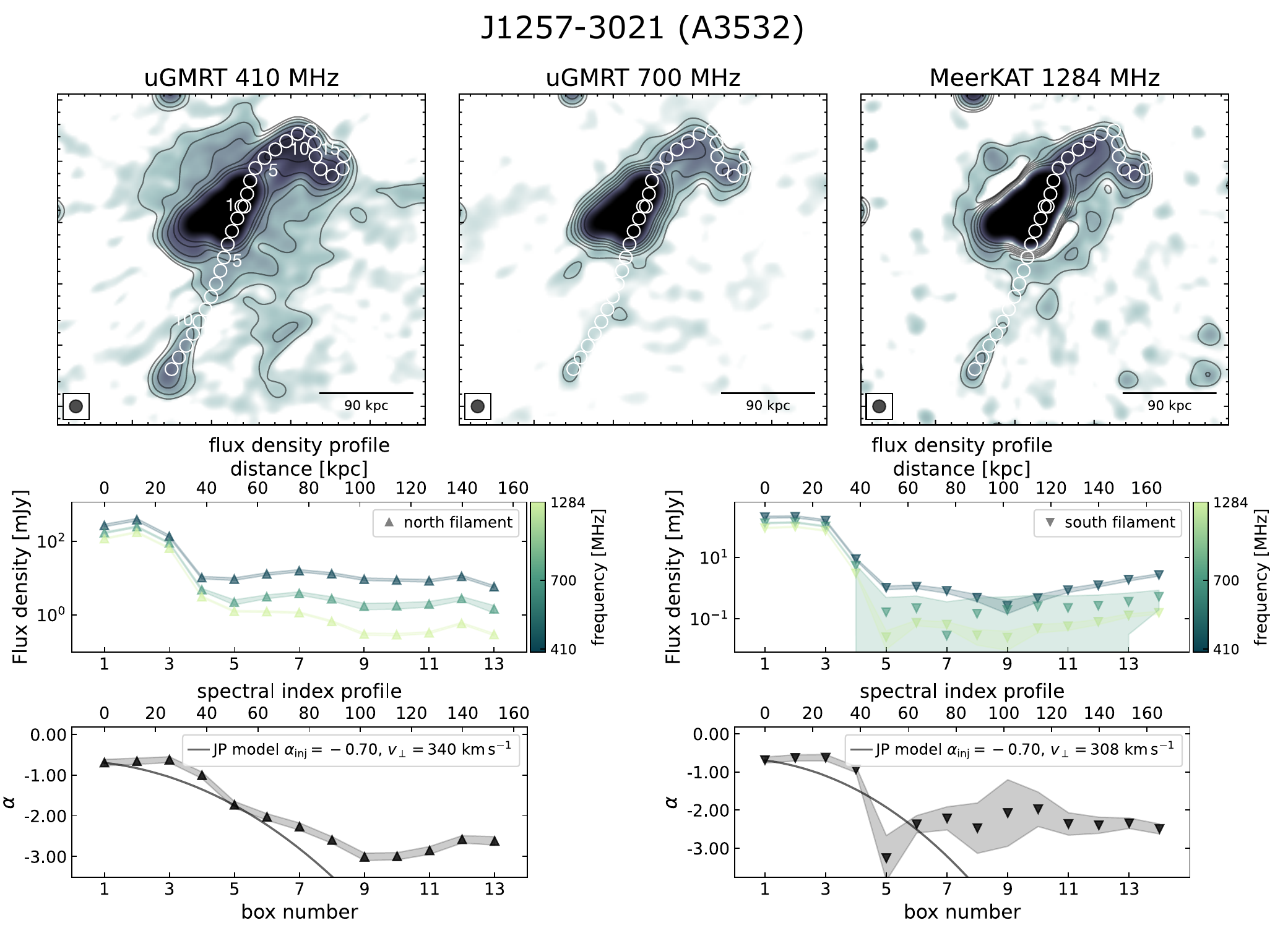}
\caption{Flux density and spectral index profiles of the two filaments in A3532. Top row: 12\arcsec-resolution images of the uGMRT Band 3 (left), uGMRT Band 4 (central) and MeerKAT L-band (right). Bottom left: profiles for the northern filament; bottom right: profiles for the southern filament.}\label{fig:profiles_A3532_filament}
\end{figure*}

The filaments in A3532 also have complex morphology. The BCG, J1257-3021, is supposedly moving towards west, as also suggested by the direction of the two lobes (see bottom right panels in Fig.~\ref{fig:band5}). Although the two filaments do not start from the outermost edge of the two lobes, we extracted the profile starting from the location of the core. For both filaments the first three boxes, i.e. those tracing the $\sim30$ kpc of the two lobes, display a flat spectral index and flux density profiles, then steepening towards the external edges of the lobes (Fig. \ref{fig:profiles_A3532_filament}). These are well-described by a JP model with $\alpha_{\rm inj}=-0.7$ and $\varv_\perp\sim320-340$ \kms. The two filaments then start to deviate from this model, presenting a plateau in the flux density profiles. The spectral index profile in the northern filament still shows a gradient, although it steepens more slowly than the JP model, and it flattens around $\alpha\sim -2.7$, as in the comma feature. For the southern filaments, the spectral index profiles are less trivial to interpret, as it is undetected in the uGMRT Band 4 and, partly, in the MeerKAT observations. If we only take into account boxes \#10-14, where we detect the filament at least at two observing frequencies, we measure a constant spectral index of $\alpha\sim-2.4$.

The spectral index profile in the radio filaments in A3528S and A3532 suggests that those are unlikely simply aged plasma from the AGN. Assuming these are moving at the same velocity as the galaxies (i.e. $\varv_\perp\sim140$~\kms\ for J1254-2913 and $\varv_\perp\sim300-340$~\kms\ for J1257-3021), it would imply a age of the plasma of about 1.2 Gyr and 400 Myr for the northern and southern filaments in A3528S, and about 300 Myr for the northern and southern filaments in A3532. These plasma ages - and particularly that for the northern filament in A3528S - are unrealistic. This, and the spectral properties of the two filaments, suggest that some form of re-acceleration must have occurred in these two clusters.

Shocks, cold fronts and adiabatic compression have been observed to play a role in (re-)accelerating electrons, and displace the relativistic plasma in the ICM \citep[e.g., radio phoenices;][]{ensslin+gopal-krishna01}. The coincidence with these kinds of discontinuities, and their interplay with the cluster magnetic field lines, were a possible explanation for the brightness and spectral profiles in the two jets in the BCG in A3376 \citep{chibueze+21}, and in the radio filament detected at the edge of the HT J1333-3141 in another cluster in the Shapley Supercluster, i.e A3562 \citep{giacintucci+22}. For these cases, the radio emission associated with the radio galaxies is explained with magnetic reconnection (and therefore in-situ re-acceleration) between the radio jets and the cold front or with fast diffusion of cosmic-ray electrons along magnetic field lines stretched under the sloshing cold front, respectively.
However, these processes are unlikely to explain steep and uniform spectral indices in region of radio emission that extend for a few hundreds of kpc, as they would require a simultaneous particle re-acceleration by the shock for the full extension of the filaments. One clear example are the so-called GReETs \citep[i.e. Gently Re-Energized Tails, see][]{degasperin+17}, where a mild and continuous re-acceleration of aged (fossil) relativistic electrons in the tails and lobes of radio galaxies (and balancing their radiative losses) is proposed to be induced by merger-induced turbulence. A case similar to this was also reported in A1550 \citep{pasini+22}.

In order to investigate the possible presence of disturbed ICM in the clusters, we compared the thermal emission from new eROSITA observations \citep[Sanders et al., in prep.]{merloni+24,bulbul+24,kluge+24} with our radio images (Fig. \ref{fig:xray}). Although it is clear that two peaks (A3528N and A3528S) are in an off-axis merging state (see Appendix \ref{apx:betamodel} and Fig. \ref{fig:A3528_beta}) -- with A3528N falling from NW to SE and A3528S moving from NE to SW \citep[as also shown by][]{gastaldello+03} -- the two radio filaments in A3528S and the long tail in J1254-2916 do not coincide with any clear thermal discontinuities (e.g. cold front or shock front). It is important to note that, contrary to the cases described above, we only have in hands shallow X-ray observations of the \target, which can hide potential faint thermal discontinuities. Using the residual maps shown in the bottom row in Fig. \ref{fig:xray}, we searched for hints of a correlation between X-ray fluctuations and the radio emission. 
The fluctuations in the X-ray surface brightness were quantified as the sum of the residuals squared within cluster-centred annuli, i.e. $\Delta S_X=\sum \delta S_X^2/A$, where $\delta S_X$ is the residual after subtracting the $\beta$-model (see Appendix \ref{apx:betamodel}) and $A$ is the number of pixel in the annulus. We normalised the fluctuations to the number of pixels in the region, to take the different areas into account. 
Specifically, we investigated the case of A3528S as it represents the clearest example of extended emission covering strong X-ray fluctuations (bottom central panel in Fig. \ref{fig:xray}). In this case, we derived $\Delta S_X\sim0.25$ for radius $r<5^\prime$ (i.e. $r\lesssim350$ kpc), i.e. where there is the radio emission, and $\Delta S_X\sim0.07$, for $5^\prime<r<9^\prime$ (i.e. $350<r<600$ kpc), see red annuli in Fig. \ref{fig:fluctuations} in Appendix \ref{apx:betamodel}. Although this is a simplified approach, this finding supports the idea that the radio emission in the \target\ is associated with disturbed ICM, i.e. where the highest fluctuations are. In Appendix \ref{apx:betamodel}, we also show the correlation between the radio flux density and the X-ray surface brightness fluctuations from the {\tt taper=10\arcsec} beam-size annuli (Fig. \ref{fig:fluctuations}).
The estimation of the power spectra, $P(k)$, which is also linked to the X-ray surface brightness fluctuations, is beyond the scope of this paper.

\begin{table*}[]
\caption{Summary of the {\tt Halo-FDCA} elliptical fitting results, as well as the ``manual'' flux densities.}\label{tab:minihalo}
\vspace{-5mm}
\begin{center}
\resizebox{\textwidth}{!}{
\begin{tabular}{ccccccccc}
\hline
\hline
Frequency & Central brightness & Major {\it e}-folding radius & Minor {\it e}-folding radius & Goodness fit & Fitted flux density & Image flux density & Fitted radio power & Image radio power \\
$\nu$ [MHz] & $I_0~[\mu{\rm Jy\,arcsec}^{-2}]$ & $r_{e,1}$ [kpc] & $r_{e,2}$ [kpc] & $\chi^2/{\rm dof}$ & $S_{<3re}$ [mJy]  &  $S_{\rm img}$ [mJy] &  $P_{<3re}~[{\rm W\,Hz^{-1}}]$  &  $P_{\rm img}~[{\rm W\,Hz^{-1}}]$ \\
\hline
410 & $6.2\pm0.4$ & $83.9^{+5.1}_{-4.8}$ & $57.6^{+3.4}_{-3.2}$ & $922.13/891$ & $158.9\pm7.8$ & $185.7\pm15.1$ & $(9.5\pm0.5)\times10^{23}$ & $(1.1\pm0.1)\times10^{24}$ \\
700 & $3.4\pm0.2$ & $127.6^{+9.6}_{-8.9}$ & $48.6\pm4.0$ & $742.65/683$ & $111.9\pm8.2$ & $66.1\pm3.6$ & $(6.7\pm0.5)\times10^{23}$ & $(4.1\pm0.2)\times10^{23}$\\
1284 & N/A & N/A & N/A & N/A & $76.3\pm15.0^\dagger$ & $8.4\pm1.0$ & $(4.1\pm0.2)\times10^{23}$ & $(5.2\pm0.6)\times10^{22}$\\
\hline
\end{tabular}}
\end{center}
\vspace{-5mm}
\tablefoot{$^\dagger$Flux density extrapolated from the flux densities at 410 MHz and 700 MHz. The image flux density was calculated subtracting the flux density within the white solid lines in Fig. \ref{fig:minihalo} from a region surrounding the candidate mini-halo (dashed line in Fig. \ref{fig:minihalo}). Uncertainties were estimated summing in quadrature the uncertainties of each flux desnity (i.e. `total' and 'masked') following Eq. \ref{eq:fluxerr}.}
\end{table*}

\begin{figure*}
\centering
\includegraphics[width=\textwidth]{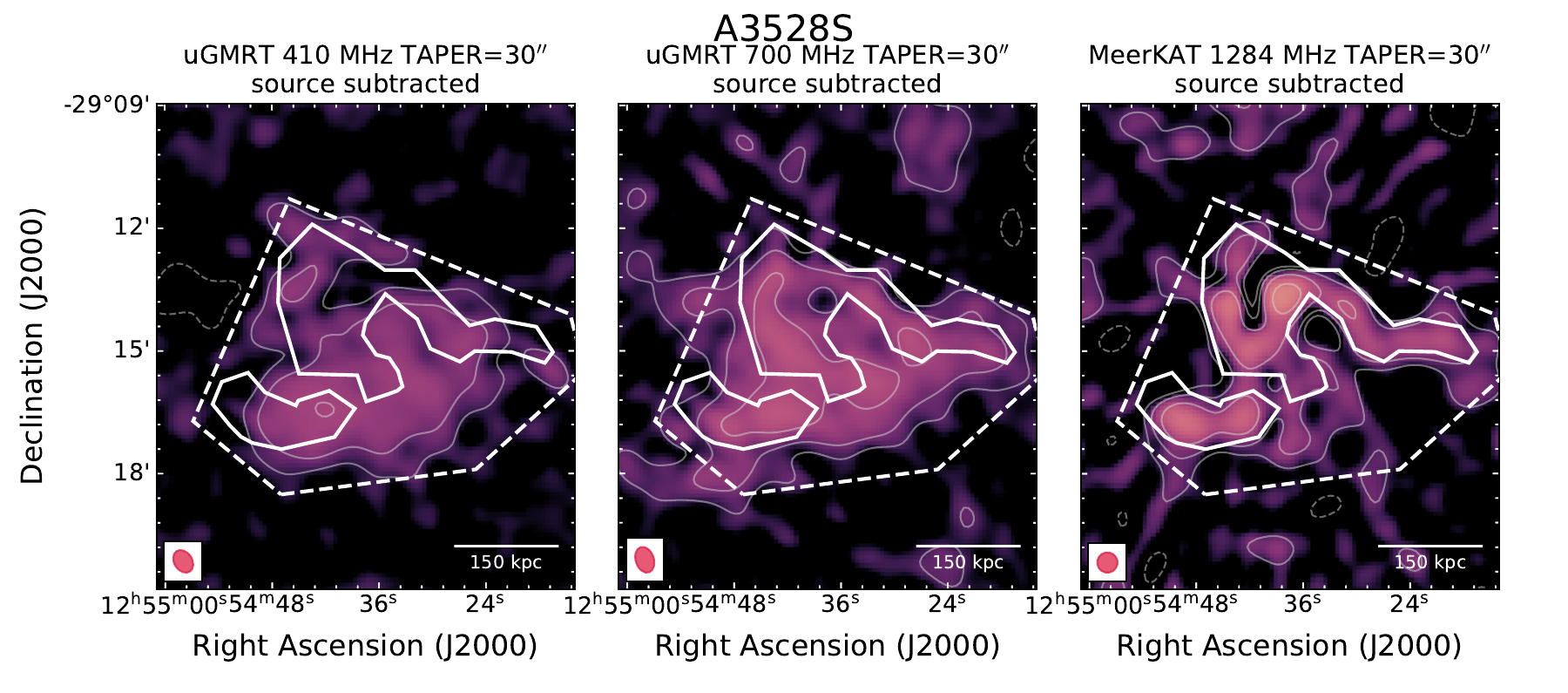}
\vspace{-8mm}
\caption{Source subtracted (with inner $uv$-cut of $2000\lambda$, i.e. 100 kpc) images of A3528S with a taper of 30\arcsec. Radio contours are drawn at $3\sigma_{\rm rms}\times[-1,1,2,4,8,16,32]$, with $\rm \sigma_{rms,410}=260~\mu Jy\,beam^{-1}$, $\rm \sigma_{rms,700}=144~\mu Jy\,beam^{-1}$ and $\rm \sigma_{rms,1280}=48~\mu Jy\,beam^{-1}$. 
Solid white regions show the location of the radio galaxies from the same-resolution image, which are masked for the {\tt Halo-FDCA} fitting \citep[see Appendix \ref{apx:minihalo}]{boxelaar+21}. The dashed region is the area used to calculate the flux density of the candidate radio minihalo from the image.}\label{fig:minihalo}
\end{figure*}

Finally, it is also unclear whether these filaments are polarised, as no polarisation signal has been recorded in our MeerKAT observations. Similarly collimated radio filaments have been observed outside of radio galaxies, and have been shown to have high polarisation fraction (up to 50\%) and small Faraday Rotation variation \citep{rudnick+22}.
An estimate of the magnetic fields in these filaments can be obtained assuming equipartition arguments:

\begin{equation}
B_{\rm eq} = \sqrt{ \frac{24\pi}{7} \frac{E_{\rm tot,min}}{V} } \quad [{\rm Gauss}]\, ,
\end{equation}
where $E_{\rm tot,min}/V=8.6\times10^{-24}(1+k)^{4/7}(P_{\rm1.4GHz}/V)^{4/7}$. In this calculation, we assume a cylindrical geometry ($V=\pi r^2L$, where $r$ is the width and $L$ the length of the filaments in kpc, respectively) and an electron-to-ion ratio of $k=1$. This latter assumption is probably an underestimation, as it refers to the ratio at the time of the injection when the electrons have not yet suffered energy losses, whilst the measured from spectral indices in the filaments suggest old plasma. For A3528S, we obtain $B_{\rm eq}\sim1~\mu$Gauss and $B_{\rm eq}\sim1.5~\mu$Gauss for filament N ($r=12$ kpc and $L=350$ kpc) and S ($r=12$ kpc and $L=100$ kpc), respectively. For A3532, we obtain $B_{\rm eq}\sim1.7~\mu$Gauss and $B_{\rm eq}\sim0.9~\mu$Gauss for filament N and S ($r=12$ kpc and $L=100$ kpc), respectively. These values would increase of a factor 3 at least, if we would assume $k\gg1$ \citep{rybicki+ligthman86}.

\subsection{A displaced radio mini-halo?}
The low resolution images of A3528S (see Appendix \ref{apx:lowres}, Fig. \ref{fig:images_A3528Salltelescope_allres}) also revealed the presence of additional diffuse radio emission with physical size of about 200 kpc in-between the two filaments (Figs. \ref{fig:images_alltelescope_fullres} and \ref{fig:images_A3528Salltelescope_allres}). 
At 410 MHz and 700 MHz, the emission becomes more prominent after the subtraction of compact sources (i.e. with the inner {\it uv}-cut of $\sim2000\lambda$ to exclude all the emission below 100 kpc), after convolving to 30\arcsec\ resolution (see Fig. \ref{fig:minihalo}). At MeerKAT frequencies, this radio emission disappears, either because it becomes too faint at 1.28 GHz to be detected, or because the size of the emission is included in the subtracted visibilities.
This emission seems to extend in the north-west/south-east direction, and does not seem to coincide with the peak of the thermal emission (see middle panel in Fig. \ref{fig:xray}).

Following \cite{murgia+09} for the description of the radial brightness profile of radio (mini-)halos, we assumed that the extended diffuse radio emission in-between the two filaments in A3528S could be modelled as $I(\vec{r})=I_0\exp(-|\vec{r}|/r_e)$, where $I_0$ is the central surface brightness and $r_e$ the $e$-folding radius (i.e. the radius at which the brightness drops to $I_0/e$), which relates with the physical size of the extended diffuse emission according to $R=3r_e$.
For this analysis, we used the Halo-Flux Density CAlculator \citep[{\tt Halo-FDCA}\footnote{\url{https://github.com/JortBox/Halo-FDCA}};][]{boxelaar+21} fitting procedure, assuming elliptical geometry due to the elongated shape of the emission. We used the source-subtracted 30\arcsec\ uGMRT Band 3 and uGMRT Band 4 images, after masking the region where the diffuse emission associated with J1254-2913 and J1254-2916 were located (see white masks in Fig. \ref{fig:minihalo}), and then integrating up to $3r_e$. The fitting results at the two frequencies, are shown in Appendix \ref{apx:minihalo}.
For comparison, we also estimated the flux density from the radio images ($S_{\rm img}$). In this case, we defined a region around the diffuse radio emission in the 30\arcsec\ uGMRT Band 3, uGMRT Band 4 and MeerKAT L-band images (see dashed region in Fig. \ref{fig:minihalo}), and then we manually subtracted the flux density associated with the diffuse emission in J1254-2913 and J1254-2916 (see solid regions in Fig. \ref{fig:minihalo}). The results of the two analyses are listed in Tab.~\ref{tab:minihalo}. While for the uGMRT Band 3 measurements we find a good agreement (i.e. $S_{<3r_e}=158.9\pm7.8$ and $S_{\rm img}=185.7.6\pm15.1$), the flux densities at 700 MHz have a large discrepancy (i.e. $S_{<3r_e}=111.9\pm8.2$ and $S_{\rm img}=66.1\pm3.6$). This could be due to a less accurate source subtraction outside the diffuse radio emission and could also explain why the major axis of the ellipse (i.e. $r_{e,1}$) is larger at 700 MHz than at 410 MHz. 
Using the flux densities at 410 MHz and 700 MHz, we obtain $\alpha_{\rm fit}=-0.66\pm0.16$ and $\alpha_{\rm img}=-1.9\pm0.2$. The spectral index value from the fit is flatter than that usually reported from mini-halos \citep{giacintucci+19}. However, contrary to the image flux density measurements where the halo region is fixed, the {\tt Halo-FDCA} fit is sensitive to different region (see $r_e$ in Tab. \ref{tab:minihalo}). 

The flux density from the 1280 MHz image is $S_{\rm img}=8.4\pm1.0$ mJy, corresponding to a radio luminosity of $P_{\rm 1.4GHz}=(4.4\pm0.6)\times10^{22}~{\rm W\,Hz^{-1}}$, assuming the $\alpha=-1.9\pm0.2$. This radio power and the eROSITA bolometric luminosity $L_{X,500}=(1.8\pm0.2)\times10^{44}~{\rm erg\,s^{-1}}$ would be in line with the correlation found for literature sample of mini-halos, although in the low (radio and X-ray) luminosity regime \citep{giacintucci+19}. 
We also compared this extended radio emission with the largest sample of giant-/mini-halos in the framework of the LOFAR Two-Meter Sky Survey \citep[LoTSS;][]{shimwell+19} sample. Scaling the radio power to 150 MHz (assuming $\alpha$ in the range $[-1.9,-0.6]$), the candidate mini-halo would agrees well the $P_{\rm150MHz}-k_BT^2\,M_{\rm gas}$ relation presented in \cite{zhang+23}, which unifies the diffuse radio emission over the small (i.e. mini-halos) and large (i.e. giant-halos) scales.

Despite the unusual location with respect to the ICM, and the challenges to completely remove the radio emission associated with the radio galaxies, the shape and the spectral properties support the scenario where this radio source is a candidate mini-halo.

\section{Conclusion}
\label{sec:concl}

In this paper, we presented a multi-frequency study of the \target\ in the Shapley Concentration Core, which includes four clusters, namely A3528N, A3528S, A3532 and A3530. Particularly, we focus on the radio emission associated with the brightest cluster galaxies (BCGs) and tailed radio galaxies, using data from the uGMRT Band 3 (250--500 MHz), Band 4 (550--900 MHz) and Band 5 (1000--1460 MHz), and MeerKAT L-band (900--1670 MHz) observations. These are complemented with the thermal information given by five observing cycles of the eROSITA All-Sky Survey (eRASS:5). Below, we summarise the results of our work.

\begin{itemize}
\item[$\bullet$] We detect a plethora of diffuse radio emission in the three radio-loud clusters in the complex, namely A3528N, A3528S and A3532. Part of this radio emission was never detected in previous radio observations \citep{venturi+01,digennaro+18b}. Given its radio-quiteness nature, A3530 was excluded by the observations.

\smallskip
\item[$\bullet$] The morphological parameters of the intracluster medium (ICM), as well as the residual maps after subtracting a standard $\beta$-model for the ICM surface brightness, as inferred from the eROSITA images \citep[Sanders et al. in prep.]{merloni+24,bulbul+24,kluge+24} indicate that the three clusters are mildly disturbed, in agreement with a previous analysis with XMM-Newton \citep{gastaldello+03} and Chandra \citep{lakhchaura+13}.

\smallskip
\item[$\bullet$] We find different morphologies and spectral characteristics in the BCGs of the three clusters. J1254-2900 in A3528N has a S-shape and no further diffuse radio emission. J1254-2913 in A3528S is a wide-angle tail with opening angle of $\sim120^\circ$. The highest-resolution image ($\Theta_{1260}=2.9''\times2.2''$) reveals the presence of two collimated unresolved jets, with the southern one being ejected towards south-east and then turning west. Additionally, around this radio source we detect two {\it mushrooms} north-east and south-east of it, respectively, and two radio filaments in the south-west. Finally, J1257-3021 in A3532 is also a wide-angle tail with opening angle of $180^\circ$ and with the two lobes pointing towards east. As for J1254-2913, we detect the presence of two radio filaments, in the north and in the south of the radio source. All these filaments extend over large scales, ranging from $\sim200$ up to $\sim400$ kpc. No particular thermal features are revealed from the eRASS:5 image, at the location of this diffuse radio emission.

\smallskip
\item[$\bullet$] The tailed radio galaxies in A3528N present a standard spectral index profile from the head ($\alpha\sim-0.4$) to the end of the tail ($\alpha\sim-1.0$). This profile can be explained by standard ageing models, i.e. a Jaffe-Perola (JP) model. On the contrary, the long head-tail in A3528S can be explained by a JP ageing model only for the first $\sim 60$ kpc. The spectral index profile of the remaining part of the tail can be either explained by re-acceleration due to sloshing core or by a change in the projected velocity of the galaxy within the ICM.

\smallskip
\item[$\bullet$]
We confirm the presence of two ``{\it mushrooms}'' around in J1254-2913, with spectral index values of $\sim-2$. Assuming buoyancy arguments, we obtain rising velocities of $\sim1000$ \kms.

\smallskip
\item[$\bullet$] We measure steep spectra for the radio filaments in J1254-2913 and in J1257-3021, with values of $\alpha\sim-2,-2.5$ that remain almost constant over the filament length. If we assume that these filaments are ejected from the radio galaxies, the spectral index profile strongly deviates from the standard JP models. Assuming equipartition arguments, we also estimated a lower limit of $B_{\rm eq}\sim1-2~\mu$Gauss for these filaments.

\smallskip
\item[$\bullet$] We also find hints of further diffuse emission in-between the two filaments of A3528S at all radio frequencies. Despite the uncertainties due to the subtraction of the extended emission associated with the radio galaxies and despite the displacement of this emission with respect to the thermal footprint of the cluster, we classified this emission as a candidate mini-halo. The position of this radio source in the standard $P_{\rm 1.4GHz}-L_{X,500}$ diagram, and in the $P_{\rm 150MHz}-(k_BT)^2M_{\rm gas}$ are consistent with literature studies \citep[respectively]{giacintucci+19,zhang+23}.

\end{itemize}

Our results show that the presence of diffuse radio emission in the form of filaments and bubbles of plasma can be common in undisturbed or mildly disturbed galaxy clusters. The spatial proximity with radio galaxies strongly supports the scenario where these are the main sources of populations of fossil plasma in clusters.
Despite the lack of clear thermal discontinuities for the X-ray images in the \target, the flatter spectral index of these diffuse radio emission than that expected from ageing models and the hints of correlations with increasing X-ray fluctuations suggest some form of particle (mild) re-acceleration. This could also explain the origin of additional diffuse radio emission, such as mini-halos.

\begin{acknowledgements}
We thank the referee for the suggestions which improved the quality of the manuscript.
GDG and MB acknowledges funding by the DFG under Germany's Excellence Strategy -- EXC 2121 ``Quantum Universe'' --  390833306.
EB, AL, and XZ acknowledge financial support from the European Research Council (ERC) Consolidator Grant under the European Union’s Horizon 2020 research and innovation program (grant agreement CoG DarkQuest No 101002585).
KT acknowledges financial assistance by the SARAO (\url{www.sarao.ac.za}).
This paper is based on data obtained with the Giant Metrewave Radio Telescope (GMRT). We thank the staff of the GMRT that made these observations possible. GMRT is run by the National Centre for Radio Astrophysics of the Tata Institute of Fundamental Research.
The MeerKAT telescope is operated by the South African Radio Astronomy Observatory, which is a facility of the National Research Foundation, an agency of the Department of Science and Innovation.
This work is based on data from eROSITA, the soft X-ray instrument aboard SRG, a joint Russian-German science mission supported by the Russian Space Agency (Roskosmos), in the interests of the Russian Academy of Sciences represented by its Space Research Institute (IKI), and the Deutsches Zentrum f\"ur Luft- und Raumfahrt (DLR). The SRG spacecraft was built by Lavochkin Association (NPOL) and its subcontractors, and is operated by NPOL with support from the Max Planck Institute for Extraterrestrial Physics (MPE). The development and construction of the eROSITA X-ray instrument was led by MPE, with contributions from the Dr. Karl Remeis Observatory Bamberg \& ECAP (FAU Erlangen-Nuernberg), the University of Hamburg Observatory, the Leibniz Institute for Astrophysics Potsdam (AIP), and the Institute for Astronomy and Astrophysics of the University of T\"ubingen, with the support of DLR and the Max Planck Society. The Argelander Institute for Astronomy of the University of Bonn and the Ludwig Maximilians Universit\"at Munich also participated in the science preparation for eROSITA.
The data published here have been reduced using the CARACal pipeline, partially supported by ERC Starting grant number 679627 ``FORNAX'', MAECI Grant Number ZA18GR02, DST-NRF Grant Number 113121 as part of the ISARP Joint Research Scheme, and BMBF project 05A17PC2 for D-MeerKAT. Information about CARACal can be obtained online under the URL: \url{https://caracal.readthedocs.io}.
This research made use of {\tt APLpy}, an open-source plotting package for Python \citep{aplpy}. Basic research in radio astronomy at the Naval Research Laboratory is supported by 6.1 Base funding.
\end{acknowledgements}

\bibliographystyle{aa}
\bibliography{biblio.bib}

\clearpage
\onecolumn
\begin{appendix}
\clearpage

\section{Low-resolution images}\label{apx:lowres}
\begin{figure*}[h!]
\includegraphics[width=\textwidth]{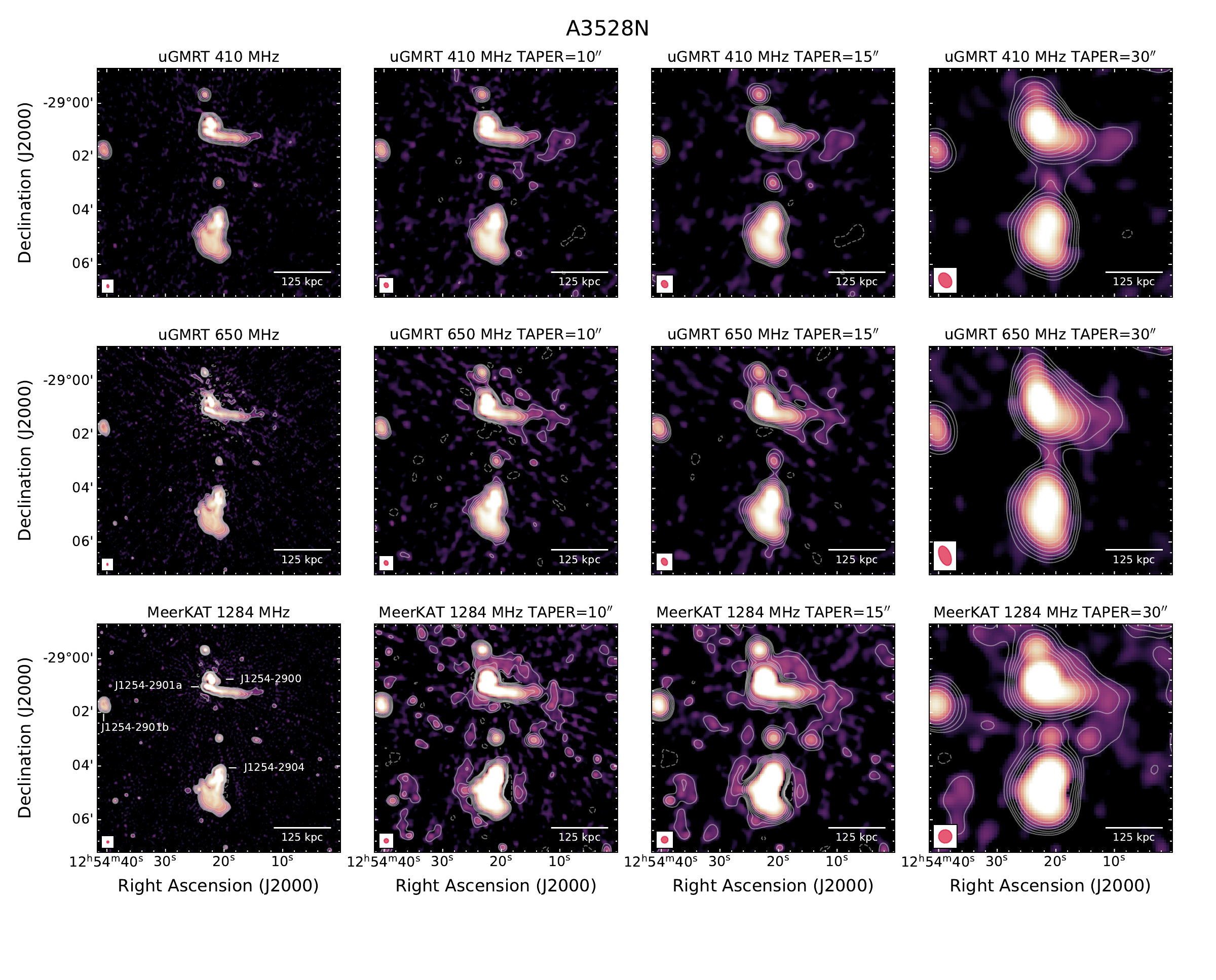}
\vspace{-15mm}
\caption{Full- (left column) and low-resolution (centre-left, centre-right and right columns) images of A3528N. Top row: uGMRT Band 3; central row: uGMRT Band 4; bottom row: MeerKAT L-band.}\label{fig:images_A3528Nalltelescope_allres}
\end{figure*}

\begin{figure*}[h!]
\includegraphics[width=\textwidth]{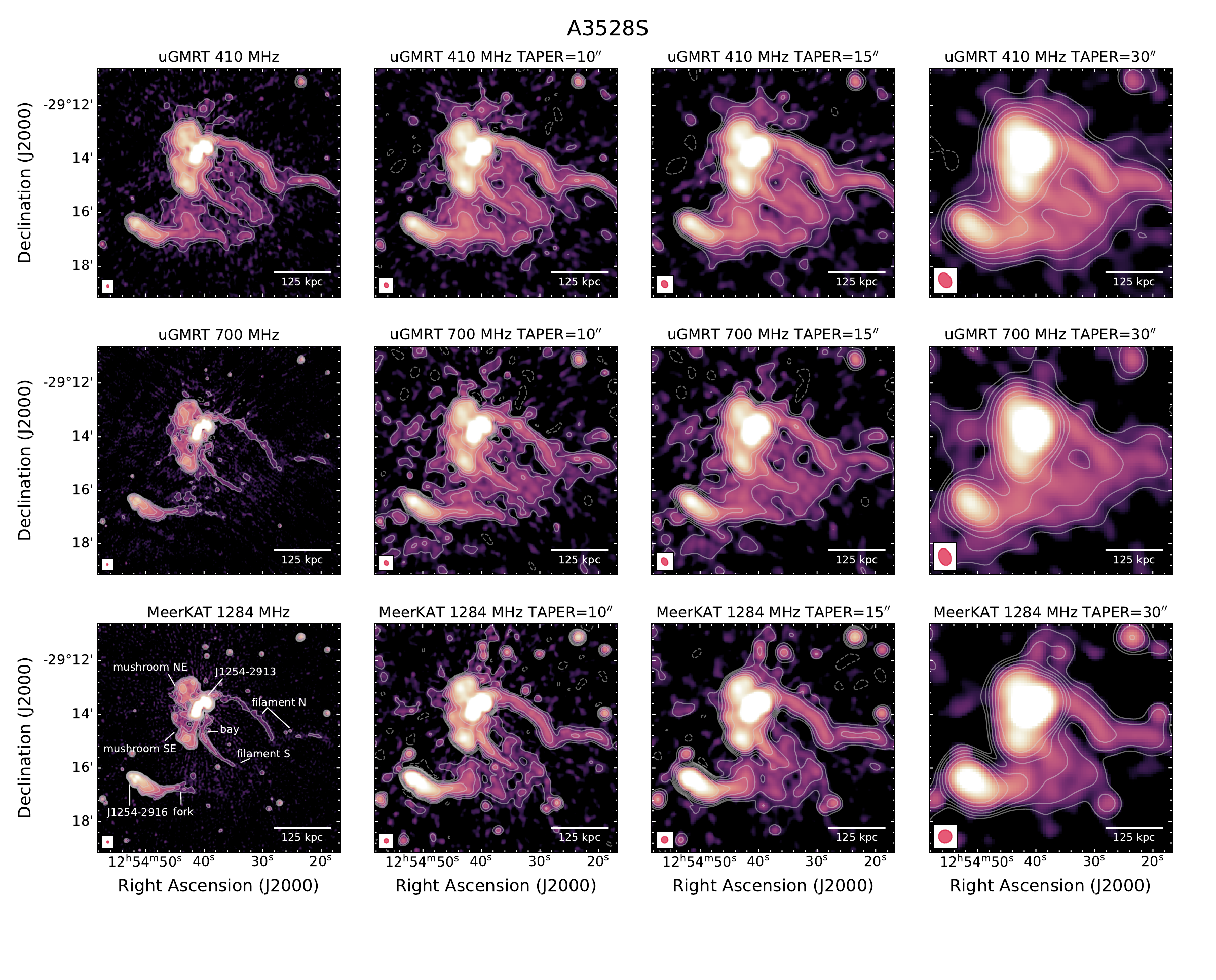}
\vspace{-15mm}
\caption{Full- (left column) and low-resolution (centre-left, centre-right and right columns) images of A3528S. Top row: uGMRT Band 3; central row: uGMRT Band 4; bottom row: MeerKAT L-band.}\label{fig:images_A3528Salltelescope_allres}
\end{figure*}

\begin{figure*}[h!]
\includegraphics[width=\textwidth]{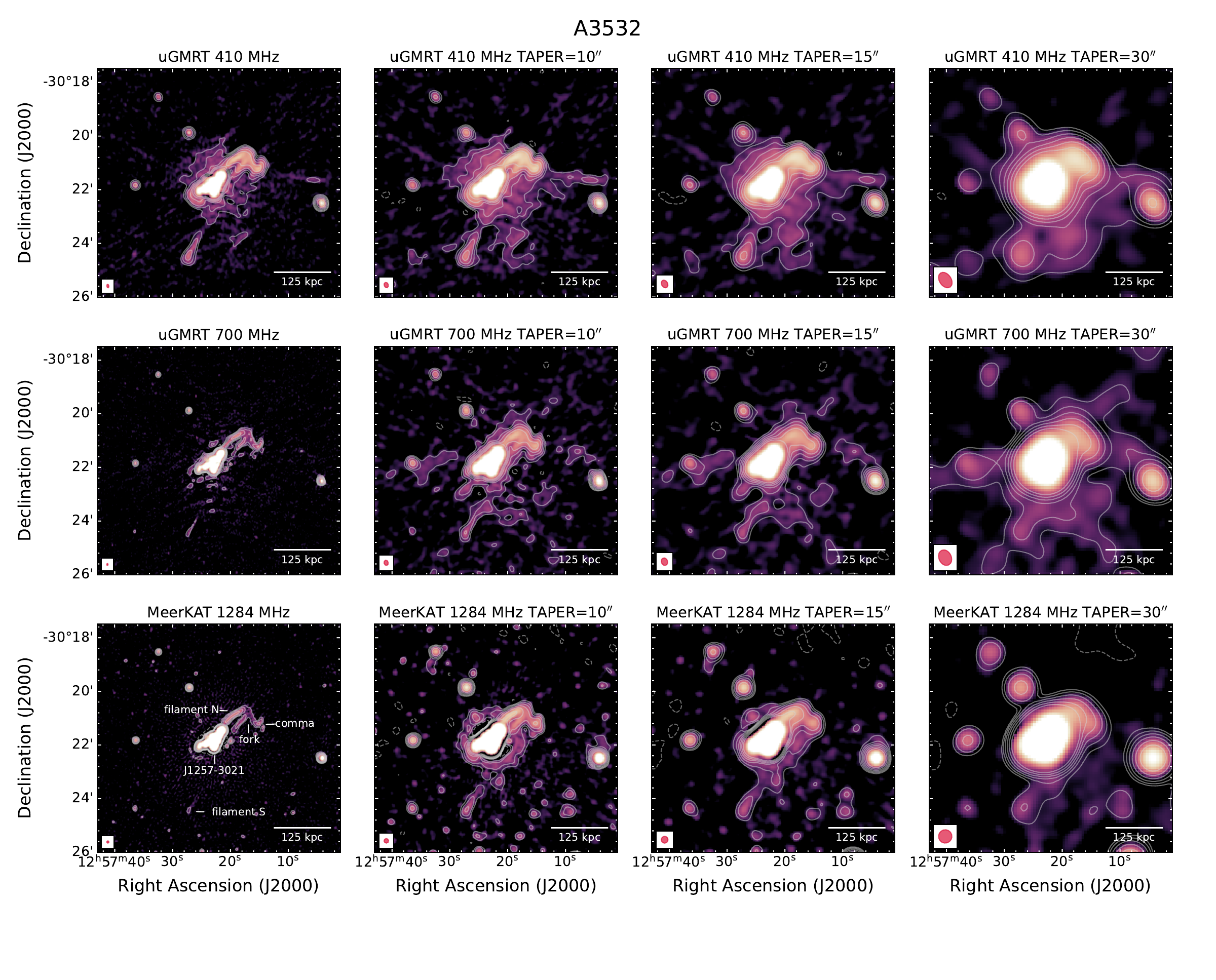}
\vspace{-15mm}
\caption{Full- (left column) and low-resolution (centre-left, centre-right and right columns) images of A3532. Top row: uGMRT Band 3; central row: uGMRT Band 4; bottom row: MeerKAT L-band.}\label{fig:images_A3532alltelescope_allres}
\end{figure*}

In this section, we show images at all the resolution available (i.e., from left to right columns: full resolution, {\tt taper=10\arcsec}, {\tt taper=15\arcsec} and {\tt taper=30\arcsec}) at each observing frequencies (from top to bottom: uGMRT Band 3, uGMRT Band 4 and MeerKAT) for A3528N (Fig. \ref{fig:images_A3528Nalltelescope_allres}), A3528S (Fig. \ref{fig:images_A3528Salltelescope_allres}) and A3532 (Fig. \ref{fig:images_A3532alltelescope_allres}).

\clearpage
\newpage

\section{Surface brightness $\beta$-model}\label{apx:betamodel}
\begin{figure}[h!]
\centering
\includegraphics[width=\textwidth]{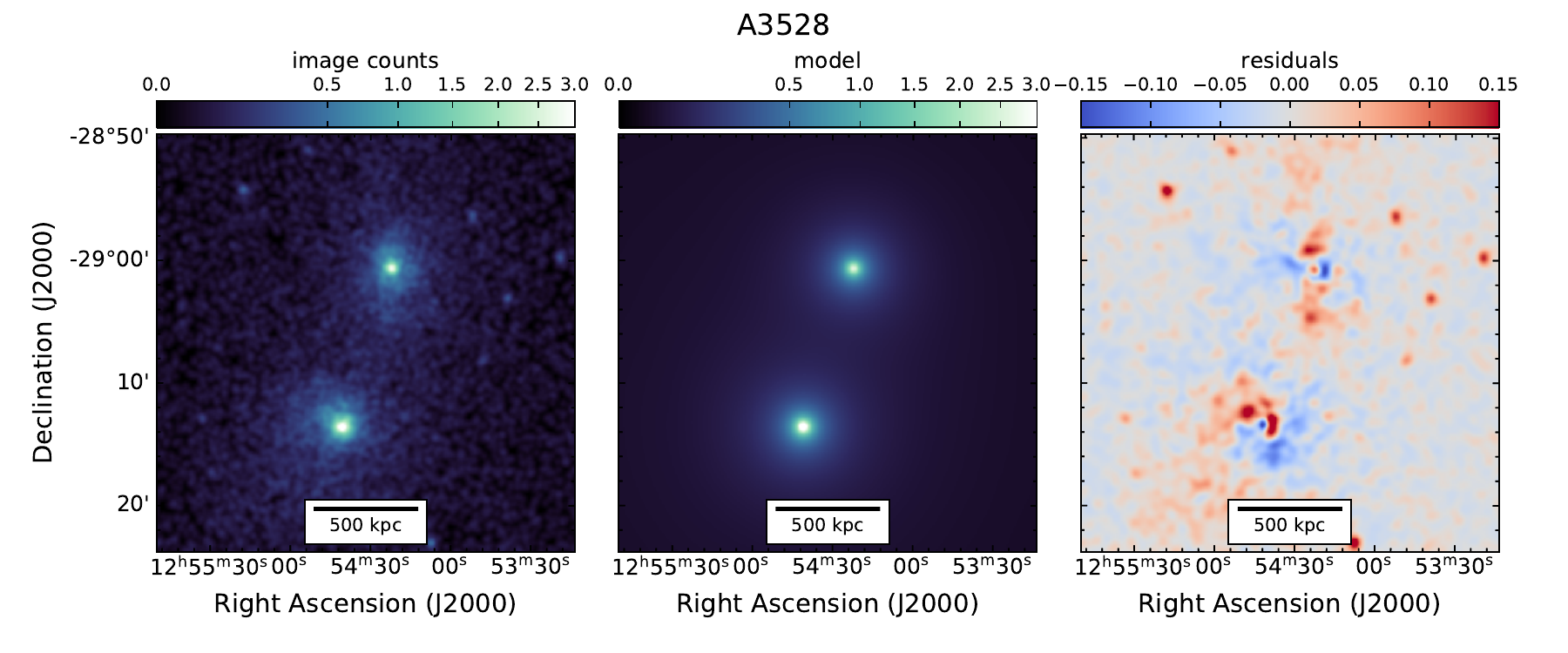}
\caption{eROSITA count image (left), $\beta$-model image (middle), and residual map (right) for A3528. Each of the two sub-clusters (A3528N and A3528S) have been modelled with two different $\beta$-models.}
\label{fig:A3528_beta}
\end{figure}

\begin{figure}[h!]
\centering
\includegraphics[width=\textwidth]{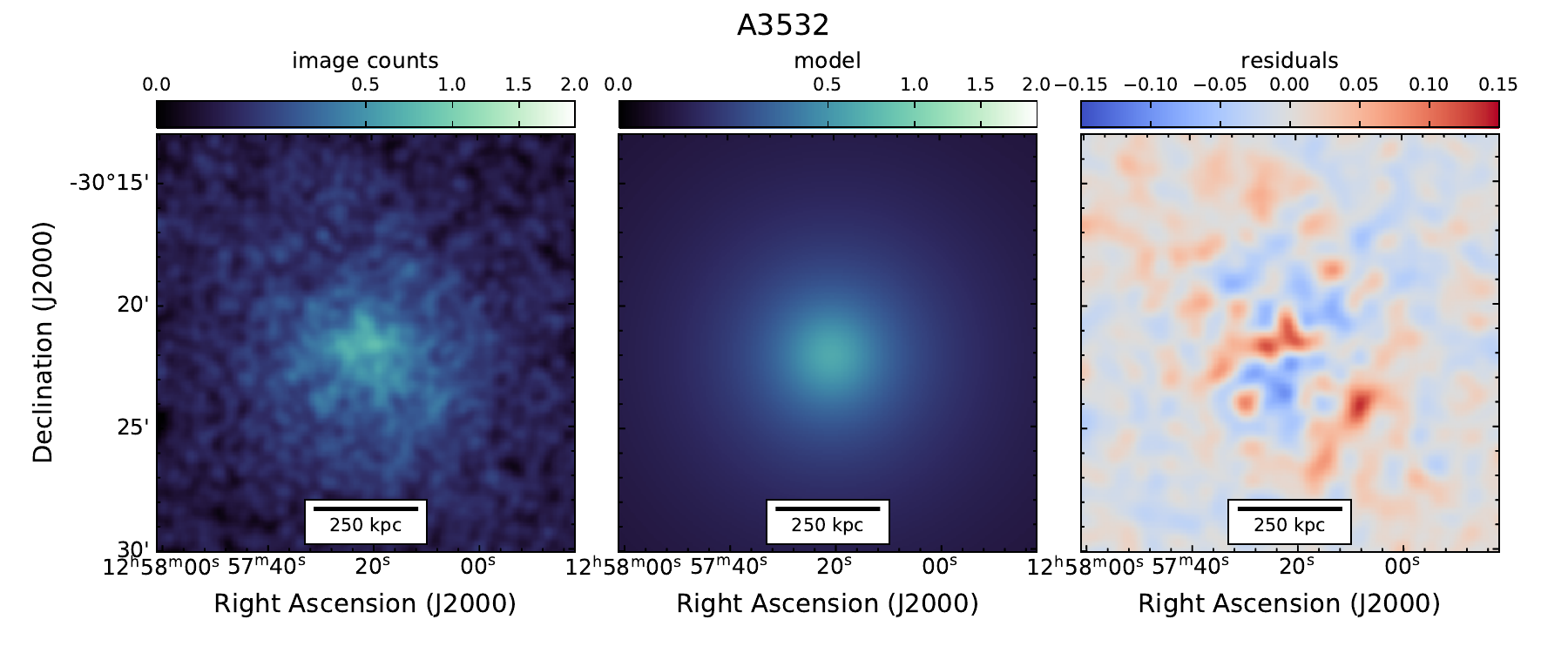}
\caption{eROSITA count image (left), $\beta$-model image (middle), and residual map (right) for A3532.}
\label{fig:A3532_beta}
\end{figure}

In this section, we show the eROSITA images, the assumed $\beta$-model and residual maps for A3528 (Fig. \ref{fig:A3528_beta}) and A3532 (Fig. \ref{fig:A3532_beta}).

According to the $\beta$-morel \citep{cavaliere+fusco-femiano76}, the cluster surface brightness $S_X$ is described as:
\begin{equation}
S_X = S_0 \left [ 1 + \left ( \frac{r}{r_c} \right )^2 \right ]^{-3\beta + 0.5} \, ,
\end{equation}
where $r_c$ is the core radius, and $S_0$ is the central surface brightness. The $\beta$ parameter determines the ratio of specific kinetic energies of galaxies and gas \citep{gitti+12}, and is defined as $\beta=\frac{\sigma^2_{r}}{k_BT/\mu m_p}$, where $\sigma_{r}$ is the line-of-sight velocity dispersion, $k_B$ the Boltzmann parameter, $T$ the ICM temperature, $\mu$ the mean molecular weight, and $m_p$ the proton mass.

\begin{figure}[h!]
\centering
\includegraphics[width=\textwidth]{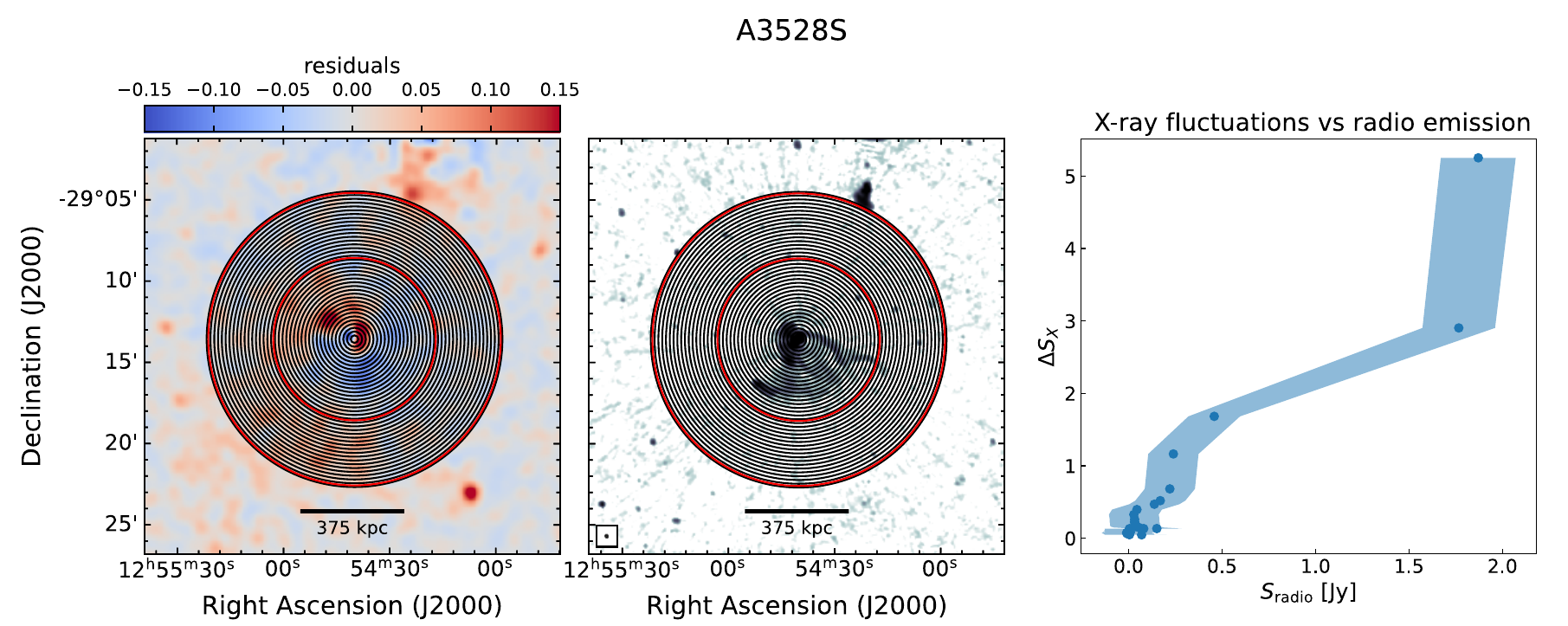}
\caption{Comparison of the X-ray surface brightness fluctuations and the radio uGMRT Band 3 emission in A3528S. Red annuli show the two regions, based on the radio emission, where we calculated the X-ray surface brightness residuals. Each black annulus has a width of $\sim12''$, i.e. the beam size of the radio observations.}
\label{fig:fluctuations}
\end{figure}

\newpage

\section{Spectral index uncertainty maps}
\label{apx:spixerr}

In this section, we present the spectral index uncertainty maps for the clusters in the \target.

\begin{figure*}[h!]
\centering
\includegraphics[height=0.32\textwidth]{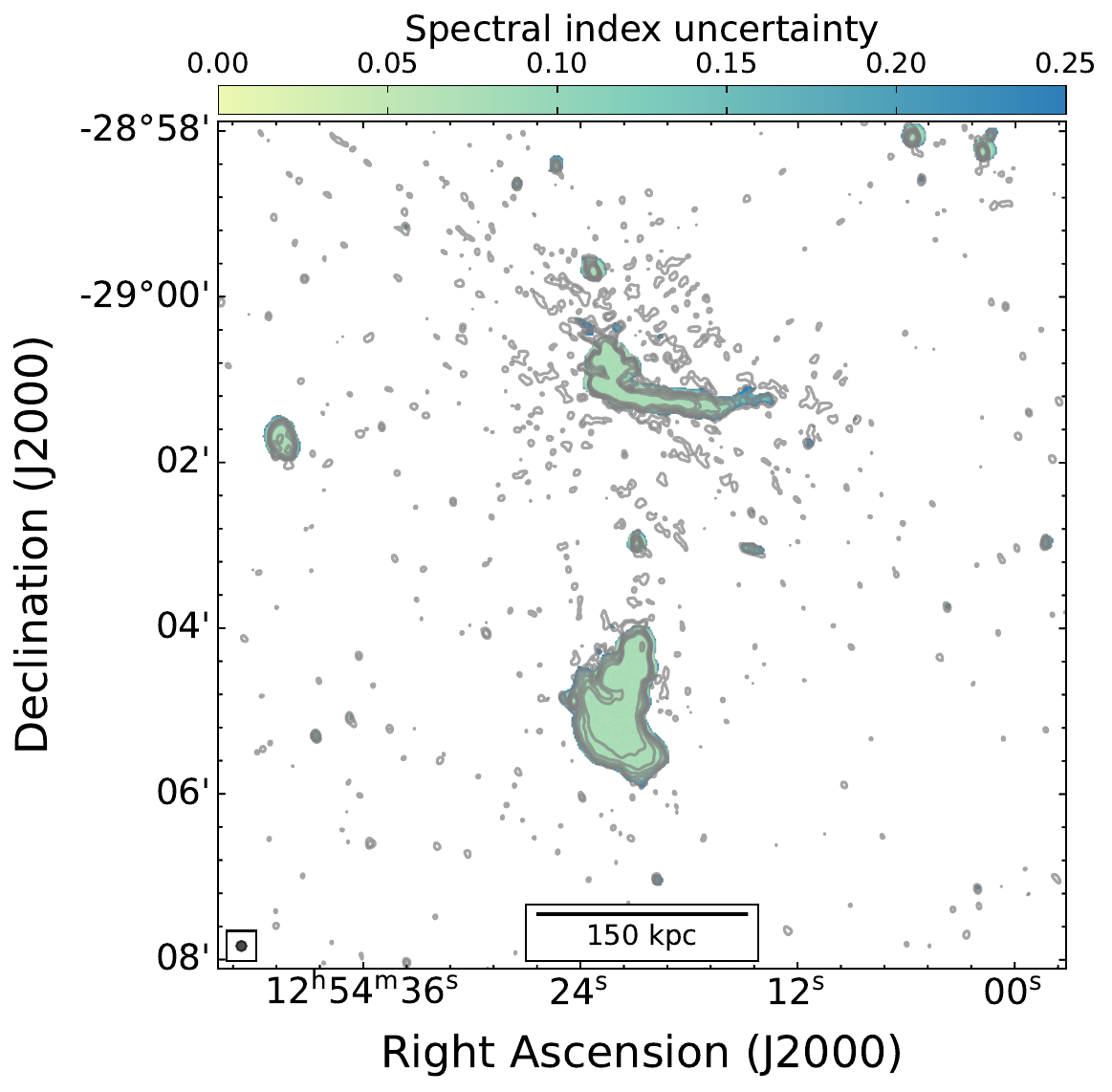}
\includegraphics[height=0.32\textwidth]{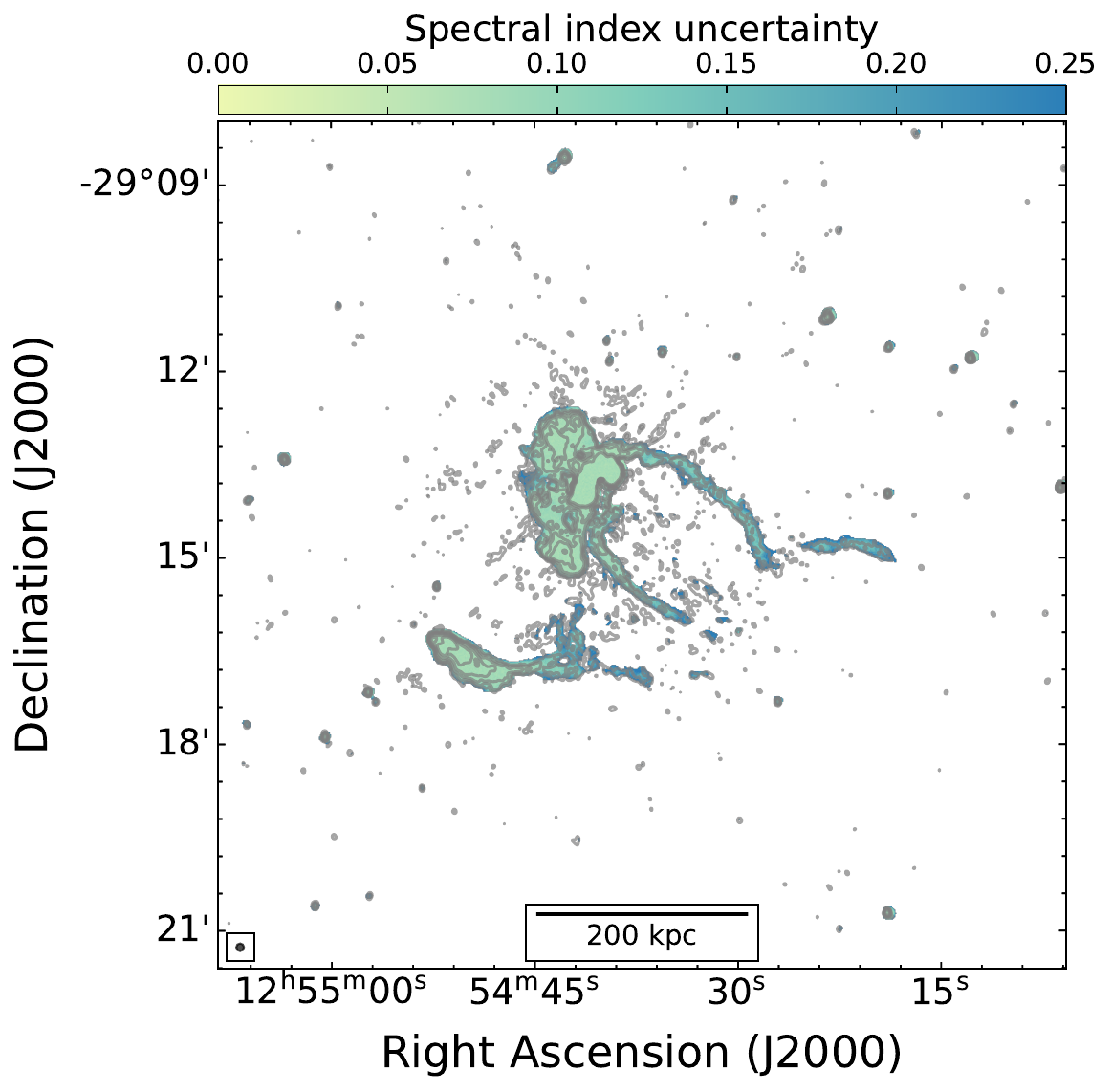}
\includegraphics[height=0.32\textwidth]{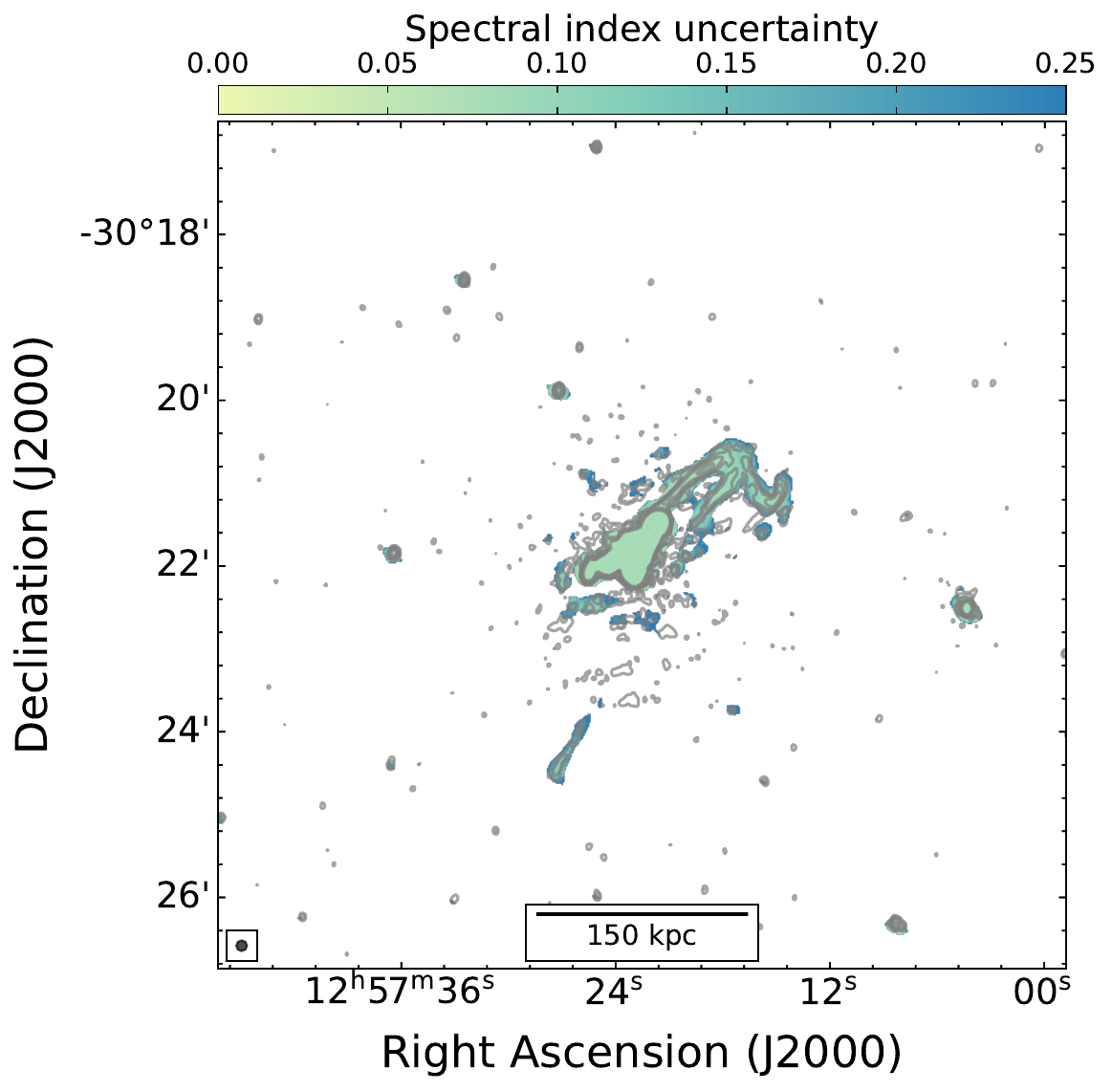} \\
\includegraphics[height=0.32\textwidth]{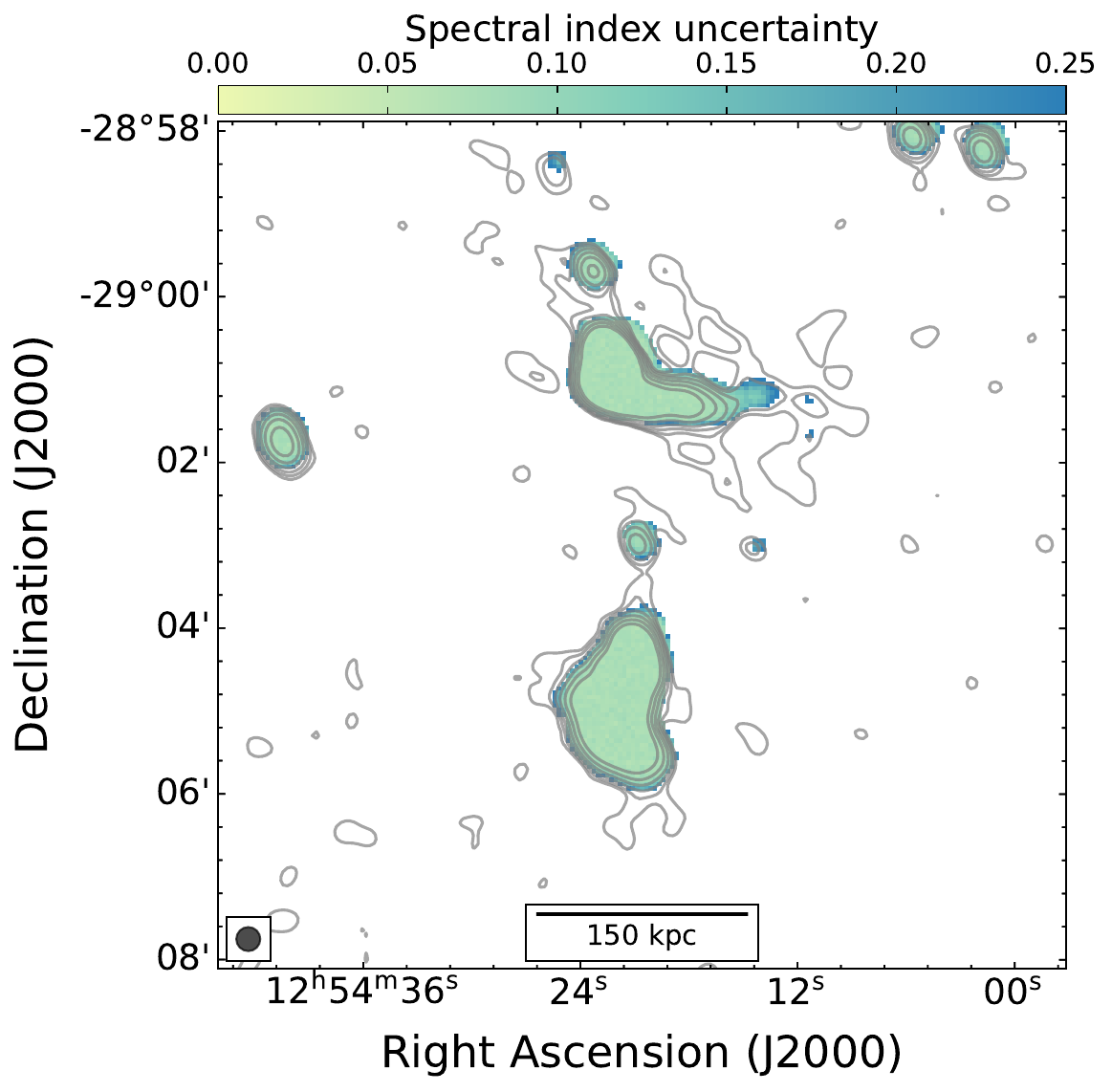}
\includegraphics[height=0.32\textwidth]{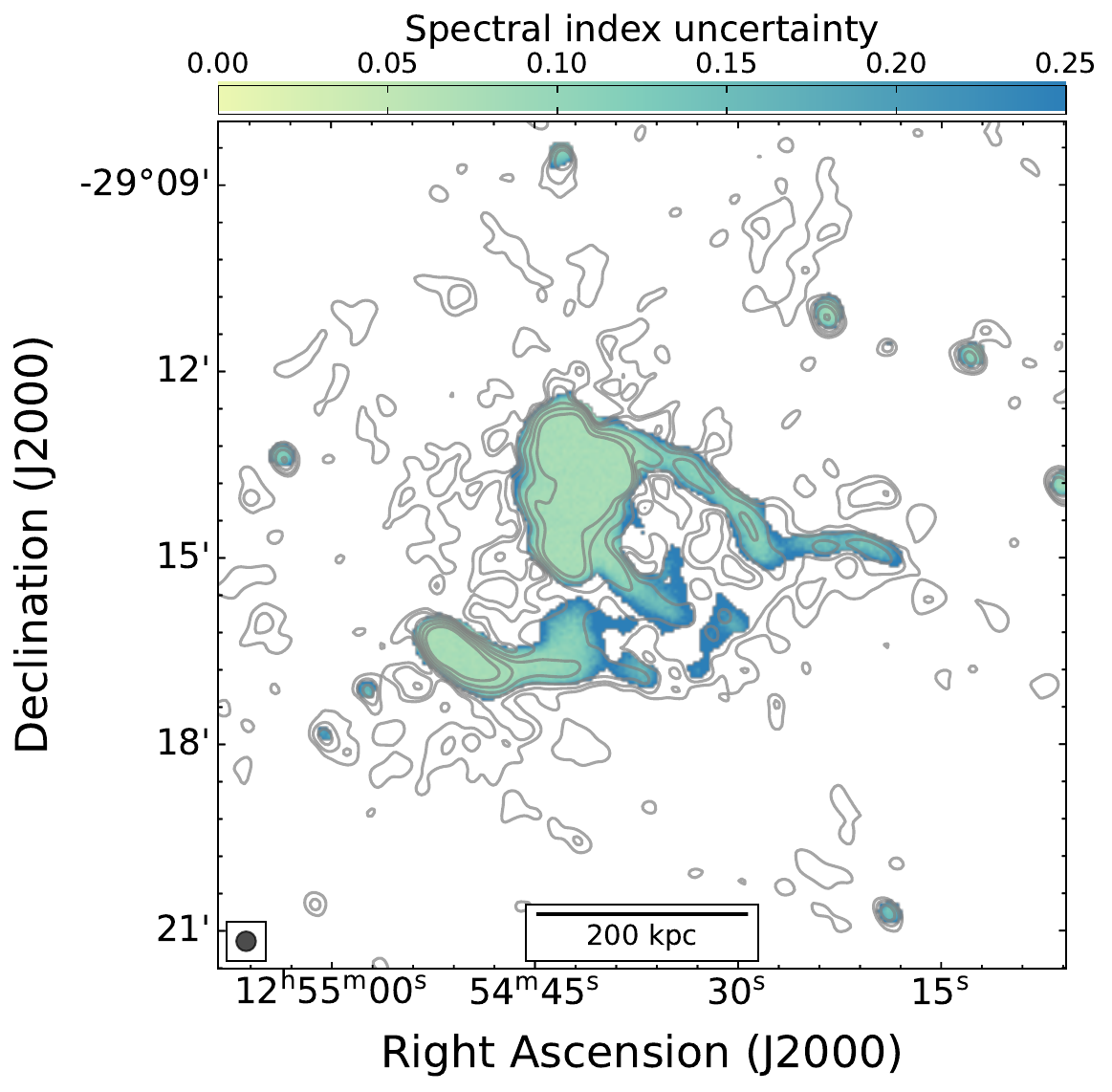}
\includegraphics[height=0.32\textwidth]{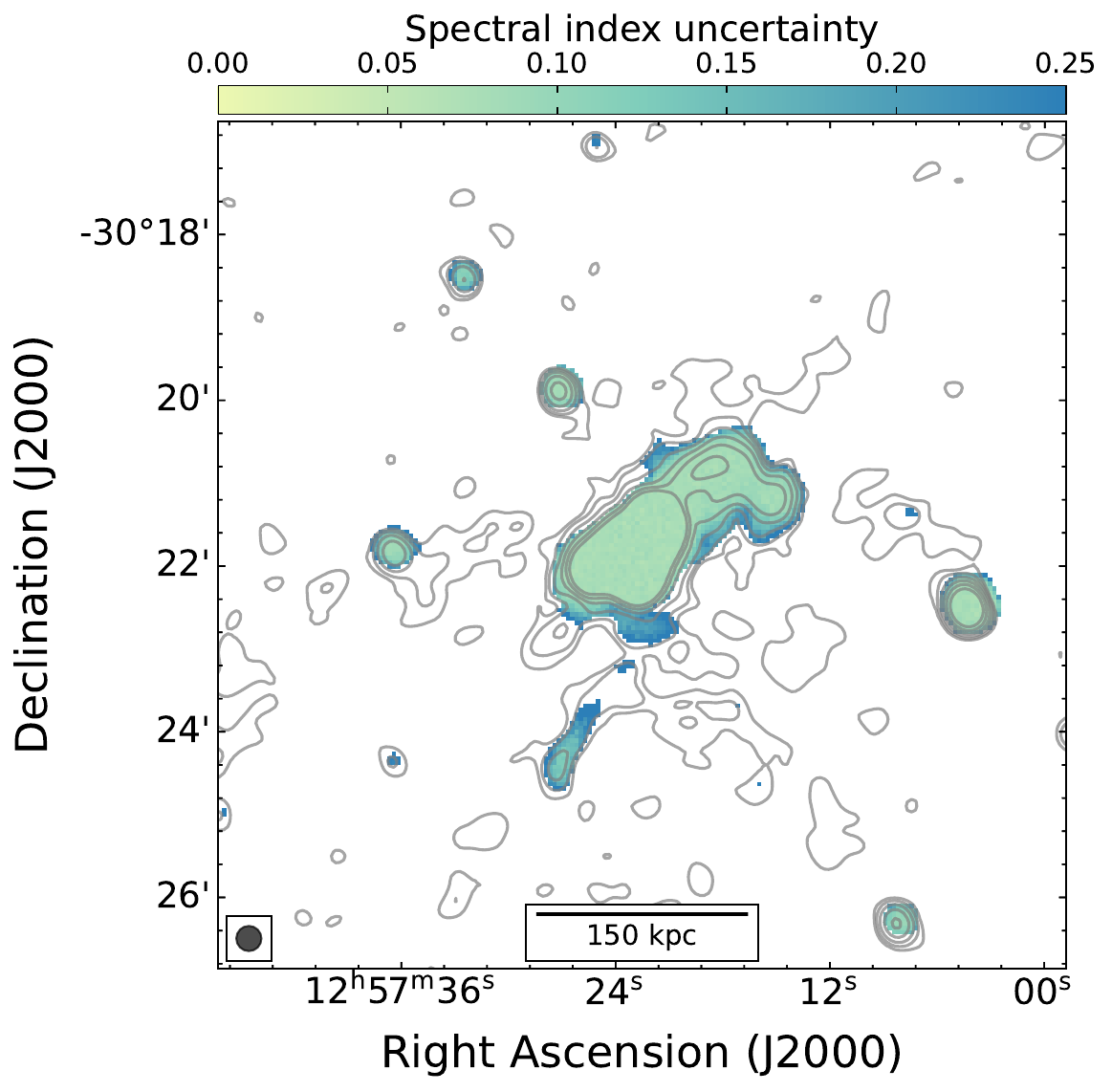}

\caption{Spectral index uncertainty maps for Fig. \ref{fig:spix}}\label{fig:spixerr}
\end{figure*}

\begin{figure*}
\centering
\includegraphics[height=0.32\textwidth]{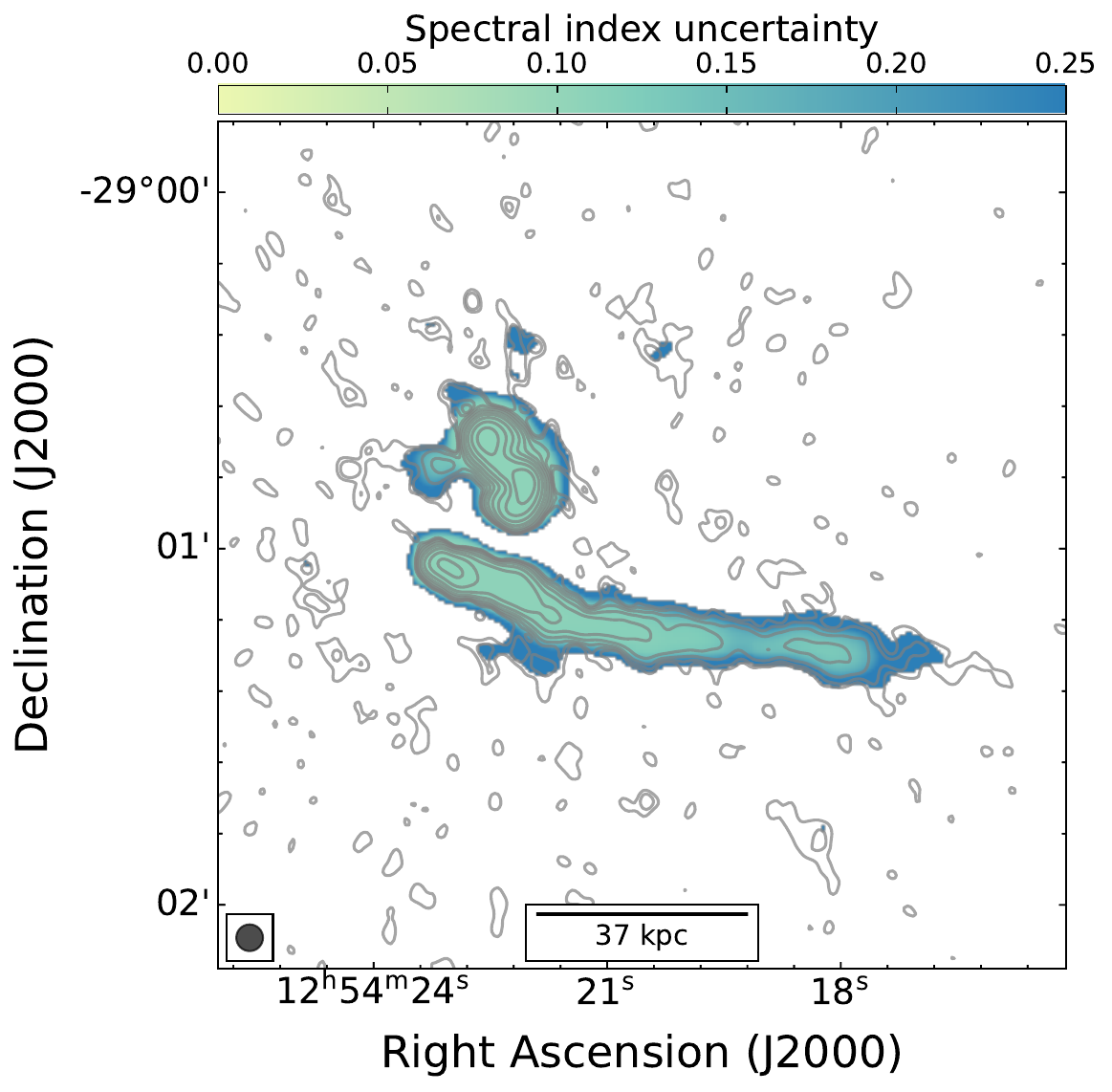}
\includegraphics[height=0.32\textwidth]{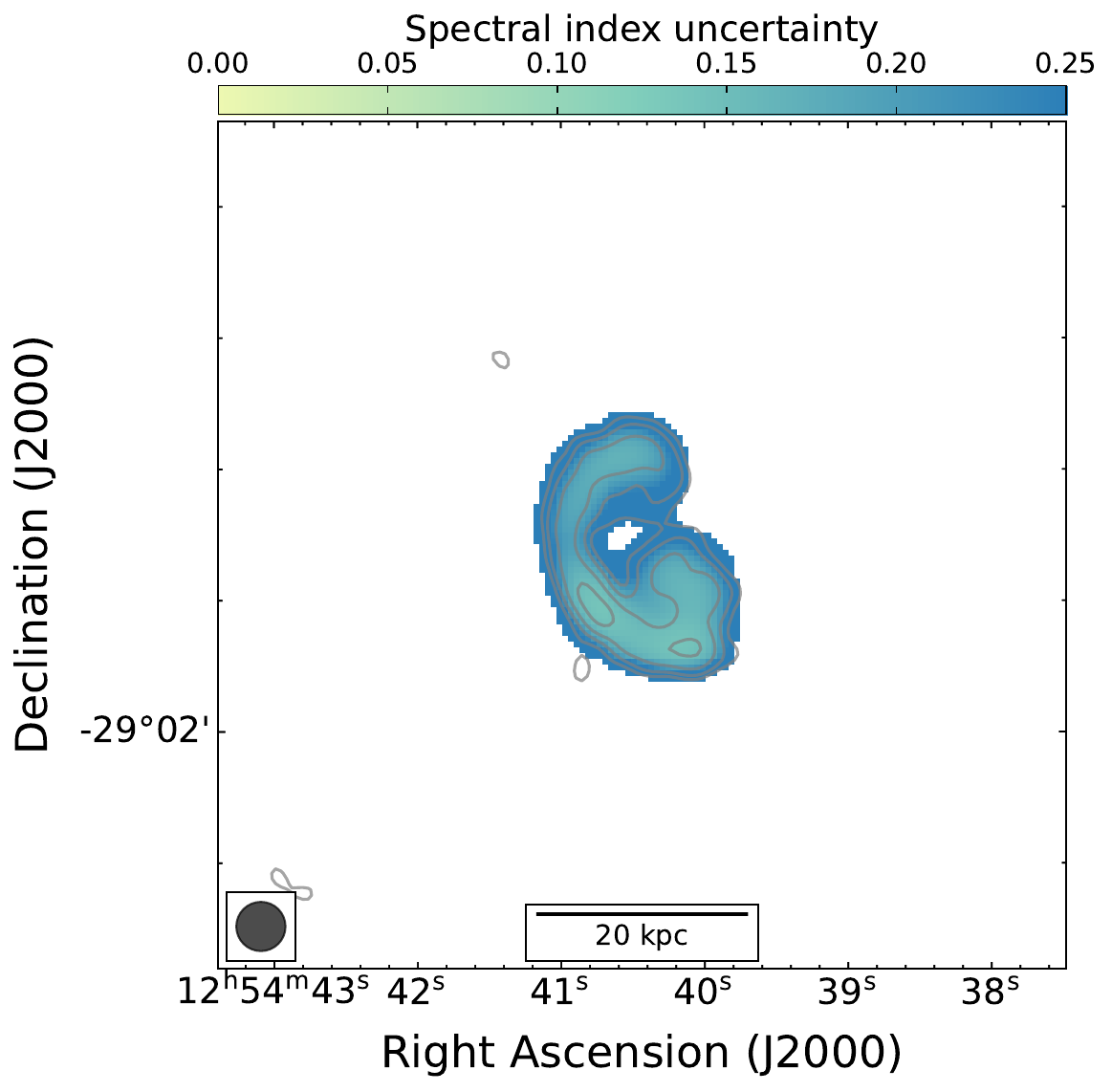}
\includegraphics[height=0.32\textwidth]{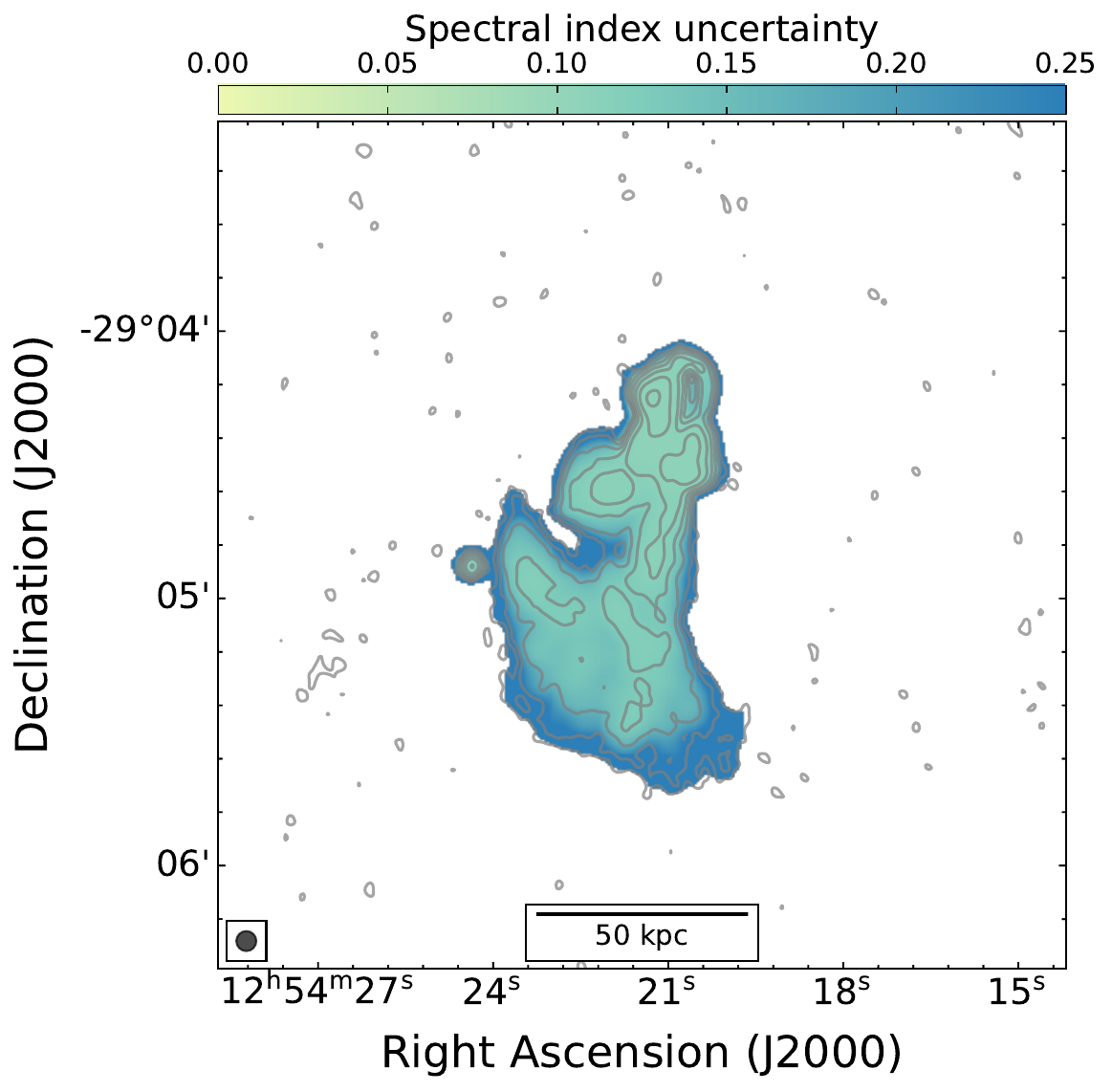} \\
\includegraphics[height=0.32\textwidth]{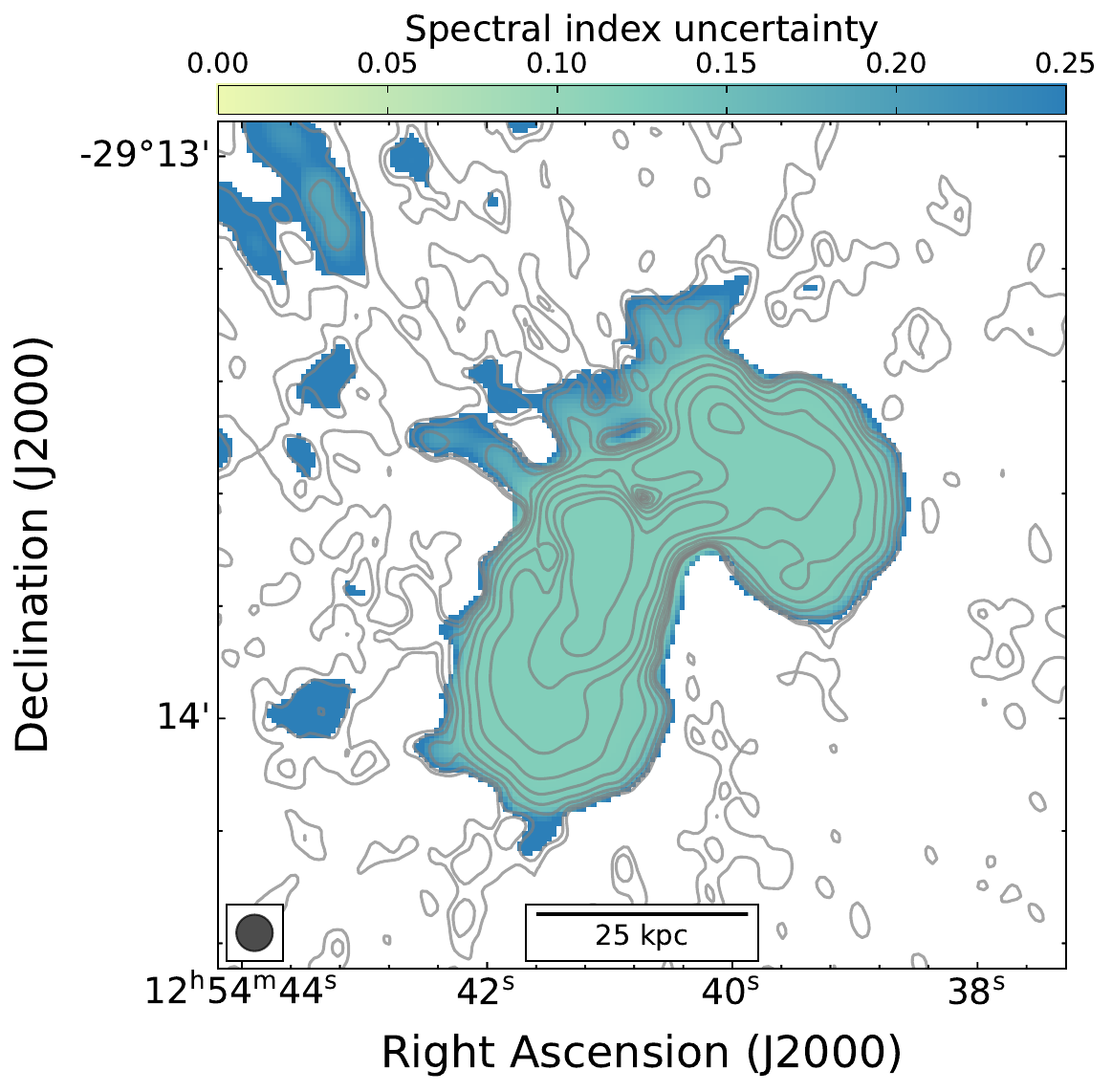}
\includegraphics[height=0.32\textwidth]{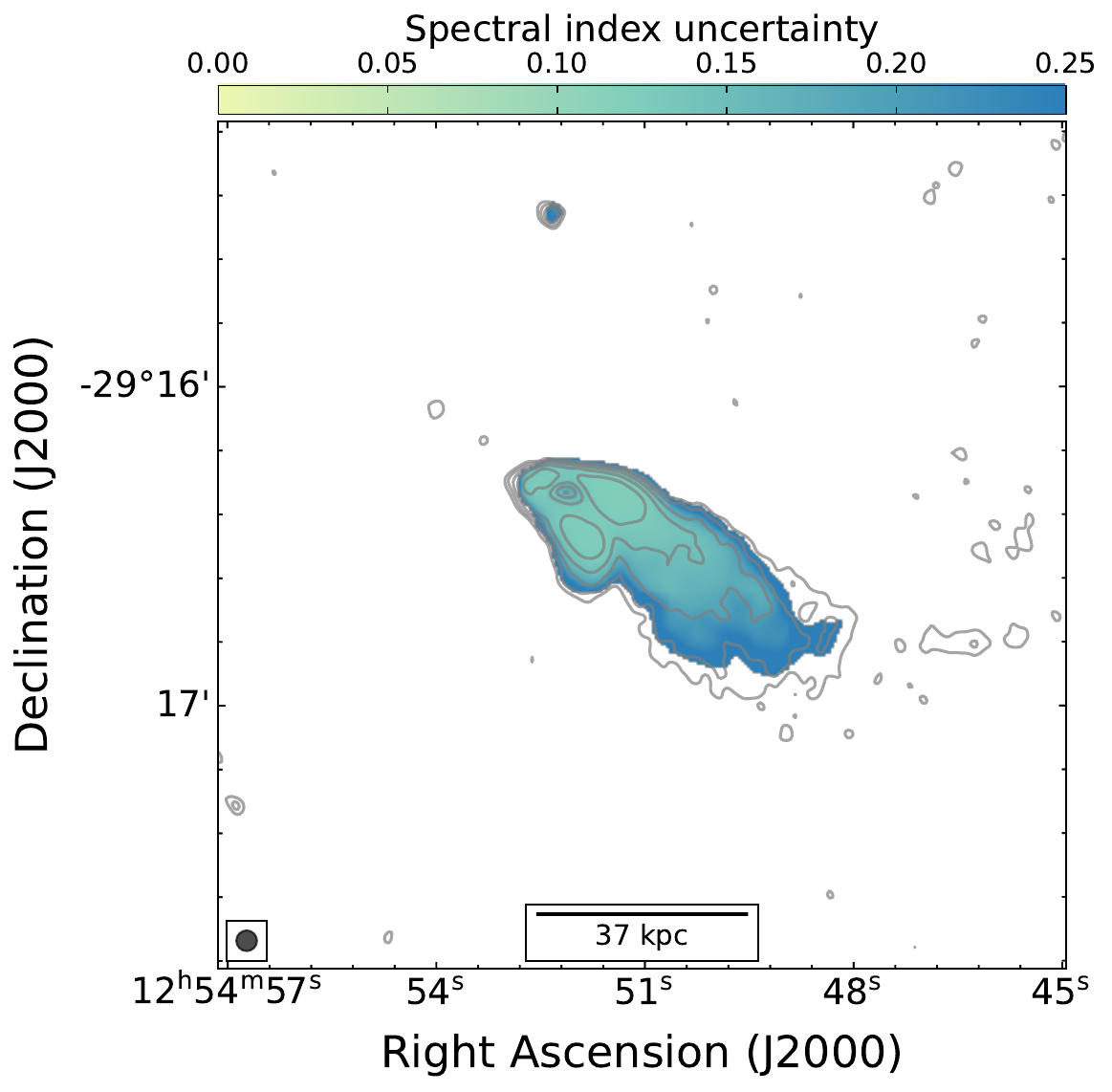}
\includegraphics[height=0.32\textwidth]{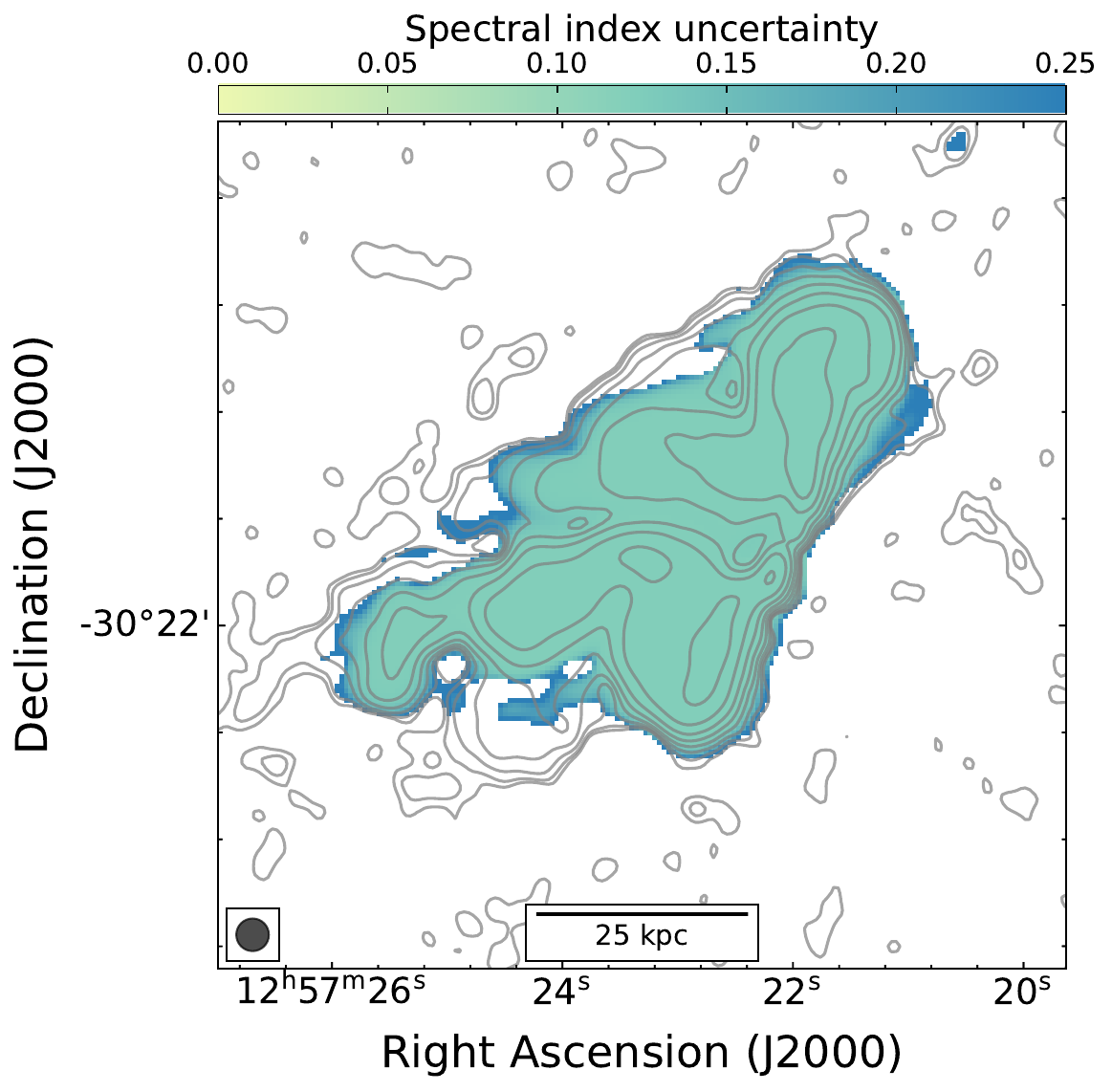}
\caption{Spectral index uncertainty maps for Fig. \ref{fig:spixRG}.}\label{fig:spixmaperrRG}
\end{figure*}

\section{{\tt Halo-FDCA} fitting results}\label{apx:minihalo}
In this section, we show the fitting result of the candidate mini-halo in A3528S, using {\tt Halo-FDCA} \citep{boxelaar+21}.

\begin{figure*}
\centering
\includegraphics[width=\textwidth]{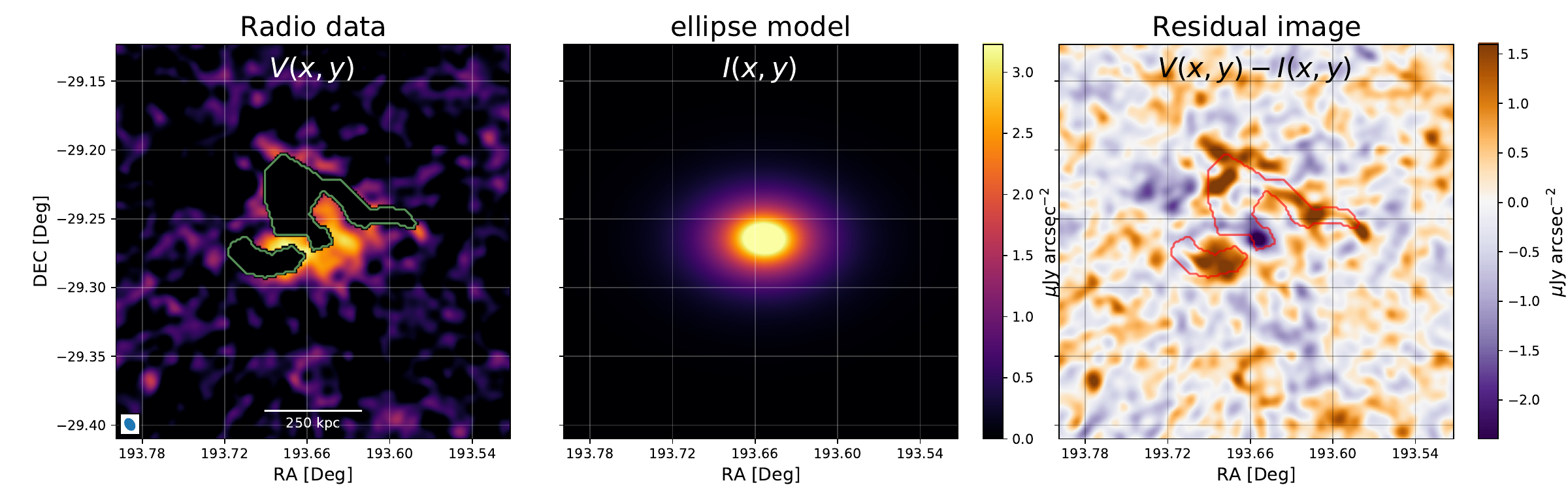}
\includegraphics[width=\textwidth]{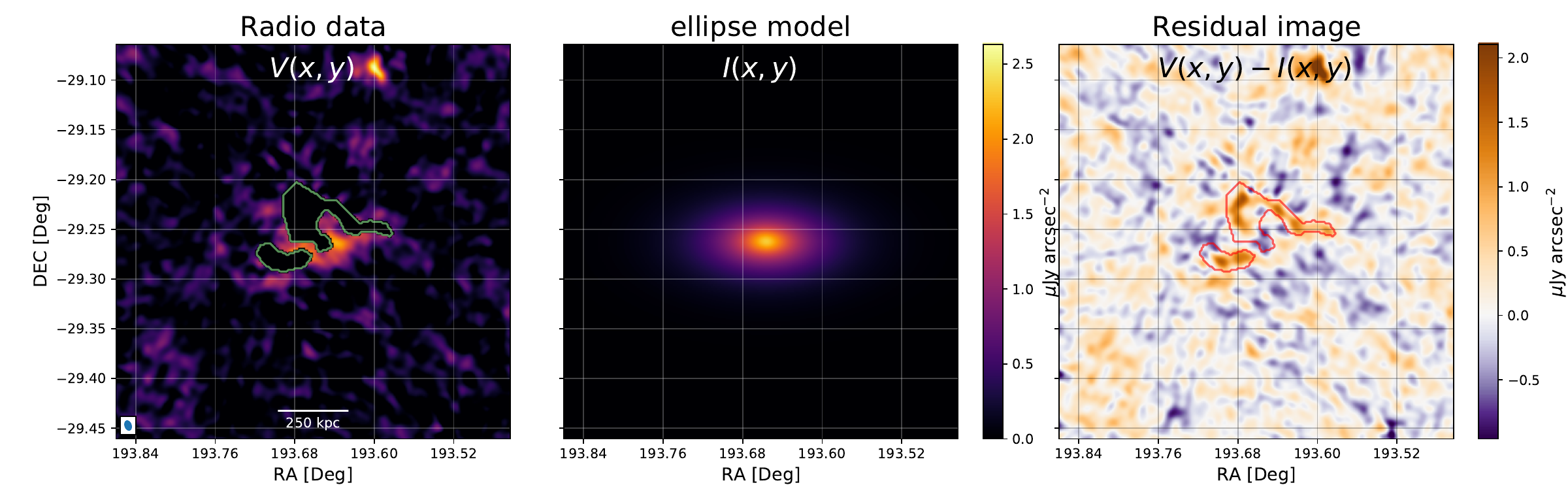}
\caption{Results of the mini-halo fitting (from left to right: radio data, model assumed for the radio emission, residual image). Top panels: uGMRT Band 3. Bottom panels: uGMRT Band 4. The green and red regions in the left and right columns are excluded from the fitting.}
\end{figure*}

\end{appendix}

\end{document}